\documentclass[11pt,letterpaper]{article}

\usepackage[margin=1.0in]{geometry} %
\usepackage{setspace}
\usepackage[english]{babel}
\usepackage[utf8x]{inputenc}
\usepackage{amsmath}
\usepackage{mathtools}
\mathtoolsset{showonlyrefs}
\usepackage{amssymb} 
\usepackage[retainorgcmds]{IEEEtrantools}
\usepackage{graphicx}
\usepackage{here} %
\usepackage{tabularx}
\usepackage{subfig}
\usepackage[T1]{fontenc} %
\usepackage{microtype} 
\usepackage{booktabs}   %
\usepackage{bm}         %
\usepackage{listings}   %
\usepackage{verbatim}   %
\usepackage{float}    %
\usepackage{color}  

\usepackage[hyphens]{url}
\usepackage[colorlinks=true, allcolors=blue]{hyperref}
\usepackage[colorinlistoftodos]{todonotes}
\usepackage{natbib}

\usepackage{dcolumn}
\usepackage{tikz}
\usepackage{authblk}
\usepackage{color,soul}
\usepackage{hyperref}
\usepackage{subfig}
\usepackage{natbib}
\usepackage{footmisc} %
\usepackage{url} %
\usepackage{soul,color}

\newenvironment{figurenotes}[1][Note]{%
  \vspace{0.5ex}%
  \begin{minipage}[t]{\textwidth}%
  \raggedright
  \footnotesize\textit{#1:}\ %
}{%
  \end{minipage}%
}

\usepackage{subfig}

\usepackage{amsthm}
\newtheorem{Proposition}{Proposition} %
\theoremstyle{definition}
\newtheorem{definition}{Definition}
\newcommand{\wtilde}[1]{\widetilde{#1}} %

\definecolor{mygreen}{rgb}{0.0, 0.62, 0.42}
\definecolor{myblue}{rgb}{0.16, 0.32, 0.75}
\definecolor{mygreen}{RGB}{28,172,0} 
\definecolor{mylilas}{RGB}{170,55,241}
\newcommand{\prn}[1]{\left({#1}\right)}
\newcommand{\mprn}[1]{\{{#1}\}}

\definecolor{mygreen}{rgb}{0.0, 0.62, 0.42}
\definecolor{myblue}{rgb}{0.16, 0.32, 0.75}
\definecolor{mypurple}{rgb}{0.44, 0.07, 0.99}
\definecolor{myred}{rgb}{0.81, 0.01, 0.11}

\definecolor{matlabblue}{rgb}{0,0.4470,0.7410}
\definecolor{matlaborange}{rgb}{0.8500,0.3250,0.0980}
\definecolor{matlabyellow}{rgb}{0.9290,0.6940,0.1250}

\begin{document}
\title{Trade Policy and Structural Change\thanks{We would like to thank Daisuke Adachi, Md Moshi Ul Alam, Cheng Chen, Terry Cheung, Jonathan Eaton, Markus Epp,  Marten Hillebrand, Julian Hinz, Hirokazu Ishise, Danial Lashkari, Oliver Landmann, Andrea Lassman, Kiminori Matsuyama, Javier Mendoza, Sanghwa Moon, Philip Saur\'e, Yuta Suzuki, Han Yang, Joschka Wanner, Yuta Watabe and seminar participants at Clark University, JGU Mainz, Bielefeld University, University of Freiburg, Nagoya University, College of the Holy Cross, G\"ottinger Workshop, Midwest Economic Association Annual Conference, Midwest International Trade Conference, CITP Conference, Kansai Macro Workshop, Yonsei University, APTS (Chulalongkorn, Tokyo, and HKUST), Academia Sinica, the German Council of Economic Experts,  ETSG (Athens), FIW-Conference International Economics, ATW (Brisbane), IEFS Japan, Midwest Macroeconomics Meeting, and Danish International Economics Workshop for their helpful comments. This work is supported by the Joint Usage/Research Center, Institute of Economic Research, Hitotsubashi University (IERPK2321, IERPK2417, IERPK2527); the Japan Society for the Promotion of Science (JP20H01495, JP24H00014); and Clark University Start-Up Research Fund. In the early stage of this project, we made use of the Basic Survey on Overseas Business Activities, the Basic Survey of Japanese Business Structure and Activities, and the Census of Manufacture, all conducted by the Ministry of Economy, Trade and Industry (METI), although the present paper does not report any results based on these data. We are grateful to Edward Nyarko, Zeyi Qian, and Li Shen for the outstanding research assistance.}}
\author{Hayato Kato\thanks{Institute of Developing Economies (IDE-JETRO); CESifo Research Network. Email: \href{mailto:hayato_kato@ide.go.jp}{hayato\_kato@ide.go.jp}} \qquad Kensuke Suzuki\thanks{Department of Economics, Clark University; Economic Research Center, Nagoya University. Email: \href{mailto:KSuzuki@clarku.edu}{KSuzuki@clarku.edu}} \qquad Motoaki Takahashi\thanks{Graduate School of Economics, the University of Osaka. Email: \href{motoaki.takahashi@econ.osaka-u.ac.jp}{motoaki.takahashi@econ.osaka-u.ac.jp}}}
\date{\today}

\maketitle

\begin{abstract}
    We study how tariffs affect industrial structure and welfare in an economy where sectors are complements and preferences are nonhomothetic---two drivers of structural change. Tariffs reshape sectoral composition through relative prices, income effects, and sectoral net exports. We characterize these mechanisms analytically and quantify them in a dynamic multi-country model with capital accumulation and input-output linkages. A counterfactual 20-percentage-point increase in U.S. manufacturing tariffs raises the manufacturing value-added share by about one percentage point and increases U.S. welfare by 0.43 percent, while lowering welfare abroad; retaliation would make all countries worse off. The optimal unilateral U.S. manufacturing tariff is 20.9 percent. Comparing different preference specifications, we show that homothetic preferences without income effects overstate the U.S. welfare gains from unilateral tariff increases.
({\it JEL} {\normalfont F11, F13, O41}) 

\end{abstract}

\let\oldaddcontentsline\addcontentsline
\renewcommand{\addcontentsline}[3]{} %
\section{Introduction}\label{sec: intro}

Trade policy as a means of protecting domestic industries has attracted renewed attention. %
On February 1, 2025, the U.S. President signed executive orders imposing additional tariffs of 25 percentage points on most imports from Canada and Mexico and 10 percentage points on imports from China. A subsequent proclamation on April 2, 2025, introduced a 10 percent baseline additional tariff on imports from virtually all countries.\footnote{See the White House statements, ``Fact Sheet: President Donald J. Trump Imposes Tariffs on Imports from Canada, Mexico, and China'': \url{https://www.whitehouse.gov/fact-sheets/2025/02/fact-sheet-president-donald-j-trump-imposes-tariffs-on-imports-from-canada-mexico-and-china/} and ``Fact Sheet: President Donald J. Trump Declares National Emergency to Increase our Competitive Edge, Protect our Sovereignty, and Strengthen our National and Economic Security'': \url{https://www.whitehouse.gov/fact-sheets/2025/04/fact-sheet-president-donald-j-trump-declares-national-emergency-to-increase-our-competitive-edge-protect-our-sovereignty-and-strengthen-our-national-and-economic-security/} (accessed November 11, 2025).}
These measures reflect the current U.S. administration's attempts to reshape trade relations in favor of American workers and manufacturers, who have faced growing competition from developing countries in the global market (\citealp{GoldbergReed2023}).\footnote{In the U.S. context, Mexico and China played a major role. For example, \cite{HakobyanMcLaren2016} find that the North American Free Trade Agreement (NAFTA) significantly reduced wage growth for blue-collar workers in industries and regions most exposed to Mexican import competition from 1990 to 2000 (see also \citealp{Choietal2024} for the negative effect of NAFTA on employment growth). \cite{Acemogluetal2016} report that 2.0 to 2.4 million U.S. manufacturing workers lost their jobs due to Chinese import competition from 1999 to 2011 (see also \citealp{PierceSchott2016}).}

However, the shrinking of manufacturing is not unique to the U.S.; it is ubiquitous in major developed countries. From a long-term perspective, the average share of manufacturing in domestic value-added (manufacturing value-added share, henceforth) among G7 nations declined from 20.1 to 14.6 percent between 1965 and 2014, while the service value-added share increased from 59.1 to 78.2 percent over the same period.\footnote{Over the period from 1965 to 2014, the average manufacturing value-added share among four major non-resource-dependent emerging economies, China, India, South Korea, and Taiwan, increased from 16.3 to 24.1 percent (the authors' calculations using the World Input-Output Database). The sectoral value-added we refer to throughout this paper is the sectoral nominal GDP adjusted by taxes less subsidies on products (\citealp[pp.7--8]{Woltjer2021}).} The shift of resources from manufacturing to services within a country occurs even in the absence of international trade, a process known as structural change (\citealp{Kuznets1973}).

Can tariffs effectively mitigate the decline of manufacturing in a country undergoing structural change? In addition to their conventional role in trade protection, how do tariffs interact with key drivers of structural change such as sectoral complementarity (\citealp{NgaiPissarides2007}) and nonhomothetic preferences (\citealp{Kongsamutetal2001})?
Tariffs on manufacturing raise its relative price relative to agriculture and services. If sectoral goods are gross complements in consumption, the higher relative price of manufacturing may bias expenditure toward manufacturing and shift resources away from agriculture and services. On the other hand, tariffs generate government revenue, bring terms-of-trade gains, and may raise overall income.\footnote{Following the current U.S. administration's tariff increases on nearly all imports, U.S. tariff revenue rose sharply in fiscal year 2025, reaching about 210 billion USD, about 2.8 times its fiscal year 2024 level of about 76 billion USD (\citealp[pp.61--62]{USTreasury2026}). U.S. tariff revenue in 2025 was about 44 percent of its corporate income tax revenue and about five percent of its individual income tax revenue.} This shifts demand from less income-elastic sectors, such as agriculture and manufacturing, toward more income-elastic sectors, such as services. The two effects of tariffs on manufacturing---through changes in relative prices (\textit{relative price effect}) and through changes in income (\textit{income effect})---may work in opposite directions.

Our primary goal is to qualitatively and quantitatively evaluate the relative price and income effects of tariffs on sectoral composition and welfare.
Our static model for theoretical analysis consists of two countries and three sectors and features trade based on Ricardian comparative advantage (\citealp{Eaton2002}) and the (isoelastically) nonhomothetic constant elasticity of substitution (CES) preferences (\citealp{CLM}; \citealp{Matsuyama2019}). The nonhomothetic CES neatly delineates the income effect from the relative price effect of tariffs.\footnote{As we will see, nonhomothetic CES preferences allow for sector-specific parameters ($\epsilon^j$) capturing differences in income elasticity across sectors, while treating separately the parameter ($\sigma$) that captures the (constant) elasticity of substitution. Another advantage of nonhomothetic CES preferences is that, unlike Stone-Geary preferences, they allow income elasticity differences across sectors to persist even among rich countries or households. This feature helps explain empirical observations (\citealp{BueraKaboski2009}; \citealp{CLM}; \citealp{Alderetal2022}).}

In the presence of sectoral complementarity alone, a marginal increase from zero in a country's manufacturing tariff raises the expenditure and value-added shares of manufacturing, while lowering those of agriculture and services, through the relative-price effect.\footnote{In our two-country model, a sector's value-added share equals its expenditure share plus sectoral net exports, minus sectoral tariff revenues, with the latter two terms normalized by labor income.} The real income of the tariff-imposing country rises, while that of the other country falls.

Nonhomotheticity makes the effects of a manufacturing tariff more nuanced by introducing the income effect. In particular, if the tariff-imposing country is at an advanced stage of structural transformation,\footnote{More specifically, this corresponds to country $n$ having an expenditure-weighted income elasticity above one, $\overline\epsilon_n = \sum_{j=a,m,s} \omega_n^j \epsilon^j>1$, where $\omega_n^j$ is country $n$'s expenditure share in sector $j$, and $\epsilon^j$ is the parameter governing the income elasticity of demand in sector $j$ (Proposition~\ref{prop: proposition 1}).} the manufacturing expenditure and value-added shares respond less positively, because an increase in real income shifts expenditure away from agriculture and manufacturing toward services. Moreover, the real-income gains from the manufacturing tariff are smaller under nonhomothetic CES preferences, where the income effect is present, than under homothetic CES preferences, where it is absent. The reason is that an increase in real income raises service demand disproportionately, thereby increasing average expenditure per unit of real consumption, i.e., the aggregate price index.\footnote{The aggregate price index is defined as aggregate expenditure $E$ divided by real consumption $C$, i.e., $P=E/C$. Under homothetic CES preferences, $P$ is independent of $C$. Under nonhomothetic CES preferences, however, $P$ may increase with $C$, because higher $C$ following a tariff hike can raise spending on more income-elastic sectors and thereby increase aggregate expenditure $E$ more than proportionately. The details are discussed in Section~\ref{sec: tariff&cons}.} In sum, ignoring nonhomotheticity may lead to an overstatement of the tariff-induced expansion of protected sectors and of welfare gains.

To quantitatively evaluate whether these qualitative mechanisms are plausible, we extend the two-country static model to a multi-country dynamic framework incorporating endogenous capital accumulation and input-output linkages. We bring the model to the data for the world economy, encompassing 24 countries over the period from 1990 to 2014. We calibrate the model's fundamentals, such as sectoral productivity and non-tariff trade barriers, which allows us to solve for the transition paths of the economy in terms of \textit{levels}, not in \textit{relative changes} typically referred to as the exact hat-algebra method \citep{Dekle2008, CaliendoParro2015, Caliendoetal2019}. We then conduct a counterfactual experiment of a 20-percentage-point increase in U.S. tariffs applied to manufacturing imports from all countries, effective starting in 2001.\footnote{This exercise of a permanent tariff increase reflects the persistence of such policies, e.g., tariffs imposed during the first term of the Trump administration (2017--2020) remained in place throughout the Biden administration (2021--2024) (\citealp{Huetal2025}).} We select the year 2001 because the U.S. manufacturing value-added share experienced its largest annual decline in that year over our sampled period, dropping sharply by over two percentage points (from 14.8 percent in 2000 to 12.4 percent in 2001).\footnote{Additionally, 2001 marked China's accession to the World Trade Organization (WTO) and the beginning of George W. Bush's first term.}

We find that tariffs alter sectoral composition in a manner consistent with the standard trade protection argument and the relative price effect. Specifically, compared with the baseline equilibrium, a 20-percentage-point increase in the U.S. manufacturing tariffs since 2001 leads to a 6.4--11.2 percent (0.9--1.4 percentage points) increase in the manufacturing value-added share in each year from 2001 to 2014, and service value-added share falls by 1.1--1.4 percent (0.8--1.1 percentage points).\footnote{We conduct the same counterfactual exercise under homothetic CES preferences and find that the increase in U.S. manufacturing value-added share is 0.02--0.04 percentage points higher than the one in the benchmark case of nonhomothetic CES, due to the absence of the income effect.} Although these results might seem encouraging for trade protectionists, the U.S. welfare increases by only 0.43 percent. The U.S. welfare gains, however, come at the expense of the other countries. Canada experiences the largest loss, with a 1.58 percent decline in welfare. Furthermore, if trading partners retaliate with equally high tariff increases on U.S. manufacturing exports, U.S. welfare drops by 0.17 percent.

We also find that nonhomothetic preferences matter for evaluating the welfare impact. When conducting the same counterfactual exercise under homothetic CES preferences, the U.S. welfare increases by 0.49 percent, 14 percent (0.06 percentage point) higher than in the benchmark case of nonhomothetic CES. In addition to considering a marginal tariff increase from zero, we quantitatively compute the unilateral optimal time-invariant tariff rate for the U.S. under the two preference specifications, obtaining 20.4 percent under homothetic CES preferences and 20.9 percent under nonhomothetic CES preferences.
We find a similar pattern at the optima: the welfare gain for the U.S. is 14 percent larger under homothetic CES preferences than under nonhomothetic CES preferences.
As suggested by our theoretical analysis, this disparity arises because homothetic CES preferences shut down the income effect and therefore understate the negative effect of a higher aggregate price index following a tariff increase in countries at an advanced stage of structural transformation. The quantitative assessment based on widely used homothetic CES preferences may overestimate the gains from unilateral tariff policy. Although nonhomotheticity may not matter for whether tariffs protect manufacturing, it matters for assessing the welfare gains from such protection.

Our paper relates to two strands of the literature. The first is a growing body of studies on structural change and trade (\citealp{Alessandriaetal2023}; \citealp{GollinKaboski2023} for recent surveys).\footnote{\cite{Herrendorfetal2014} and \cite{DonovanSchoellman2023} survey the literature on structural change in the closed-economy context. Recent studies highlight the roles of domestic transportation infrastructure (\citealp{FajgelbaumRedding2022}; \citealp{CheungYang2024}; \citealp{Kaboskietal2026}), population aging (\citealp{Cravinoetal2022}), and schooling (\citealp{Porzioetal2022}; \citealp{Cheung2023}).} These studies offer a number of new insights such as the impact of trade on the skill premium (\citealp{CravinoSotelo2019}), the deepening intermediate-input intensity as economies develop (\citealp{Sposi2019}; \citealp{FarrokhiPellegrina2023}), the decomposition of different mechanisms for declining manufacturing share (\citealp{Swiecki2017}; \citealp{Smitkova2024}), the interaction between host and home countries of multinationals (\citealp{Alviarezetal2022}), and the joint evolution of service expenditure and trade openness (\citealp{Lewisetal2022}; \citealp{Bonadioetal2025}; \citealp{Hanetal2025}; \citealp{Lee2026}).

Closest to our paper are \cite{Matsuyama2019} and \cite{Sposietal2026}. \cite{Matsuyama2019} analytically characterizes how trade cost reductions affect sectoral composition and welfare using a two-country model of intra-industry trade. \cite{Sposietal2026} develop a quantitative multi-country model embedding capital accumulation and input-output linkages to explore the determinants of the hump-shaped path of manufacturing share in the course of development.\footnote{See also \cite{FujiwaraMatsuyama2024} for an analytical approach to this issue.} While these studies treat tariffs as a part of trade costs, our contribution is to explicitly analyze the role of tariffs distinct from trade costs in general. Unlike trade costs, tariffs bring about government revenue, thereby leading to richer welfare implications.
Quantitatively, we use tariff data and separately calibrate sector-specific non-tariff trade barriers using a structural gravity equation.

Our paper is also related to the broad literature on (optimal) trade policy through both quantitative and theoretical approaches.
A quantitative strand evaluates the welfare consequences of tariff policies (\citealp{Ossa2016}; \citealp{CaliendoParro2020} for surveys).\footnote{Recent work, such as \cite{LashkaripourLugovskyy2023}, \cite{Juetal2024} and \cite{BCDR_2025_JPE}, also studies domestic industrial policies, including output taxes and subsidies, alongside trade policy.} Many studies confirm the possibility of welfare gains from unilaterally increasing tariffs from a low level, unless other countries retaliate (\citealp{CostinotRodriguezClare2014}; \citealp{Ossa2014}; \citealp{Ignatenkoetal2025}; \citealp{Balistrerietal2026}). \cite{CostinotRodriguezClare2014} report in their Table 4.2 that U.S. welfare gains from imposing a 40 percent tariff on all imports in 2008 would be at most 0.63 percent, which is close in magnitude to our main result of 0.43 percent, despite differences in model setup.\footnote{In a partial-equilibrium framework, \cite{Amitietal2019} find that the 2018 U.S. tariffs reduced U.S. welfare, based on the estimates of the complete pass-through rate of tariffs to domestic consumer prices. However, because their regression framework does not incorporate country-level wage responses in foreign economies, their estimates may not fully capture the associated terms-of-trade effects (\citealp[Section 2.1]{CaliendoParro2020}; \citealp[Section VI]{Fajgelbaumetal2020}).} However, most of these studies assume homothetic preferences and/or an elasticity of substitution across sectors being greater than one. Among others, two papers are worth mentioning. \cite{FajgelbaumKhandelwal2016} emphasize the importance of nonhomothetic preferences (Almost-Ideal Demand System in their case) in evaluating the distributional impact of trade-cost reductions. \cite{Spearot2016} highlights the role of nonhomothetic preferences (quadratic preferences in his case) in evaluating tariff policies.\footnote{More broadly, some studies introduce nonhomothetic preferences to improve the explanatory power of the gravity equation (\citealp{Fieler2011}; \citealp{Caronetal2014}). An alternative approach is to allow for more flexible productivity distributions across countries (\citealp{LindRamondo2023}).} Although their models do not address structural change, we share their insight that overlooking nonhomothetic preferences can lead to different conclusions: particularly, we find an overestimation of welfare gains from unilateral tariffs in the U.S.\footnote{See \cite{BagwellLee2020} for detailed analytical expositions comparing how trade policies operate under CES and quadratic preferences, and \cite{Bertolettietal2018} for welfare gains from trade cost reductions in a broader class of preferences that encompasses CES and quadratic preferences as special cases. In the context of U.S. solar panel tariffs during 2014--2020, \cite{Gerardenetal2025} also caution against unilateral tariffs, highlighting their negative environmental impacts and the resulting decline in demand from downstream manufacturers.}

The theoretical strand of this literature characterizes optimal trade policies in general equilibrium. While existing work has emphasized how optimal trade policies depend on supply-side channels including comparative advantage \citep{Costinotetal2015}, goods-market power \citep{Demidovaetal2024}, and labor market conditions \citep{Baietal2024, Baietal2025}, we highlight a demand-side channel that has received less attention. In our model, nonhomothetic preferences change the expression of the welfare effect of a marginal tariff increase from zero, attenuating the welfare gain from such a small tariff in countries at an advanced stage of structural transformation. By contrast, nonhomothetic preferences do not change the inverse-elasticity expression for the optimal tariff.\footnote{In our two-country qualitative model, the optimal tariff for a sector is given by the inverse of the product of the Foreign export supply elasticity and the sectoral import share; see \eqref{eq: optimal tariff} and the discussion following it.} Nevertheless, they matter through the import shares, which depend on the expenditure shares, entering that expression. Our results complement existing work by highlighting the role of nonhomotheticity that shapes how unilateral tariffs operate and how their welfare effects should be evaluated.

The remainder of this paper is structured as follows. Section~\ref{sec: qual} presents a two-country model of Ricardian trade and nonhomothetic preferences. Section~\ref{sec: model} extends it to a full-fledged quantitative model. Section~\ref{sec: calib} explains the calibration of the model, the solution algorithm, and the model fit. Section~\ref{sec: counterfactuals} presents the counterfactual results, and Section~\ref{sec: concl} concludes.

\section{Two-country Model}\label{sec: qual}

To highlight the role of tariffs in shaping a country's sectoral composition, we first present a simple trade model \textit{\`{a} la} \cite{Eaton2002}, incorporating essential features for structural change: nonhomothetic CES preferences and a less-than-unity elasticity of substitution across sectors. Consider a static economy with two countries, Home ($H$) and Foreign ($F$). There are three tradable sectors, agriculture ($a$), manufacturing ($m$), and services ($s$).\footnote{Sectors and industries are synonymous in this paper. Although we state Propositions in the case of three sectors, the analytical results of this section hold in the case of general $J$ sectors (see Supplemental Appendices A to C).}

\vspace{0.1cm}

\textbf{Demand Side:} \ \ Let $C_n$ be the aggregate consumption of country $n \in \{ H, F\}$ and $L_n$ be the mass of workers. The representative household minimizes its expenditure given a certain level of $C_n$, by choosing consumption for sectoral composite goods, $C_n^j$ for $j \in \{a, m, s \}$:
\begin{align*}
    &\min_{ \{C_n^j \}_j } \ \sum_{j=a,m,s} P_n^j C_n^j, \qquad \text{s.t.} \         \sum_{j=a,m,s}\left(\dfrac{C_n}{L_{n}} \right)^{\frac{\epsilon^{j}(1-\sigma)}{\sigma}}\left(\dfrac{C_{n}^{j}}{L_{n}}\right)^{\frac{\sigma-1}{\sigma}}=1,
\end{align*}
where $P_n^j$ is the price index of the sectoral composite good in sector $j$ in country $n$. The aggregate consumption $C_n$ is \textit{implicitly} defined by the constraint. Two key parameters governing structural change are $\epsilon^{j}>0$, capturing the degree of nonhomotheticity, and $\sigma \in (0,1)$ measuring the elasticity of substitution across sectoral composite goods. We assume the parameter ranges such that $0 < \epsilon^a < \epsilon^m =1 < \epsilon^s$ and $\sigma \in (0,1)$ hold. If $\epsilon^j=1$ for all $j$, the utility function reduces to a standard CES aggregator of sectoral composite goods, $C_n = \left[\sum_j (C_n^j)^{\frac{\sigma-1}{\sigma}} \right]^{\frac{\sigma}{\sigma-1}}$. If we let $\sigma$ approach one under appropriate normalization, the utility function reduces to the Cobb-Douglas one, $C_n = \prod_{j} (C_n^j)^{\frac{1}{3}}$.

Letting $E_n = \sum_j P_n^j C_n^j$ be the (minimized) total expenditure, the Hicksian demand function for the sectoral composite good is obtained as
\begin{align}\label{eq: C_n^j}
    C_n^j = L_n \left( \frac{P_n^j}{E_n/C_n}\right)^{-\sigma} \left( \frac{C_n}{L_n}\right)^{\epsilon^j(1-\sigma)+\sigma}.
\end{align}
Substituting this into the budget constraint, we solve for the expenditure function:
\begin{align} \label{eq: E_n}
    E_n = L_{n}\left[\sum_{j=a,m,s} \left\{ \left(\frac{C_{n}}{L_{n}}\right)^{\epsilon^{j}}P_{n}^{j}\right\}^{1-\sigma} \right]^{\frac{1}{1-\sigma}}.
\end{align}
Letting $P_n = E_n/C_n$ be an aggregate price index, we have\footnote{Unlike the CES case, the price index cannot be separated from the aggregate consumption $C_n$. }
\begin{align} \label{eq: P_n}
    P_{n}=\left[\sum_{j=a,m,s} \left\{ \left(\frac{C_{n}}{L_{n}}\right)^{\epsilon^{j}-1}P_{n}^{j}\right\}^{1-\sigma} \right]^{\frac{1}{1-\sigma}}.
\end{align}
Using the relation $C_n/L_n = (E_n/L_n)/P_n$, we interpret $C_n/L_n$ as the real per capita consumption, which in equilibrium corresponds to real per capita income. This measure is distinct from the nominal aggregate expenditure $E_n$.

From these results, we can see the income (expenditure) and the relative price effects. The income effect is seen from the non-unitary income (expenditure) elasticity of sectoral demand:\footnote{We rearrange \eqref{eq: C_n^j} and take its log to obtain
\begin{align}\label{eq: dlnC/dlnE}
    \ln C_n^j = (1-\sigma)(1-\epsilon^j)\ln L_n -\sigma \ln P_n^j + \sigma \ln E_n + (1-\sigma)\epsilon^j \ln C_n.
\end{align}
To obtain the income elasticity of sectoral demand, we take the derivative of this with respect to $E_n$, considering its effect on the real consumption (i.e., utility) by noting $\partial \ln C_n/\partial \ln E_n = 1/\overline{\epsilon}_n$ from \eqref{eq: E_n}.}
\begin{align} \label{eq: dC_n^j/dE_n}
    \frac{\partial \ln C_n^j}{\partial \ln E_n} = \sigma + (1-\sigma) \frac{\epsilon^j}{\overline{\epsilon}_n}> 0,
\end{align}
where $\overline{\epsilon}_n$ captures the average of country $n$'s nonhomotheticity parameters $\epsilon^j$ weighted by the sectoral expenditure shares $\omega_n^j$:
\begin{align} \label{eq: omega_n^h}
    \overline{\epsilon}_n = \sum_{h=a,m,s} \omega_n^h \epsilon^h, \qquad \omega_n^h = \frac{P_n^h C_n^h}{\sum_k P_n^k C_n^k} = \frac{(P_n^h)^{1-\sigma}(C_n/L_n)^{\epsilon^h(1-\sigma)}}{\sum_k (P_n^k)^{1-\sigma}(C_n/L_n)^{\epsilon^k(1-\sigma)}}.
\end{align}
Similarly, the real income elasticity of sectoral demand also varies with sector: $\partial \ln C_n^j/\partial \ln C_n = \sigma \overline{\epsilon}_n + (1-\sigma)\epsilon^j$. In the case of CES preferences with $\epsilon^j=1$ for all $j$, the sectoral demand elasticity with respect to real income is unity across all sectors.
As we assume $0<\epsilon^a<\epsilon^m=1<\epsilon^s$, agriculture has the lowest elasticity, and services have the highest one. Therefore, $\epsilon^j$ can be interpreted as the degree of income elasticity of demand for sector $j$. With this interpretation in mind, we can view $\overline\epsilon_n$ as an expenditure-weighted income elasticity. Since $\overline\epsilon_n$ is higher as country $n$ spends less on lower-income-elastic sectors and more on higher-income-elastic sectors, it roughly captures the stage of structural transformation in country $n$.

The relative price effect results from the sectoral demands that are gross complements:\footnote{Using $C_n = E_n/P_n$, we have
\begin{align*}
    \ln C_n^j = (1-\sigma)(1-\epsilon^j)\ln L_n -\sigma \ln P_n^j + [\sigma + (1-\sigma)\epsilon^j]\ln E_n - (1-\sigma)\epsilon^j \ln P_n.
\end{align*}
Differentiating this with respect to $P_n^h$ while keeping $E_n$ fixed (i.e., $C_n^j$ here being a Marshallian demand) yields the expression in the text. In doing so, we use $\partial \ln P_n/\partial \ln P_n^h = \omega_n^h/\overline{\epsilon}_n$ from \eqref{eq: P_n}.}
\begin{align*}
    \frac{\partial \ln C_n^j}{\partial \ln P_n^h} = -(1-\sigma) \frac{\omega_n^h \epsilon^j}{\overline{\epsilon}_n} < 0, \qquad j\neq h.
\end{align*}
It also reflects the less-than-unity elasticity of substitution across sectors in the absence of the income effect:
\begin{align*}
    -\frac{\partial \ln(C_n^j/C_n^h)}{\partial \ln (P_n^j/P_n^h)} = \sigma \in (0, 1) \qquad \text{if \ $\epsilon^j = 1$ for all $j$}.
\end{align*}

\vspace{0.1cm}

\textbf{Supply Side:} \ \
The production side follows \cite{Eaton2002}. Producers of sectoral composite goods in sector $j$ in country $n$ are perfectly competitive and bundle input varieties $z \in [0,1]$ using a CES technology, with $\eta>0$ being the elasticity of substitution. While the sectoral composite goods are not tradable, the producers source tradable input varieties from the lowest-cost country.

The variety producers are also perfectly competitive and produce $y_n^j(z)$ using a linear technology such that $y_n^j(z) = a_n^j(z) l_n^j(z)$, where $a_n^j(z)$ and $l_n^j(z)$ are respectively the labor productivity and the labor input in sector $j$ in country $n$ for producing variety $z$. The labor productivity follows the Fr\'echet distribution with the cumulative distribution function (CDF): $
        \text{Pr}[a_{n}^{j}\le a]=\exp [-(a/(\widetilde{\gamma} A_{n}^{j}) )^{-\theta} ]$.
Here, the shape parameter $\theta>1$ governs the dispersion of productivity shocks, and the location parameter $A_{n}^{j}$ governs the average productivity. $\widetilde{\gamma} = [\Gamma( (\theta+1-\eta)/\theta )]^{-\frac{1}{1-\eta}}$ is a normalizing constant, where $\Gamma(\cdot)$ is the Gamma function. We assume $\theta+1-\eta > 0$ to ensure that the expectation of prices is finite. In this section, we assume that the average productivity is the same across sectors, $A_n^j=A_n$ for all $j$, to highlight the role of sector-specific tariffs.

\vspace{0.1cm}

\textbf{International Trade:} \ \ Trade in varieties is subject to tariffs, $\tau_{ni}^j \ge 0$ for $n \neq i \in \{ H,F\}$ and $\tau_{nn}^j=0$, and non-tariff trade barriers, $d_{ni}^j > 1$ for $n \neq i \in \{ H, F\}$ and $d_{nn}^j=1$. Thus, the total bilateral trade costs per shipment of a variety in sector $j$ from country $i$ to $n$ are $b_{ni}^j = d_{ni}^j \left( 1+\tau_{ni}^j \right)$. Letting $w_n$ be the wage in country $n$, the unit cost of variety $z$ produced in country $n$ and shipped to $i$ is $w_n b_{in}^j/a_n^j(z)$. As a result of the cost-minimization of sectoral composite good producers, the price of the variety available in country $n$ becomes $p_n^j(z) = \min_i \{ w_i b_{ni}^j/a_i^j(z) \}$. With these results and the Fr\'echet-distributed productivity shocks, the price of the sectoral composite good is obtained by
\begin{align} \label{eq: P_n^j}
    P_n^j = \left[ \sum_{i=H,F} \left( w_i b_{ni}^j/A_i \right)^{-\theta} \right]^{-\frac{1}{\theta}}.
\end{align}

Let $X_{ni}^j$ be country $n$'s expenditure on sector $j$ varieties from $i$. The share of country $i$'s varieties in country $n$'s sectoral expenditure becomes
\begin{align} \label{eq:pi_{ni}^j}
    \frac{X_{ni}^j}{P_n^j C_n^j} = \pi_{ni}^j = \frac{\left( w_i b_{ni}^j/A_i \right)^{-\theta}}{ \sum_{i'=H,F} \left( w_{i'} b_{ni'}^j/A_{i'} \right)^{-\theta} },
\end{align}
which we call the trade share of country $i$ in the market of country $n$.

\vspace{0.1cm}

\textbf{Market Clearing:} \ \ As labor is the only factor of production, the sectoral value added is the sectoral labor income, which comes from sales in the domestic and foreign markets:
\begin{align*}
    VA_n^j = w_n L_n^j
    = \sum_{i=H,F} \frac{X_{in}^j}{1+\tau_{in}^j} = \sum_{i=H,F} \frac{\pi_{in}^j P_i^j C_i^j}{1+\tau_{in}^j},
\end{align*}
where $L_n^j$ is sector $j$ employment in country $n$, and the export sales are divided by gross tariff rates, $1+\tau_{in}^j$. This is because the price index is tariff inclusive (c.i.f.), while sales for producers are based on the tariff-exclusive (f.o.b.) price. Summing this condition over sectors gives the labor market clearing condition:
\begin{align} \label{eq: w_n L_n}
    w_n L_n = \sum_{j=a,m,s} \sum_{i=H,F} \frac{\pi_{in}^j P_i^j C_i^j}{1+\tau_{in}^j}.
\end{align}
One can check that this is equivalent to the trade balance condition.\footnote{The trade balance condition is $\sum_j \pi_{HF}^j P_H^j C_H^j/(1+\tau_{HF}^j)=\sum_j \pi_{FH}^j P_F^j C_F^j/(1+\tau_{FH}^j)$.}
The aggregate expenditure must be equal to the aggregate income consisting of labor income and tariff revenues, $\widetilde{T}_n$:
\begin{align} \label{eq: income}
    E_n = w_n L_n + \widetilde{T}_n,
\end{align}
where $\widetilde{T}_n = \sum_j \widetilde{T}_n^j = \sum_j \tau_{ni}^j IM_n^j$, $IM_n^j = X_{ni}^j/(1+\tau_{ni}^j) =  \pi_{ni}^j P_n^j C_n^j/(1+\tau_{ni}^j)$ for $i\neq n$, and $\widetilde{T}_n^j$ is country $n$'s tariff revenues in sector $j$ and $IM_n^j$ is country $n$'s imports in sector $j$. This completes the model. With a choice of num\'eraire such that $w_H=1$, the equilibrium wage $w_F$ and the aggregate consumption $\{C_i \}_{i=H,F}$ satisfy equilibrium conditions \eqref{eq: C_n^j}, \eqref{eq: E_n}, \eqref{eq: P_n^j}, \eqref{eq: w_n L_n}, and \eqref{eq: income}.

\medskip

\subsection{Tariffs and Real Consumption}\label{sec: tariff&cons}

Before examining the tariff impact on sectoral composition, let us first see how tariffs affect real per capita consumption $C_n/L_n$.\footnote{As tariffs do not affect population $L_n$, the following discussion holds both in terms of aggregate and per capita consumption.} In the following subsections, we consider a unilateral increase in Home's tariffs on the manufacturing sector, $d\tau_{HF}^m = d\tau^m > 0$, starting from a tariff-free world. That is, tariffs both in Home and Foreign are initially zero, $\tau_{HF}^j=0$ and $\tau_{FH}^j=0$ for all $j \in \{a,m,s \}$.

Whether preferences are nonhomothetic CES or homothetic CES, the real per capita consumption in Home is expressed as $C_H/L_H = (E_H/P_H)/L_H$, where $E_H = \sum_j P_H^j C_H^j = L_H + \tau^m IM_H^m = L_H + \tau^m \pi_{HF}^m \omega_H^m E_H/(1+\tau^m)$. The welfare effect of the tariff is
\begin{align}
    \frac{d\ln (C_H/L_H)}{d\ln (1+\tau^m)}\bigg|_{\tau^m=0} = \pi_{HF}^m \omega_H^m -\frac{d\ln P_H}{d\ln (1+\tau^m)}. \label{eq: dln(C/L)}
\end{align}
The first term captures revenue gains, proportional to the share of Home's expenditure spent on Foreign varieties in the manufacturing sector. The second term captures the welfare loss from a higher aggregate price index. Using \eqref{eq: P_n}, we can further decompose it as follows: {\small
\begin{align} \label{eq: dlnP}
    \frac{d\ln P_H}{d\ln (1+\tau^m)}\bigg|_{\tau^m=0}
    &= \frac{d\ln (E_H/L_H)}{d\ln (1+\tau^m)} - \frac{d\ln (C_H/L_H)}{d\ln (1+\tau^m)} \notag \\
    &= \sum_{j=a,m,s} \omega_H^j \frac{d\ln P_H^j}{d\ln (1+\tau^m)} + \sum_{j=a,m,s} \omega_H^j \frac{d\ln (C_H^j/L_H)}{d\ln (1+\tau^m)} - \frac{d\ln (C_H/L_H)}{d\ln (1+\tau^m)} \notag \\
    &= \sum_{j=a,m,s} \omega_H^j \frac{d\ln P_H^j}{d\ln (1+\tau^m)} + \left( \overline{\epsilon}_H - 1 \right)\frac{d\ln (C_H/L_H)}{d\ln (1+\tau^m)} \notag \\
    &= \underbrace{\left( \sum_{j=a,m,s} \pi_{HF}^j \omega_H^j \right)\frac{d\ln w_F}{d\ln (1+\tau^m)}}_{\text{(a) Terms-of-trade gain}< 0} + \underbrace{\pi_{HF}^m \omega_H^m}_{\text{(b) Direct tariff effect}>0} +  \underbrace{\left( \overline{\epsilon}_H - 1 \right)\frac{d\ln (C_H/L_H)}{d\ln (1+\tau^m)}}_{\text{(c) Income effect}\gtrless 0},
\end{align} }
\noindent where $\overline{\epsilon}_H = \sum_j \omega_H^j \epsilon^j$. In moving from the second to the third line, we use $d\ln C_H^j = [\sigma \overline{\epsilon}_H + (1-\sigma)\epsilon^j]d\ln C_H$ (see \eqref{eq: dC_n^j/dE_n} and the discussion therein). The first two terms in the last line, (a) and (b), capture changes in the sectoral price indices. An increase in tariffs in Home is likely to reduce import demand there and raise Home's export price relative to its import price, or equivalently lower Foreign's export price relative to its import price, to maintain balanced trade. With our choice of num{\'e}raire such that $w_H=1$, this adjustment leads to a lower relative wage in Foreign, $d\ln w_F/d\ln (1+\tau^m)<0$, which \citet{Matsuyama2019} refers to as the terms-of-trade effect in terms of production factors. We can show that this is indeed the case and thus term (a) is negative under symmetric sectoral productivities and non-tariff trade barriers, $A_H^j=A_H$, $A_F^j=A_F$ and $d_{ni}^j=d_{ni}$ for all $j$.

Term (b) captures the direct effect of the manufacturing tariff on the consumer price of imported manufacturing varieties.

Term (c) arises only under nonhomothetic CES preferences, while under homothetic CES preferences, it vanishes since we have $\epsilon^j=1$ for all $j$ and thus $\overline{\epsilon}_H = \sum_j \omega_H^j \cdot 1 = 1$. In the case of homothetic CES, a one percent increase in real consumption raises sectoral demands by one percent in every sector, $d\ln C_H^j/d\ln C_H = 1$ for all $j$. Higher real consumption (or utility) is attained by scaling up the same consumption basket, so that nominal expenditure is proportional to real consumption, $E_H(\{P_H^j \}_j, C_H) = P_H(\{P_H^j \}_j)C_H$. The aggregate price index defined as the average expenditure per unit of real consumption $P_H(\{P_H^j \}_j)=E_H(\{P_H^j \}_j, C_H)/C_H$ depends only on sectoral price indices.

In the case of nonhomothetic CES, however, the sectoral demand responds to changes in real consumption disproportionately as $d\ln C_H^j/d\ln C_H = \sigma \overline{\epsilon}_H + (1-\sigma)\epsilon^j$ (from \eqref{eq: dlnC/dlnE}). The composition of the consumption basket therefore depends on the level of real consumption, so the aggregate price index depends not only on sectoral price indices, but also on $C_H$. A one percent increase in $C_H$ changes sectoral demands unevenly, according to the nonhomotheticity parameters $\epsilon^j$. If the expenditure-weighted income elasticity is sufficiently high such that $\overline{\epsilon}_H>1$, higher real consumption shifts the basket sufficiently toward services and raises aggregate expenditure more than proportionately. This increases the aggregate price index, making term (c) positive. As the derivation of \eqref{eq: dlnP} shows, the result that the response of the aggregate price index depends on whether $\overline{\epsilon}_H \gtrless 1$ holds independently of the number of countries, sectors, or any symmetry assumptions.

Combining \eqref{eq: dln(C/L)} with \eqref{eq: dlnP}, we can compute the welfare effect of increasing the tariff from the tariff-free world as
\begin{align} \label{eq: Prop01}
    \frac{d\ln (C_H/L_H)}{d\ln (1+\tau^m)}\bigg|_{\tau^m=0} = - \frac{\Lambda_{HF} \varepsilon_{w, F}}{\overline{\epsilon}_H},
\end{align}
where $\Lambda_{HF} = \sum_j \pi_{HF}^j \omega_H^j$ is the fraction of total Home expenditure on Foreign
produced varieties, and $\varepsilon_{w,F} = d\ln w_F/d\ln (1+\tau^m)$ is the elasticity of Foreign's wage with respect to Home's manufacturing tariff. Under sectoral symmetry, we have $\varepsilon_{w,F}<0$, so the welfare effect is positive. The formula says that a small Home tariff affects welfare only through its terms-of-trade effect: by lowering Foreign's wage, $\varepsilon_{w,F}<0$, it reduces the cost of Foreign-produced varieties for Home. The size of this gain is proportional to Home's exposure to imports, $\Lambda_{HF}$, since the wage decline matters more when Home spends more on Foreign varieties. The gain is divided by the expenditure-weighted income elasticity $\overline\epsilon_H$ since, under nonhomothetic CES preferences, a given change in real consumption requires a larger expenditure adjustment when $\overline\epsilon_H$ is high.

Because higher real income may raise the aggregate price index, as captured by the positive term (c) in \eqref{eq: dlnP}, the welfare gain from a manufacturing tariff may be smaller under nonhomothetic CES preferences than under homothetic CES preferences. We can show that this is indeed the case if $\overline{\epsilon}_H>1$ and $\omega_H^m<1/3$. This corresponds to Home being at an advanced stage of structural transformation, which is true for most advanced countries such as the U.S.\footnote{The average U.S. tariff on manufacturing products was around three percent between 2000 and 2014 (see ``inward'' tariffs in Figure~\ref{fig: tradecost_tau} in Supplemental Appendix~\ref{ap: baseline parameters} for the U.S. and other developed countries). Given our calibrated values in Section~\ref{sec: calib} ($(\epsilon^a, \epsilon^m, \epsilon^s)=(0.05, 1, 1.2)$ as in \citealp{CLM}), the expenditure-weighted average of nonhomotheticity parameters in the U.S., $\overline{\epsilon}_{US}$, consistently exceeded one and increased in most years from 1990 to 2014.\label{ft: epsilon_bar}}

The discussion is summarized in the following proposition. The welfare formula \eqref{eq: Prop01} holds for an arbitrary sector $j$ in the $J$ sector model. We provide the proofs of this and the subsequent propositions, along with other results, in Supplemental Appendices~\ref{ap: proof of prop 1} to \ref{ap: proof of prop 3}.

\medskip

\medskip

\begin{Proposition}\phantomsection\label{prop: proposition 1}
\noindent \textbf{The Welfare Effect of Tariffs}
\vspace{.2cm}

\noindent We consider an economy with two countries and three sectors. Suppose Home imposes a tariff in the manufacturing sector, starting from a tariff-free world ($d\tau_{HF}^m = d\tau^m > 0$; $\tau_{HF}^j = 0$ and $\tau_{FH}^j = 0$ for $j \in \{ a,m,s\}$).
\begin{itemize}
    \item[(i)] The effect of an increase in the Home tariff from zero on its real per capita consumption is
\begin{align}
\left.\frac{d\ln (C_H/L_H)}{d\ln(1+\tau^m)}\right|_{\tau^m=0} = \begin{cases}
    -\dfrac{\Lambda_{HF} \varepsilon_{w, F}}{\overline\epsilon_H} &\text{under nonhomothetic CES preferences} \\
    -\Lambda_{HF} \varepsilon_{w, F} &\text{under homothetic CES preferences}
\end{cases},
\label{app: dln(uH)_v1}
\end{align}
where $\Lambda_{HF}=\sum_j \pi_{HF}^j \omega_H^j$ is the fraction of total Home expenditure on Foreign produced varieties; $\varepsilon_{w,F} = d\ln w_F/d\ln (1+\tau^m)$ is the elasticity of Foreign's wage with respect to manufacturing tariff; and $\overline{\epsilon}_H=\sum_{j} \omega_H^j \epsilon^j$ is the expenditure-weighted income elasticity. The Foreign wage elasticity $\varepsilon_{w,F}$ is negative and thus the welfare effect is positive, if sectoral productivities and non-tariff trade barriers are symmetric, $A_H^j=A_H$, $A_F^j=A_F$, and $d_{ni}^j=d_{ni}$ for all $j$.
    \item[(ii)] If sectors are symmetric, the welfare gain in (i) under nonhomothetic CES preferences is smaller than under homothetic CES preferences, if $\omega_H^m < \overline{\epsilon}_H/3$. A sufficient condition for this is $\overline{\epsilon}_H > 1$ and $\omega_H^m < 1/3$, i.e., Home's expenditure has shifted sufficiently from manufacturing toward services.
    \item[(iii)] A mirror-image result holds for Foreign. If sectors are symmetric, Foreign loses from a marginal increase in Home's manufacturing tariff from zero, $d\ln (C_F/L_F)/d\ln (1+\tau^m)<0$ at $\tau^m=0$. If in addition $\omega_H^m<\overline{\epsilon}_F/3$, a sufficient condition for which is $\overline{\epsilon}_F>1$ and $\omega_H^m<1/3$, this welfare loss is smaller in magnitude under nonhomothetic CES preferences than under homothetic CES preferences.
\end{itemize}
\end{Proposition}

\medskip

So far, we have examined the effect of a marginal tariff increase from zero. One may also be interested in the optimal level of tariff that maximizes Home's welfare measured by its real per capita consumption.\footnote{\citet{Brodaetal2008} empirically test the inverse-elasticity rule for optimal tariffs by examining non-WTO-member countries.} The optimal manufacturing tariff is given by
\begin{align} \label{eq: optimal tariff}
    \tau^m
    &= \frac{1}{s_{HF}^m \varepsilon_{X,F}},
\end{align}
where $\varepsilon_{X,F}$ is the Foreign export supply elasticity with respect to the net-of-tariff world price of Foreign manufacturing varieties, and $s_{HF}^m$ is the share of manufacturing imports out of Home's total imports. The detailed derivation is given in Supplemental Appendix~\ref{ap: optimal tariff}. The expression in \eqref{eq: optimal tariff} is consistent with the standard inverse-elasticity rule: the optimal tariff is higher when Foreign export supply is less elastic, since Home can then lower the net-of-tariff price received by Foreign exporters, and hence improve its terms of trade, without inducing a large reduction in import volume. In our context, however, the tariff lowers the net-of-tariff world price only for manufacturing imports. If the manufacturing sector accounts for only a small share of Home's total imports, even a sizable terms-of-trade improvement in that sector has only a limited effect on Home's total imports. This is why the relevant elasticity is effectively $s_{HF}^m\varepsilon_{X,F}$.

The optimal tariff in \eqref{eq: optimal tariff} has the same form under homothetic and nonhomothetic CES preferences. However, the two cases differ through the sectoral import share $s_{HF}^m$. Holding the Foreign export supply elasticity $\varepsilon_{X,F}$ fixed, the optimal tariff is higher when $s_{HF}^m$ is smaller. Under sectoral symmetry, $s_{HF}^m$ increases with the sectoral expenditure share $\omega_H^m$, so nonhomotheticity affects the optimal tariff through its effect on $\omega_H^m$. Thus, if the manufacturing expenditure share under nonhomothetic CES preferences is sufficiently lower than under homothetic CES preferences, $\omega_H^{m, \text{nh}} < \omega_H^{m,\text{h}}$, then the optimal manufacturing tariff is higher under nonhomothetic CES preferences, $\tau^{m,\text{nh}} > \tau^{m,\text{h}}$. These inequalities of the optimal tariffs and the expenditure shares at the optimal tariffs hold in our quantitative model as in Figures~\ref{fig: optimal tariffs}, \ref{fig:optimal-tariffs-omega1-zeta1}, and \ref{fig: expshare_optimum}.

\medskip

\subsection{Tariffs and Expenditure Shares}\label{sec: tariff&exp}

Let us turn to the impact of the manufacturing tariff on the sectoral expenditure share given in \eqref{eq: omega_n^h}. The logarithmic change in Home's expenditure share of sector $j \in \{ a,m,s\}$ is decomposed as
\begin{align}
    \frac{d\ln \omega_H^j}{d\ln (1+\tau^m)}\bigg|_{\tau^m=0} &= \underbrace{(1-\sigma)\biggr[\frac{d\ln P_H^j}{d\ln (1+\tau^m)} - \sum_{h=a,m,s}\omega_H^h \frac{d\ln P_H^h}{d\ln (1+\tau^m)} \biggl]}_{\text{(a) Relative price effect}} \notag \\
    &\qquad \qquad + \underbrace{ (1-\sigma)\bigl( \epsilon^j - \overline{\epsilon}_H \bigr) \frac{d\ln (C_H/L_H)}{d\ln (1+\tau^m)}}_{\text{(b) Income effect}}. \label{eq: dln(exp share)}
\end{align}
Term (a) in \eqref{eq: dln(exp share)} is the relative price effect through changes in the relative price of the sector $j$ composite good. To highlight this, we consider homothetic CES preferences by setting $\epsilon^j=1$ for all $j$. The tariff increase in a sector raises its price index relative to those in the other sectors by making the imported intermediate varieties more expensive to produce the sectoral composite good. With a less-than-unity elasticity of substitution across sectors, $\sigma \in (0, 1)$, the higher relative price of the sector's composite good leads to a higher expenditure share in that sector and lowers shares in the other sectors.

Term (b) in \eqref{eq: dln(exp share)} is the income effect through changes in the real per capita consumption and only shows up in the case of nonhomothetic CES with $\overline\epsilon_H \neq 1$. The tariff increase, starting from zero, does not just affect sectoral price indices but also raises the real per capita consumption in Home (Proposition~\ref{prop: proposition 1}). Home then shifts expenditure away from the sectors with a lower income elasticity (smaller $\epsilon^j$) toward the sectors with a higher income elasticity (larger $\epsilon^j$).

In our three-sector model with $\epsilon^a < \epsilon^m=1 < \epsilon^s$, the income effect is negative for agriculture and positive for services, while it can be negative or positive for manufacturing, depending on whether the expenditure-weighted income elasticity is greater or smaller than one ($\overline{\epsilon}_H \gtrless 1 = \epsilon^m$). In other words, compared with the case of homothetic CES, nonhomotheticity makes the negative effect of the manufacturing tariff on the agricultural expenditure share more negative, while making its negative effect on the service expenditure share less negative. In particular, when $\overline{\epsilon}_H>1$, the positive effect on manufacturing's own expenditure share becomes smaller.

We summarize the discussion in the following proposition.

\medskip

\begin{Proposition}\phantomsection\label{prop: proposition 2}
\noindent \textbf{Tariffs and Expenditure Shares}
\vspace{.2cm}

\noindent We consider an economy with two countries and three symmetric sectors. Suppose Home imposes a tariff only in the manufacturing sector, starting from a tariff-free world ($d\tau_{HF}^m = d\tau^m > 0$; $\tau_{HF}^j = 0$ and $\tau_{FH}^j = 0$ for $j \in \{ a,m,s\}$).
\begin{itemize}
    \item[(i)] Under homothetic CES preferences, a higher Home manufacturing tariff lowers the expenditure shares of agriculture and services ($d\ln \omega_H^j/d\ln (1+\tau^m)<0$ at $\tau^m=0$ for $j \in \{ a,s\}$), while raising the expenditure share of manufacturing ($d\ln \omega_H^m/d\ln (1+\tau^m)>0$ at $\tau^m=0$).
    \item[(ii)] Under nonhomothetic CES preferences and $\overline\epsilon_H>1$, relative to the case of homothetic CES, the agricultural expenditure share falls more, the manufacturing expenditure share rises less, and the service expenditure share falls less.
\end{itemize}
\end{Proposition}

\medskip

\subsection{Tariffs and Value-added Shares}\label{sec: tariff&va}

The value added of sector $j \in \{ a,m,s\}$ in the tariff-imposing Home is the labor income in that sector, which is earned from sales in the domestic and foreign markets:
\begin{align*}
    VA_H^j &= L_H^j = \sum_{i=H,F} \pi_{iH}^j P_i^j C_i^j \\
    &= P_H^j C_H^j + NX_{H}^j - \mathbf{1}_{\{ j=m\}}\widetilde{T}_H^j, \qquad NX_{H}^j = EX_H^j - IM_H^j = \pi_{FH}^j P_F^j C_F^j - \frac{\pi_{HF}^j P_H^j C_H^j}{1+\tau_{HF}^j},
\end{align*}
where the Home wage is one due to our choice of num{\'e}raire, $NX_{H}^j$ is Home's net exports in sector $j$, defined as gross exports minus gross imports, and $\mathbf{1}_{\{ j=m\}}$ is a dummy taking one if sector $j$ is manufacturing and zero otherwise. The sectoral value added in Home increases with domestic expenditure ($P_H^j C_H^j$) and net exports ($NX_H^j$), but decreases with tariff revenue ($\widetilde{T}_H^j = \tau^j IM_H^j$), since the revenue accrues from the value added generated by workers in Foreign.

The value-added share of sector $j$ in Home is then $va_H^j = VA_H^j/L_H$. Its logarithmic change is given by\footnote{The second term (b) is defined as
\begin{align*}
    \frac{NX_H^j}{L_H^j}\frac{d\ln (NX_H^j/L_H)}{d\ln (1+\tau^m)} = \frac{EX_H^j}{L_H^j}\frac{d\ln (EX_H^j/L_H)}{d\ln (1+\tau^m)} - \frac{IM_H^j}{L_H^j}\frac{d\ln (IM_H^j/L_H)}{d\ln (1+\tau^m)}.
\end{align*}}
\begin{align}
    \frac{d\ln va_H^j}{d\ln (1+\tau^m)}\bigg|_{\tau^m=0} &= \underbrace{\frac{P_H^j C_H^j}{L_H^j} \left[ \frac{d\ln \omega_H^j}{d\ln (1+\tau^m)} + \frac{1}{1+\widetilde{T}_H^m/L_H}\frac{d(\widetilde{T}_H^m/L_H)}{d\ln (1+\tau^m)} \right]}_{\text{(a) Expenditure adjusted by tariff revenue}} \\
    &\qquad + \underbrace{\frac{NX_H^j}{L_H^j}\frac{d\ln ( NX_H^j/ L_H)}{d\ln (1+\tau^m)}}_{\text{(b) Net exports}}
    - \underbrace{\frac{\mathbf{1}_{\{ j=m\}}}{L_H^j}\frac{ d\widetilde{T}_H^j}{d\ln \ (1+\tau^m)}}_{\text{(c) Tariff revenue}}.
    \label{eq: dln(va)}
\end{align}
Under autarky, the value-added shares and the expenditure shares always coincide: $va_H^j = \omega_H^j$. With international trade, however, they may differ.

The contribution of sectoral expenditure shares to value-added shares is captured by the first term (a) in \eqref{eq: dln(va)}. This expenditure channel also includes changes in aggregate tariff revenue relative to labor income---the second term within (a)---since higher revenue scales up consumption across all sectors. How the sectoral expenditure shares respond to tariffs---the first term within (a)---has already been discussed in Proposition~\ref{prop: proposition 2}. Under homothetic CES preferences, only the relative-price effect operates, so a higher Home manufacturing tariff shifts expenditure shares away from agriculture and services and toward manufacturing; through term (a), sectoral value-added shares move in the same direction. Under nonhomothetic CES preferences, the additional income effect dampens the responses for manufacturing and services, while amplifying the response for agriculture.

The last term (c) in \eqref{eq: dln(va)} is a mechanical channel from the definition of $va_H^j$. When tariffs raise revenue in a sector more than labor income, the share of contribution by Home workers in that sector falls.

The second term (b) in \eqref{eq: dln(va)} captures the standard protective role of tariffs. When only the relative-price effect operates, a higher Home manufacturing tariff reduces imports in the protected sector and raises net exports, thereby increasing Home's manufacturing value-added share. Once the income effect is also present, however, the net-export channel becomes more nuanced because changes in real income in both countries affect sectoral demand. Under sectoral symmetry, the tariff raises Home's real income and lowers Foreign's real income (Proposition~\ref{prop: proposition 1}). If the two countries are at an advanced stage of structural transformation such that $\overline{\epsilon}_H>1$ and $\overline{\epsilon}_F>1$, the resulting income expansion in Home shifts demand away from manufacturing, reducing gross imports from Foreign, while the income contraction in Foreign shifts demand toward manufacturing, increasing Home's gross exports. Both channels improve Home's manufacturing net exports.

Further assuming symmetry between countries, we can formally show that a higher Home manufacturing tariff moves the sectoral value-added shares in the same direction as their expenditure shares characterized in Proposition~\ref{prop: proposition 2}.
The results are summarized in the following proposition.

\medskip

\begin{Proposition}\phantomsection\label{prop: proposition 3}
\noindent \textbf{Tariffs and Value-added Shares}
\vspace{.2cm}

\noindent We consider an economy with two countries and three symmetric sectors. Suppose Home imposes a tariff in the manufacturing sector, starting from a tariff-free world ($d\tau_{HF}^m = d\tau^m > 0$; $\tau_{HF}^j = 0$ and $\tau_{FH}^j = 0$ for $j \in \{ a,m,s\}$).
\begin{itemize}
    \item[(i)] Under homothetic CES preferences, a higher Home manufacturing tariff lowers the value-added shares of agriculture and services ($d\ln va_H^j/d\ln (1+\tau^m)<0$ at $\tau^m=0$ for $j \in \{ a,s\}$), while raising the value-added share of manufacturing ($d\ln va_H^m/d\ln (1+\tau^m)>0$ at $\tau^m=0$).
    \item[(ii)] Under nonhomothetic CES preferences, symmetric countries, and $\overline\epsilon_H=\overline\epsilon_F>1$, relative to the case of homothetic CES, the agricultural value-added share falls more, the manufacturing value-added share rises less, and the service value-added share falls less.
\end{itemize}
\end{Proposition}

\medskip

\subsection{Toward a Full Dynamic Model}\label{sec: two-period}

As an intermediate step toward the fully quantitative dynamic model, we introduce capital into the two-country model and extend it to two periods. Using this extended model, we examine how tariff shocks affect the household's consumption-saving decision. The key point is that a permanent tariff increase reduces saving and thus investment.

Letting $K_{n,t}$ and $I_{n,t}$ denote the capital stock and investment, respectively, in country $n \in \{ H, F\}$ and period $t \in \{ 1,2\}$, the law of motion of the capital stock is $K_{n,t+1}=I_{n,t}$, where capital in the last period fully depreciates. Capital is owned by the household and rented to domestic variety producers at the rate $r_{n,t}$. Varieties are produced according to a Cobb-Douglas production technology using capital and labor. The investment good is produced from the sectoral composite goods using a CES technology with the elasticity parameter $\sigma^K \in (0,1)$ and is domestically sold at price $P_{n,t}^K$. The consumption-saving decision maximizes the discounted sum of utility derived from the aggregate consumption good: $u(C_{n,1}) +\beta u(C_{n,2})$.

\vspace{0.1cm}

\textbf{Saving Decisions and a Tariff Shock:} \ \ In each period $t \in \{ 1,2\}$, a representative household in country $H$ earns labor and capital income, $w_{H,t} L_{H,t} + r_{H,t} K_{H,t}$, as well as tariff revenue, $\widetilde{T}_{H,t}$, and spends on consumption, $E_{H,t}$, and investment, $P_{H,t}^K I_{H,t}$:
\begin{align*}
    E_{H,t} + P_{H,t}^K I_{H,t} &= w_{H,t}L_{H,t} + r_{H,t}K_{H,t} + \widetilde{T}_{H,t}.
\end{align*}
Noting $E_{H,t}=P_{H,t}C_{H,t}$, we rearrange this intertemporal budget constraint in Home in period 1 to obtain
\begin{align}
    I_{H,1} &=  \frac{P_{H,1}}{P_{H,1}^K} \biggl( \underbrace{\frac{w_{H,1}L_{H,1}+r_{H,1}K_{H,1} +\widetilde{T}_{H,1}}{P_{H,1}}}_{\text{Real income}} - \underbrace{C_{H,1}}_{\text{Real consumption}} \biggr). \label{eq: K_H1}
\end{align}
Investment occurs ($I_{H,1} > 0$) when part of real income is saved.
This helps us understand how tariffs affect the consumption-saving decision.

Let us consider an increase in the manufacturing tariff that takes effect in both periods 1 and 2.\footnote{The following discussion does not depend on which sector is subject to the tariff increase.}
The tariff increase in period 1 is a surprise shock, and the household anticipates that it will remain effective in period 2. In response to the tariff increase from a sufficiently low level, real income rises in both periods, which results in higher real consumption. As implied by \eqref{eq: K_H1}, investment in period 1 is determined by the relative changes in income and consumption. The household anticipating a higher income in the future has a weaker incentive to save for intertemporal consumption smoothing. Therefore, the real consumption may increase more than the real income, leading to a decline in investment in period 1 following the tariff shock, $dI_{H,1}<0$.

The full quantitative analysis below investigates whether the intuition obtained from the simplified setting continues to hold in a more general environment.

\section{Quantitative Model} \label{sec: model}

We extend the two-country model to a dynamic multi-country model with capital accumulation and sectoral input-output linkages. Time is discrete: $t = 0, 1, \dots$. The set of countries is $\{1, \dots, N\}$, with the number of countries being $N$. Countries are generically indexed by $i$ or $n$. We maintain three sectors as in the previous model: agriculture, manufacturing, and services.

The representative household in country $n$ as of period 0 maximizes the lifetime utility function:
\begin{equation} \label{eq: lifetime utility}
    \sum_{t=0}^{\infty}\beta^{t}\zeta_{n,t}L_{n,t}\frac{(C_{n,t}/L_{n,t})^{1-\psi}}{1-\psi},
\end{equation}
where $\beta\in(0, 1)$ is the discount factor, $\psi>1$ is the inverse of the intertemporal elasticity of substitution, $\zeta_{n,t}$ is the demand shifter in country $n$ and period $t$, and $L_{n,t}$ is the population of country $n$ in period $t$. The aggregate consumption in country $n$ and period $t$, $C_{n,t}$, is \textit{implicitly} defined by
\begin{equation} \label{eq: CLM utility}
        \sum_{j=a,m,s}(\Omega_{n,t}^{j})^{\frac{1}{\sigma}}\left(\frac{C_{n,t}}{L_{n,t}} \right)^{\frac{\epsilon^{j}(1-\sigma)}{\sigma}}\left(\frac{C_{n,t}^{j}}{L_{n,t}}\right)^{\frac{\sigma-1}{\sigma}}=1,
\end{equation}
where $C_{n,t}^{j}$ is the composite good of sector $j \in \{ a,m,s\}$ which the representative household in country $n$ and period $t$ consumes, and $\Omega^{j}_{n,t}$ is the demand shifter for sector $j$ in country $n$ and period $t$.%

Solving the intratemporal expenditure minimization problem given $C_{n,t}$, the expenditure of country $n$ in period $t$ is
    \begin{equation} \label{eq: expenditure}
        E_{n,t} = L_{n,t}\left[\sum_{j=a,m,s}\Omega_{n,t}^{j}\left\{ \left(\frac{C_{n,t}}{L_{n,t}}\right)^{\epsilon^{j}}P_{n,t}^{j}\right\}^{1-\sigma} \right]^{\frac{1}{1-\sigma}},
    \end{equation}
where $P_{n,t}^{j}$ is the price of the composite good of sector $j$ in country $n$ and period $t$. Define $P_{n,t}$ by $P_{n,t} = E_{n,t}/C_{n,t}$. Then we have
\begin{equation}
    P_{n,t}=\left[\sum_{j=a,m,s}\Omega_{n,t}^{j}\left\{ \left(\frac{C_{n,t}}{L_{n,t}}\right)^{\epsilon^j-1}P_{n,t}^{j}\right\}^{1-\sigma} \right]^{\frac{1}{1-\sigma}}.
\end{equation}
The consumption of the composite good of sector $j$ is
\begin{equation}\label{eq: sector cons}
    C_{n,t}^{j}=L_{n,t}\Omega_{n,t}^{j}\left(\frac{P_{n,t}^{j}}{P_{n,t}} \right)^{-\sigma}\left(\frac{C_{n,t}}{L_{n,t}} \right)^{\epsilon^{j}(1-\sigma)+\sigma}.
\end{equation}
Let $\omega_{n,t}^{j} = P_{n,t}^j C_{n,t}^j/E_{n,t}$ be country $n$'s consumption expenditure share on sector $j$ in period $t$. Then we have
\begin{equation}\label{eq: expenditure share}
    \omega_{n,t}^{j}=\frac{\Omega_{n,t}^{j}\left\{\left( \frac{C_{n,t}}{L_{n,t}} \right)^{\epsilon^{j}}P_{n,t}^{j} \right\}^{1-\sigma}}{\sum_{j'=a,m,s}\Omega_{n,t}^{j'}\left\{\left( \frac{C_{n,t}}{L_{n,t}} \right)^{\epsilon^{j'}}P_{n,t}^{j'} \right\}^{1-\sigma}}.
\end{equation}
By definition, we have $\sum_{j=a,m,s}\omega_{n,t}^{j}=1$.

The representative household in country $n$ is the sole owner of labor and capital there. The budget constraint of country $n$ in period $t$ is
\begin{equation} \label{eq: budget constraint}
E_{n,t}+P_{n,t}^{K}I_{n,t}\le (1-\phi_{n,t})\left( w_{n,t}L_{n,t}+r_{n,t}K_{n,t}+\widetilde{T}_{n,t} \right)+L_{n,t}T_{t}^{P},
\end{equation}
where $P_{n,t}^{K}$ is the capital good price index which will be defined later, and $I_{n,t}$ is the quantity of investment. The right-hand side represents the national income, taking into account aggregate trade imbalances. There is a global portfolio that collects an exogenous $\phi_{n,t}$ share of total value added, $w_{n,t}L_{n,t}+r_{n,t}K_{n,t}$, plus tariff revenue, $\widetilde{T}_{n,t}$, and redistributes a per capita transfer, $T_t^P$, to each country to balance the global portfolio's budget. In this way, we model trade imbalances as transfers, abstracting from cross-border borrowing and lending (\citealp{Sposietal2026}).

Let $K_{n,t}$ be the quantity of capital in country $n$ and period $t$. Then capital dynamics are
\begin{equation} \label{eq: capital dynamics}
    K_{n,t+1}=(1-\delta_{n,t})K_{n,t}+I_{n,t}^{\lambda}(\delta_{n,t}K_{n,t})^{1-\lambda},
\end{equation}
where $\delta_{n,t}$ is the capital depreciation rate in country $n$ and period $t$ and $\lambda\in[0, 1]$ is a parameter governing capital adjustment costs. Solving this for $I_{n,t}$ and viewing it as a function of $K_{n,t}$, $K_{n,t+1}$, and $\delta_{n,t}$, we have
\begin{equation}
    I_{n,t}=\Phi(K_{n,t+1}, K_{n,t}; \delta_{n,t})=\delta_{n,t}^{1-\frac{1}{\lambda}}K_{n,t}\left[\frac{K_{n,t+1}}{K_{n,t}}-(1-\delta_{n,t}) \right]^{\frac{1}{\lambda}}.
\end{equation}

The dynamic optimization problem of the representative household in country $n$ and period 0 is to maximize \eqref{eq: lifetime utility}
subject to \eqref{eq: expenditure}, \eqref{eq: budget constraint}, and \eqref{eq: capital dynamics}. Solving this problem, we obtain the Euler equation:{\small
    \begin{equation} \label{eq: Euler}
        \left( \frac{C_{n,t+1}/L_{n,t+1}}{C_{n,t}/L_{n,t}} \right)^{\psi-1} \frac{E_{n,t+1}\overline{\epsilon}_{n,t+1}}{E_{n,t}\overline{\epsilon}_{n,t}}
     = \beta \underbrace{\frac{\zeta_{n,t+1}}{\zeta_{n,t}}}_{\text{(i)}} \underbrace{\frac{L_{n,t+1}}{L_{n,t}}}_{\text{(ii)}} \underbrace{\frac{ (1-\phi_{n,t+1})r_{n,t+1} - P_{n,t+1}^K \Phi_2(K_{n,t+2},K_{n,t+1}; \delta_{n,t+1})}{P_{n,t}^K \Phi_1(K_{n,t+1},K_{n,t}; \delta_{n,t})}}_{\text{(iii)}},
    \end{equation} }
where $\Phi_l$ is the derivative of $I_{n,t}=\Phi(K_{n,t+1}, K_{n,t}; \delta_{n,t})$ with respect to its $l$-th argument,\footnote{ Specifically,
\begin{align*}
    &\Phi_{1}(K_{n,t+1}, K_{n,t})=\frac{\partial \Phi(K_{n,t+1}, K_{n,t})}{\partial K_{n,t+1}}=\frac{1}{\lambda}\delta_{n,t}^{1-\frac{1}{\lambda}}\left[\frac{K_{n,t+1}}{K_{n,t}}-(1-\delta_{n,t}) \right]^{\frac{1}{\lambda}-1}, \\
    &\Phi_{2}(K_{n,t+1}, K_{n,t}) =\frac{\partial \Phi(K_{n,t+1}, K_{n,t})}{\partial K_{n,t}} = \Phi_{1}(K_{n,t+1}, K_{n,t}) \cdot \left[ (\lambda-1)\frac{K_{n,t+1}}{K_{n,t}}-\lambda (1-\delta_{n,t}) \right].
\end{align*}
} and
    \begin{equation} \label{eq: average elasticity}
    \overline{\epsilon}_{n,t}=\sum_{j=a,m,s}\omega_{n,t}^{j}\epsilon^{j}.
    \end{equation}
    The associated transversality condition is also obtained. Both $E_{n,t+1}/E_{n,t}$ and $\overline{\epsilon}_{n,t+1}/\overline{\epsilon}_{n,t}$ are increasing in $\frac{C_{n,t+1}/L_{n,t+1}}{C_{n,t}/L_{n,t}}$. Since $\psi>1$, the left-hand side is just an increasing function of the ratio in per capita consumption between periods $t+1$ and $t$. We can see from \eqref{eq: Euler} that this per capita consumption ratio depends on the discount factor $\beta$, (i) the ratio in the intertemporal demand shifters, (ii) the ratio in populations, and (iii) the real return to capital.

We move on to producers' behavior. As in Section~\ref{sec: qual}, the sectoral composite good producers bundle input varieties $z\in[0,1]$ using a CES aggregator with elasticity $\eta>0$. The production function of variety $z$ of sector $j$ in country $n$ and period $t$ is
    \begin{equation} \label{eq: prod func}
    y_{n,t}^{j}(z)=a_{n,t}^{j}(z)\left[\frac{K_{n,t}^{j}(z)}{\gamma_{n,t}^{j}\alpha_{n,t}} \right]^{\gamma_{n,t}^{j}\alpha_{n,t}} \left[\frac{L_{n,t}^{j}(z)}{\gamma_{n,t}^{j}(1-\alpha_{n,t})} \right]^{\gamma_{n,t}^{j}(1-\alpha_{n,t})} \left[\frac{M_{n,t}^{j}(z)}{1-\gamma_{n,t}^{j}} \right]^{1-\gamma_{n,t}^{j}}.
    \end{equation}
Here $y_{n,t}^{j}(z)$ is the quantity of output, $a_{n,t}^{j}(z)$ is the productivity which will be expressed as a realization of a random variable, $K_{n,t}^{j}(z)$ is the capital, $L_{n,t}^{j}(z)$ is the labor, $\gamma_{n,t}^{j}\in (0, 1)$ is the cost share of value added (contribution by primary production factors) in total output, $\alpha_{n,t}\in (0, 1)$ is the share of capital within value added,\footnote{Note that the capital share within value added, $\alpha_{n,t}$, is common across sectors, consistent with the calibration in Section~\ref{subsec: param}.} $M_{n,t}^{j}(z)$ is the CES aggregate of sectoral intermediate inputs used for production of variety $z$, that is,
\begin{equation}
    M_{n,t}^{j}(z)=\left[\sum_{j'=a,m,s}(\kappa_{n,t}^{j,j'})^{\frac{1}{\sigma^{j}}}(M_{n,t}^{j,j'}(z))^{\frac{\sigma^{j}-1}{\sigma^{j}}} \right]^{\frac{\sigma^{j}}{\sigma^{j}-1}},
\end{equation}
where $\kappa_{n,t}^{j,j'}$ is the shifter for sector $j$'s demand for sector $j'$ composite good, $M_{n,t}^{j,j'}(z)$ is the input of sector $j'$ composite good for production of variety $z$ of sector $j$ and is produced using the sectoral composite goods, and $\sigma^{j}$ is the elasticity of substitution across sectoral composite goods for production of sector $j$ varieties. In sector $j$, the cost share of inputs from sector $j'$ \textit{within intermediate-input costs} is
\begin{equation}
    g_{n,t}^{j,j'}= \frac{P_{n,t}^{j'} M_{n,t}^{j,j'}}{\sum_{j''=a,m,s}P_{n,t}^{j''} M_{n,t}^{j,j''}}
    =\frac{\kappa_{n,t}^{j,j'}(P_{n,t}^{j'})^{1-\sigma^{j}}}{\sum_{j''=a,m,s}\kappa_{n,t}^{j,j''}(P_{n,t}^{j''})^{1-\sigma^{j}}}.
\end{equation}\par

The productivity of variety $z$ of sector $j$ in country $n$ and period $t$, $a_{n,t}^{j}(z)$, follows the Fr\'echet distribution whose CDF is given by $F_{n,t}^{j}(a)=\text{Pr}[a_{n,t}^{j}\le a]=\exp [-(a/(\widetilde{\gamma}^j A_{n,t}^{j} ))^{-\theta^{j}} ]$.
Unlike the two-country model, $\theta^j$ varies across sectors; consequently, so does $\widetilde{\gamma}^j = [\Gamma( (\theta^j+1-\eta)/\theta^j )]^{-\frac{1}{1-\eta}}$. Productivity of varieties is independent within and across sectors, countries, and periods.\par

The cost-minimization problem for the production function \eqref{eq: prod func} yields the cost of input bundle:
    \begin{equation} \label{eq: cost}
        \widetilde{c}_{n,t}^{j}=(r_{n,t})^{\gamma_{n,t}^{j}\alpha_{n,t}}(w_{n,t})^{\gamma_{n,t}^{j}(1-\alpha_{n,t})}(\xi_{n,t}^{j})^{1-\gamma_{n,t}^{j}},
    \end{equation}
where $\xi_{n,t}^{j}$ is the price index for the composite intermediate input for production of sector $j$ varieties:
\begin{equation}\label{eq: xi}
    \xi_{n,t}^{j}= \left[\sum_{j'=a,m,s}\kappa_{n,t}^{j,j'}(P_{n,t}^{j'})^{1-\sigma^{j}} \right]^{\frac{1}{1-\sigma^{j}}}.
\end{equation}
The price index of the composite good of sector $j$ in country $n$ and period $t$ is
    \begin{equation} \label{eq: price}
        P_{n,t}^{j}=\left[\sum_{i=1}^N\left(\frac{\widetilde{c}_{i,t}^{j}b_{ni,t}^{j}}{A_{i,t}^{j}} \right)^{-\theta^{j}} \right]^{-1/\theta^{j}},
    \end{equation}
where $b_{ni,t}^{j}$ is the total trade cost including tariffs and non-tariff trade barriers for sector $j$ from country $i$ to $n$. $b_{ni,t}^{j}$ is expressed as
\begin{equation}
    b_{ni,t}^{j}=d_{ni,t}^{j}(1+\tau_{ni,t}^{j}), \label{eq:b_to_tau_d}
\end{equation}
where $d_{ni,t}^{j}$ is the iceberg trade cost for sector $j$ varieties from country $i$ to $n$ in period $t$ including non-tariff barriers, and $\tau_{ni,t}^{j}$ is country $n$'s tariff rate on sector $j$ varieties from country $i$ in period $t$. For later use, define gross tariffs $\widetilde{\tau}_{ni,t}^{j}$ by $\widetilde{\tau}_{ni,t}^{j}=1+\tau_{ni,t}^{j}$.\par

The production function of capital (investment) goods in country $n$ and period $t$ is
\begin{equation} \label{eq: investment good}
    I_{n,t}= \left[\sum_{j=a,m,s}(\kappa_{n,t}^{K,j})^{\frac{1}{\sigma^{K}}}(M_{n,t}^{K,j})^{\frac{\sigma^{K}-1}{\sigma^{K}}} \right]^{\frac{\sigma^{K}}{\sigma^{K}-1}},
\end{equation}
where $M_{n,t}^{K,j}$ is the sector $j$ composite goods used for production of capital goods, and $\sigma^{K}$ is the elasticity of substitution across sectoral composite intermediates for production of capital goods. Then the cost share of sector $j$ inputs in capital goods production is
\begin{equation}
    g_{n,t}^{K,j}= \frac{P_{n,t}^{j}M_{n,t}^{K,j}}{ \sum_{j'=a,m,s}P_{n,t}^{j'}M_{n,t}^{K,j'} } =\frac{\kappa_{n,t}^{K,j}(P_{n,t}^{j})^{1-\sigma^{K}}}{\sum_{j'=a,m,s}\kappa_{n,t}^{K,j'}(P_{n,t}^{j'})^{1-\sigma^{K}}}.
\end{equation}
The ideal price index of capital goods is
\begin{equation} \label{eq: capital price index}
    P_{n,t}^{K}=\left[\sum_{j=a,m,s}\kappa_{n,t}^{K,j}(P_{n,t}^{j})^{1-\sigma^{K}} \right]^{\frac{1}{1-\sigma^{K}}}.
\end{equation}

Let $X_{ni,t}^{j}$ be country $n$'s spending on sector $j$ varieties sourced from country $i$ in period $t$. This includes spending on consumption, investment, and intermediate inputs. Summing $X_{ni,t}^{j}$ across $i$, let $X_{n,t}^{j}$ be country $n$'s spending on sector $j$ varieties in period $t$. Let $\pi_{ni,t}^{j}=X_{ni,t}^{j}/X_{n,t}^{j}$ be the share of goods sourced from country $i$ within country $n$'s expenditure on sector $j$ varieties in period $t$, or the trade share of country $i$ in the market of country $n$. Then we have
    \begin{equation} \label{eq: trade share}
        \pi_{ni,t}^{j}=\frac{(\widetilde{c}_{i,t}^{j}b_{ni,t}^{j}/A_{i,t}^{j})^{-\theta^{j}}}{\sum_{i'=1}^N(\widetilde{c}_{i',t}^{j}b_{ni',t}^{j}/A_{i',t}^{j})^{-\theta^{j}}}=
        \frac{(\widetilde{c}_{i,t}^{j}b_{ni,t}^{j} / A_{i,t}^{j})^{-\theta^{j}}}{(P_{n,t}^j)^{-\theta^j}}.
    \end{equation}

Let $Y_{n,t}^{j}$ be the value of gross production of sector $j$ in country $n$ and period $t$:
    \begin{equation} \label{eq: gross production}
        Y_{n,t}^{j}=\sum_{i=1}^N\frac{\pi_{in,t}^{j}}{\widetilde{\tau}_{in,t}^j}X_{i,t}^{j}.
    \end{equation}
Country $n$'s spending on sector $j$ varieties in period $t$ consists of the final consumption, the input for production of capital goods, and the input for production of goods and services across sectors:
    \begin{equation} \label{eq: spending}
    X_{n,t}^{j}=\omega_{n,t}^{j}E_{n,t}+g_{n,t}^{K,j}P_{n,t}^{K}I_{n,t}+
        \sum_{j'=a,m,s}(1-\gamma_{n,t}^{j'})g_{n,t}^{j',j}Y_{n,t}^{j'},
    \end{equation}
noting $\omega_{n,t}^{j}E_{n,t} = P_{n,t}^{j}C_{n,t}^j$ and $g_{n,t}^{K,j}P_{n,t}^{K}I_{n,t} = P_{n,t}^{j}M_{n,t}^{K,j}$. Country $n$'s tariff revenue is
\begin{equation} \label{eq: tariff revenue}
    \Tilde{T}_{n,t}=\sum_{j=a,m,s}X_{n,t}^{j}\sum_{i=1}^{N}\frac{\tau_{ni,t}^{j}\pi_{ni,t}^{j}}{\Tilde{\tau}_{ni,t}^{j}}.
\end{equation}

In country $n$ and period $t$, the aggregate labor income must be equal to the aggregate labor cost:
    \begin{equation} \label{eq: labor market clearing}
        w_{n,t}L_{n,t}=(1-\alpha_{n,t})\sum_{j=a,m,s}\gamma_{n,t}^{j} Y_{n,t}^{j}.
    \end{equation}
Similarly, the aggregate capital income must be equal to the aggregate capital cost:
    \begin{equation} \label{eq: capital market clearing}
        r_{n,t}K_{n,t}=\alpha_{n,t}\sum_{j=a,m,s}\gamma_{n,t}^{j}  Y_{n,t}^{j}.
    \end{equation}\par
The trade deficit, $D_{n,t}$, is imports minus exports:
\begin{equation} \label{eq: trade deficit def}
    D_{n,t}=\underbrace{\sum_{j=a,m,s}\sum_{i=1}^{N}X_{n,t}^{j}\frac{\pi_{ni,t}^{j}}{\widetilde{\tau}_{ni,t}^{j}}}_{\text{Imports}} - \underbrace{\sum_{j=a,m,s}\sum_{i=1}^{N}X_{i,t}^{j}\frac{\pi_{in,t}^{j}}{\widetilde{\tau}_{in,t}^{j}}}_{\text{Exports}}.
\end{equation}
Country $n$'s trade deficit must be equal to its net payment to the global portfolio:
\begin{equation} \label{eq: trade deficit portfolio}
    D_{n,t}=L_{n,t}T_{t}^{P}-\phi_{n,t}\left( w_{n,t}L_{n,t}+r_{n,t}K_{n,t}+\widetilde{T}_{n,t} \right).
\end{equation}
The budget balance of the global portfolio requires $\sum_{n=1}^{N} D_{n,t}=0$. Solving this for the payment from the global portfolio to each individual, $T_{t}^{P}$, we have
\begin{equation} \label{eq: payment from portfolio}
    T_{t}^{P}=\frac{\sum_{n=1}^{N}\phi_{n,t}\left( w_{n,t}L_{n,t}+r_{n,t}K_{n,t}+\widetilde{T}_{n,t} \right)}{\sum_{n=1}^{N}L_{n,t}}.
\end{equation}

We now define the equilibrium of our dynamic model. We also define the steady state because we compute transition paths that are equilibria converging to steady states. The definitions of equilibria and steady states are given below.

\

\begin{definition}[Equilibrium]\label{def: eq}
  Given the capital stocks in the initial period $\{K_{n,0}\}_n$, an equilibrium is a tuple of $\{w_{n,t}\}_{n,t}$, $\{r_{n,t}\}_{n,t}$, $\{E_{n,t}\}_{n,t}$, $\{\widetilde{c}_{n,t}^{j}\}_{n,t,j}$, $\{P_{n,t}^{j}\}_{n,t,j}$, $\{\pi_{ni,t}^{j}\}_{n,i,t,j}$, $\{Y_{n,t}^j\}_{n,t,j}$, $\{X_{n,t}^{j}\}_{n,t,j}$, $\{\overline{\epsilon}_{n,t}\}_{n,t}$, $\{\omega_{n,t}^{j}\}_{n,t,j}$, $\{C_{n,t}\}_{n,t}$, $\{K_{n,t}\}_{n,t}$, $\{I_{n,t}\}_{n,t}$, $\{\Tilde{T}_{n,t} \}_{n,t}$, $\{T_{t}^{P} \}_{t}$ satisfying a system of equations \eqref{eq: expenditure}, \eqref{eq: expenditure share}, \eqref{eq: budget constraint}, \eqref{eq: capital dynamics}, \eqref{eq: Euler}, \eqref{eq: average elasticity}, \eqref{eq: cost}, \eqref{eq: price}, \eqref{eq: trade share}, \eqref{eq: gross production}, \eqref{eq: spending}, \eqref{eq: tariff revenue}, \eqref{eq: labor market clearing}, \eqref{eq: capital market clearing}, and \eqref{eq: payment from portfolio} for $n, i \in \{1,\dots,N \}$, $j \in \{a,m,s \}$, and $t \in \{0,1,\dots \}$.\\
\end{definition}

\vspace{-0.4cm}

\begin{definition}[Steady state]\label{def: ss}
    A steady state is an equilibrium in which relevant endogenous variables are time-invariant. Specifically, a steady state is a tuple of $\{w_{n}\}_{n}$, $\{r_{n}\}_{n}$, $\{E_{n}\}_{n}$, $\{\widetilde{c}_{n}^{j}\}_{n, j}$, $\{P_{n}^{j}\}_{n,j}$, $\{\pi_{ni}^{j}\}_{n,i,j}$, $\{Y_{n}^{j}\}_{n,j}$, $\{X_{n}^{j}\}_{n,j}$, $\{\omega_{n}^{j}\}_{n,j}$, $\{C_{n}\}_{n}$, $\{K_{n}\}_{n}$, $\{\Tilde{T}_{n} \}_{n}$, $T^{P}$ satisfying a system of equations, \eqref{eq: expenditure}, \eqref{eq: expenditure share},
 \eqref{eq: cost}, \eqref{eq: price}, \eqref{eq: trade share}, \eqref{eq: gross production}, \eqref{eq: spending}, \eqref{eq: tariff revenue}, \eqref{eq: labor market clearing}, \eqref{eq: payment from portfolio},
\begin{equation} \label{eq: capital inc labor inc}
    r_{n}K_{n}=\frac{\alpha_{n}}{1-\alpha_{n}}w_{n}L_{n}, \qquad
    r_{n}=\frac{1-\beta(1-\lambda \delta_n)}{\beta(1-\phi_n)\lambda}P_{n}^{K},
\end{equation}
and
\begin{equation} \label{eq: expenditure capital}
    E_{n} = (1-\phi_n)\left( w_{n}L_{n} +r_n K_n + \widetilde{T}_n \right) - \delta_n P_n^K K_n + L_n T^P,
\end{equation}
for $n,i \in \{1,\dots,N \}$ and $j \in \{a,m,s \}$, where time subscripts $t$ are dropped from all the equations.
\end{definition}

In the following quantitative analysis, as in the qualitative analysis of Section~\ref{sec: qual}, we will focus on two measures of sectoral allocation: the sectoral consumption expenditure share $\omega_{n,t}^j = P_{n,t}^jC_{n,t}^j/E_{n,t}$ in \eqref{eq: expenditure share} and the sectoral value-added share:
\begin{align}\label{eq: va_nt^j}
    va_{n,t}^j = \frac{\gamma_{n,t}^j Y_{n,t}^j}{\sum_{j'=a,m,s}\gamma_{n,t}^{j'}Y_{n,t}^{j'}} = \frac{w_{n,t}L_{n,t}^j + r_{n,t}K_{n,t}^j}{\sum_{j'=a,m,s}\left( w_{n,t}L_{n,t}^{j'} + r_{n,t}K_{n,t}^{j'} \right)},
\end{align}
where $\gamma_{n,t}^j$ is the share of contribution by primary production factors (labor and capital) in sector $j$, $Y_{n,t}^j$ is the gross production of sector $j$ (see \eqref{eq: gross production}), and $L_{n,t}^j$ and $K_{n,t}^j$ are respectively labor and capital employed in sector $j$. Moreover, we will discuss saving decisions by looking at the saving rate $\rho_{n,t}$ defined as the ratio of capital investment to national income:
\begin{align}\label{eq: rho}
    \rho_{n,t} = \frac{P_{n,t}^K I_{n,t}}{(1-\phi_{n,t})(w_{n,t} L_{n,t} + r_{n,t} K_{n,t} + \widetilde{T}_{n,t}) + L_{n,t} T_t^P}.
\end{align}

\

\section{Calibration and Fit of the Model}\label{sec: calib}

We bring the model to the data for the global economy. We first describe our main data sources and discuss the calibration of the model. We then present the solution algorithm for computing transition paths. Lastly, we discuss the fit of the model.

\subsection{Data}\label{subsec: data}

Our primary data source is the World Input-Output Database (WIOD) Release 2016 and the Long-Run WIOD \citep{Woltjer2021, Timmer2015}, which allows us to observe the intermediate input uses across different countries and sectors of both origins and destinations. By merging the two datasets, we construct a database covering 1990--2014. Our empirical exercise encompasses 24 countries (see Table~\ref{tab: welfare main}) and the rest of the world (RoW). These are the countries listed in the Long-Run WIOD, and we classify Hong Kong as part of the RoW. We aggregate the ISIC industries into three sectors (see Table~\ref{Tab: sector_list} in Supplemental Appendix~\ref{ap: sector}).
We complement the WIOD data with the Penn World Table (PWT) 10.0 \citep{Feenstra2015} and CEPII Gravity database \citep{Mayer2011}. Bilateral tariffs on goods are sourced from the World Integrated Trade Solution (WITS).\footnote{As the bilateral tariffs are specified at each Harmonized System (HS) product level, we first map the HS products to the ISIC industries using the concordance table provided by the WITS and then compute the simple average for agriculture and manufacturing sectors.}

\subsection{Calibration of Structural Parameters and Fundamentals} \label{subsec: param}

Shares of primary production factors in the total production cost $\gamma_{n,t}^j$ are directly observed in the IO table. Cost shares of capital within value added $\alpha_{n,t}$ are calibrated as one minus labor shares, which are obtained from the PWT. Since the PWT does not provide the sectoral labor share, we apply the common value across sectors for each year and country. We set the elasticity of substitution across intermediate inputs $\sigma^j=0.38$ for all $j$ following \cite{Atlay2017}. For the capital goods production, we set the elasticity of substitution $\sigma^K=0.29$ following \citet{Sposietal2026}. Lower-than-one elasticities of substitution indicate that the relative price effect applies both to the production of goods and services and to the production of investment goods. Shape parameters of the Fr\'echet distribution are calibrated based on the estimates of \citet{CaliendoParro2015}. We choose $\theta^a =8.11$ and $\theta^m=4.55$. For the service trade elasticity, we follow \cite{Gervais2019} and set $\theta^s=0.75 \cdot \theta^m$. We set the adjustment cost elasticity in the law of motion for capital $\lambda=0.75$ following \citet{Eaton2016}, and the depreciation rate of capital $\delta_{n,t}$ is obtained from the PWT.

We calibrate the sequences of iceberg trade costs $\{d_{ni,t}^{j}\}$, productivity $\{A_{n,t}^{j}\}$, and input cost shares $\{g_{n,t}^{j,j'} \}$ and $\{g_{n,t}^{K,j} \}$ by exploiting the data from the WIOD and the PWT. Demand shifters $\{\Omega_{n,t}^{j} \}$ in nonhomothetic CES preferences \eqref{eq: CLM utility} are calibrated such that the consumption expenditure shares implied by the model (given the sectoral price indices) $\{\omega_{n,t}^{j}\}$ are matched with the data. Note that we calibrate $\{\Omega_{n,t}^{j} \}$ separately for nonhomothetic and homothetic CES preferences. See Supplemental Appendix~\ref{ap: shock} for details.

\subsection{Solution Algorithm}\label{subsec: algorithm}

We solve the equilibrium transition path backward. We first solve the model for the steady state according to Definition~\ref{def: ss}, assuming the 2014 fundamentals (e.g., productivity, trade costs, exogenous demand shifters) remain constant forever. We then suppose that the economy will reach the steady state 450 years after 2014. The solution algorithm for the transition path has two loops: the outer loop finds the sequence of investment (saving) rates, $\{\rho_{n,t} \}_{n,t}$ in \eqref{eq: rho}, that satisfies the dynamic optimality condition governed by the Euler equation \eqref{eq: Euler}, and the inner loop solves the intratemporal equilibrium for each period (i.e., solving the sectoral prices and factor prices that satisfy the equilibrium conditions listed in Definition~\ref{def: eq}). More specifically, for the given sequence of $\{\rho_{n,t} \}_{n,t}$ and the initial period capital stock, we first solve the static equilibrium period-by-period sequentially from 1990 to 2464. After we solve the periodic equilibria up to the year 2464, we update $\rho_{n,t}$ backward from 2464 to 1990 according to the Euler equation. More details are in Supplemental Appendices~\ref{ap: steady state} and~\ref{ap: transition paths} (see also \citealp{Ravikumar2019} for the details of the outer loop iteration).

\subsection{Fit of the Baseline Model}

\begin{figure}[h!]%
    \subfloat[\centering Canada ]{{\includegraphics[width=0.33\textwidth]{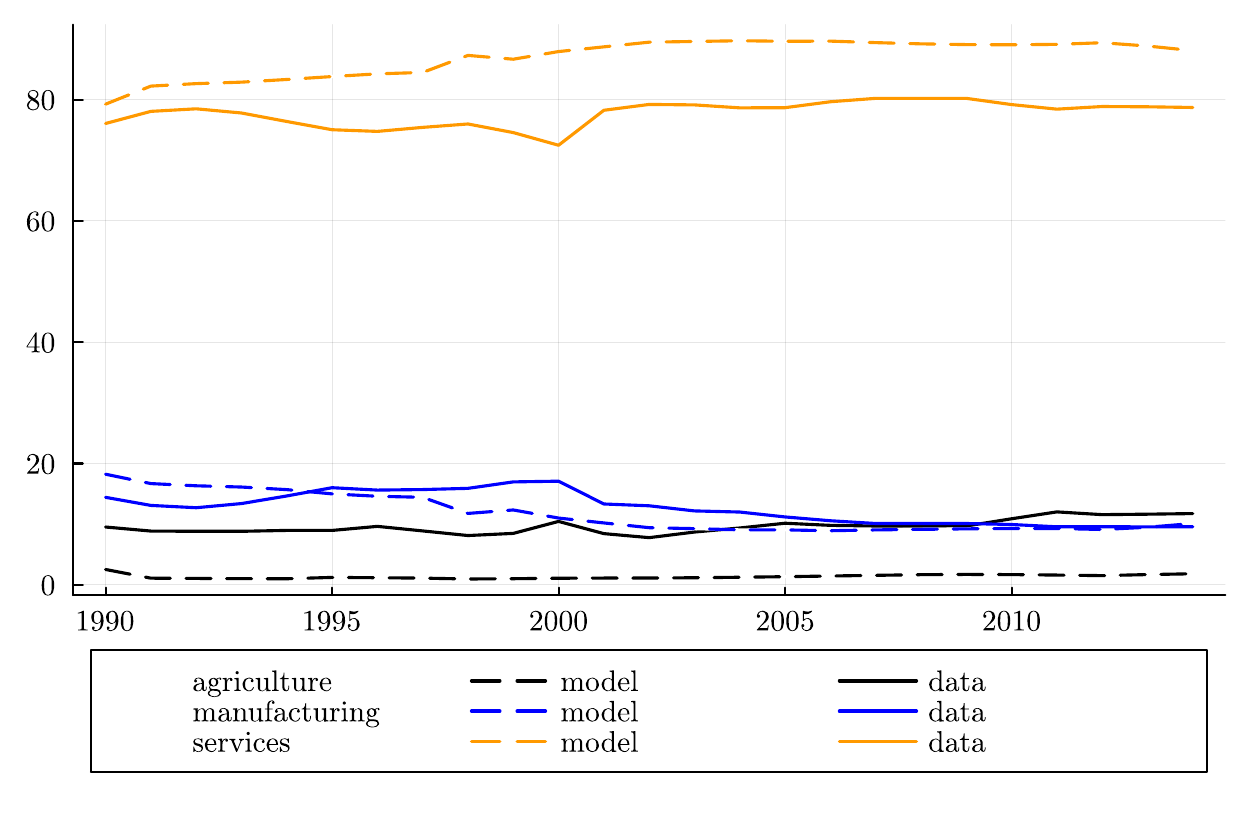} }}%
    \subfloat[\centering China ]{{\includegraphics[width=0.33\textwidth]{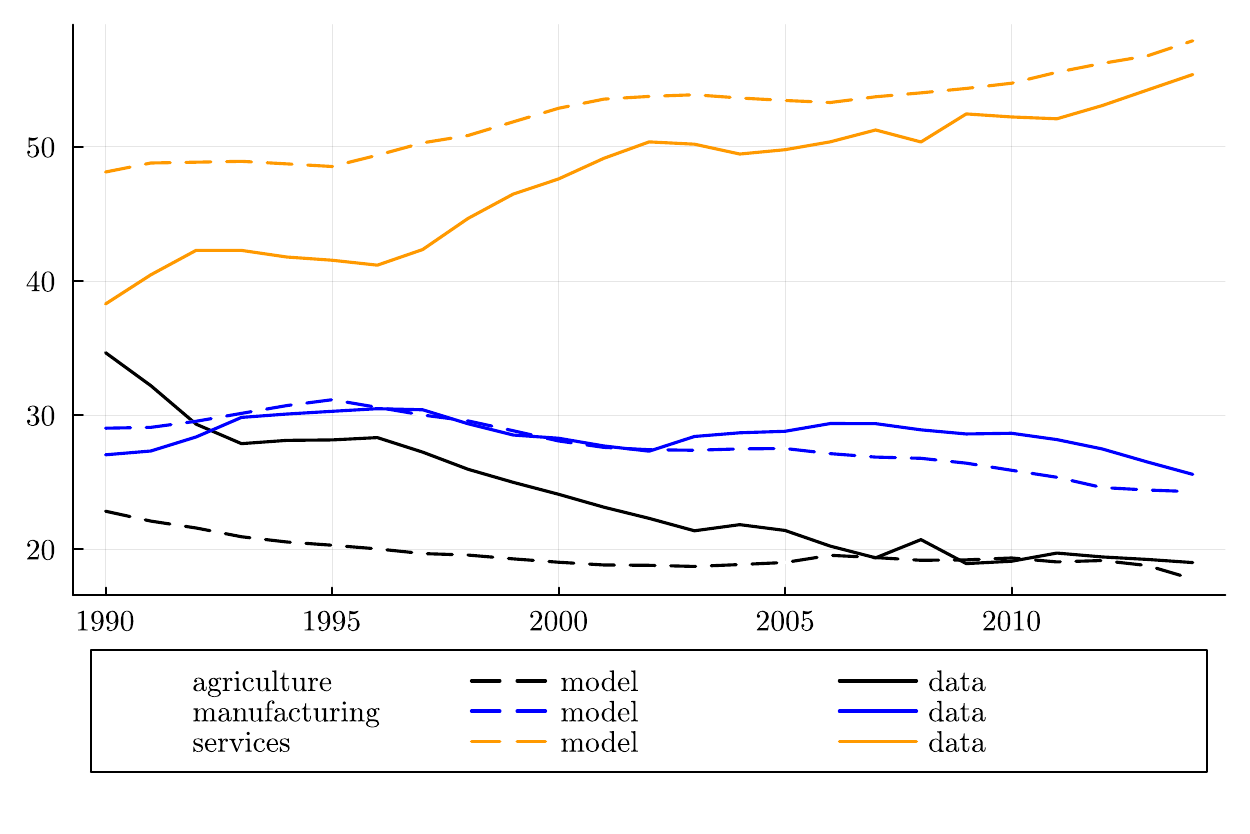} }}%
    \subfloat[\centering Germany ]{{\includegraphics[width=0.34\textwidth]{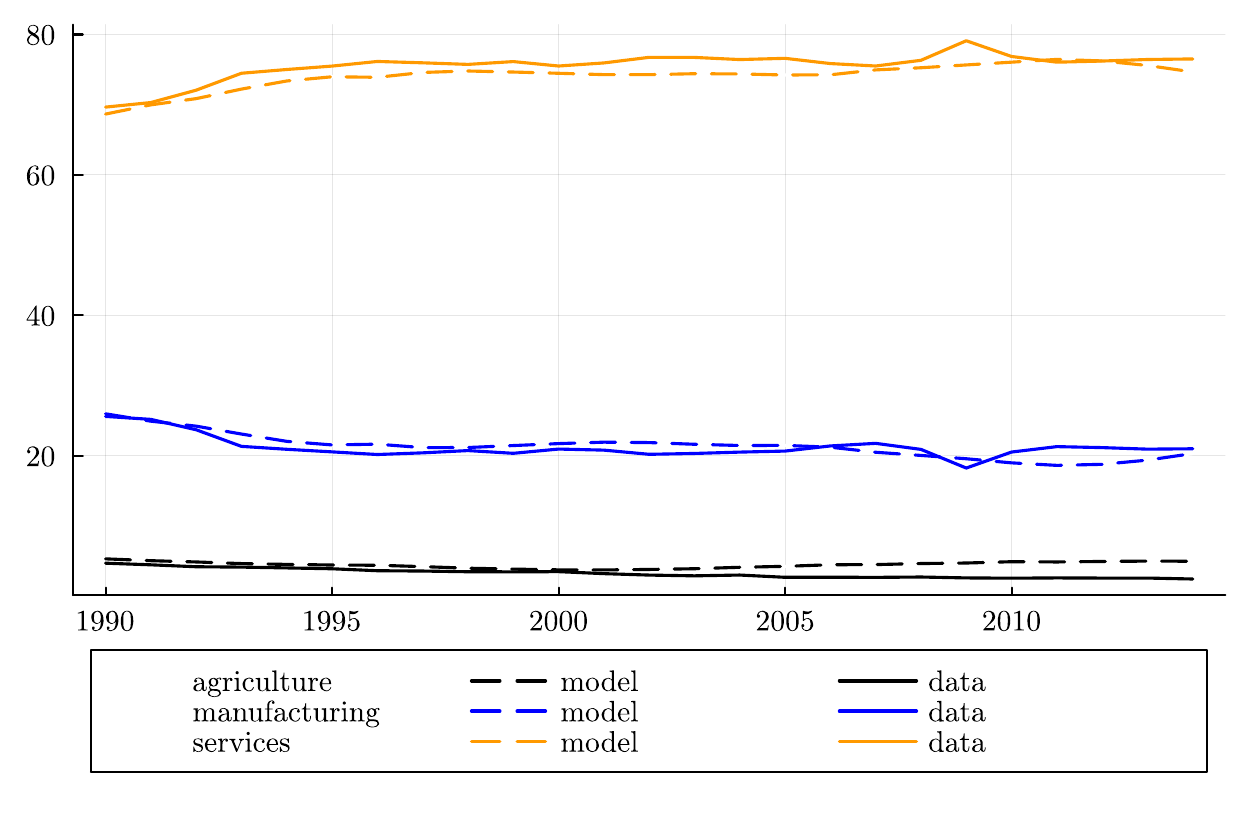} }}
    \quad
    \subfloat[\centering Japan ]{{\includegraphics[width=0.34\textwidth]{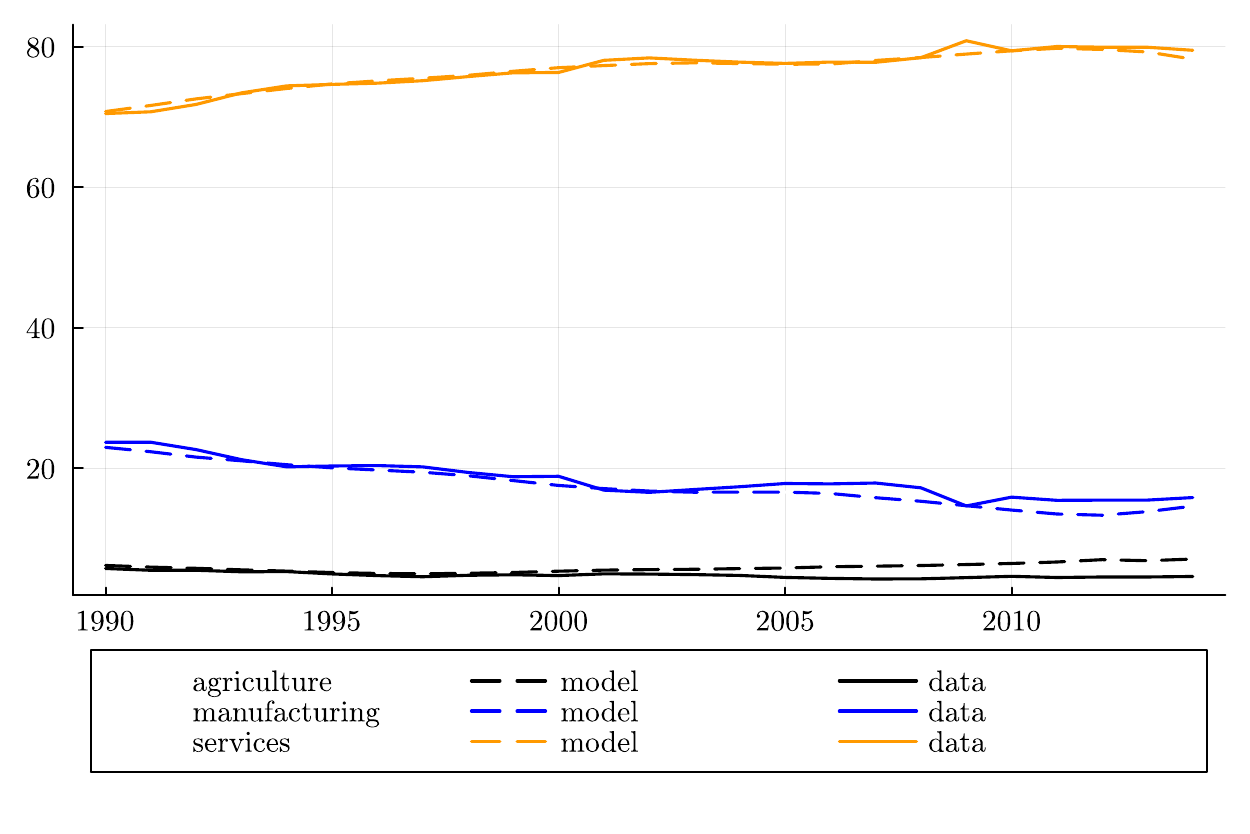} }}%
    \subfloat[\centering Mexico  ]{{\includegraphics[width=0.34\textwidth]{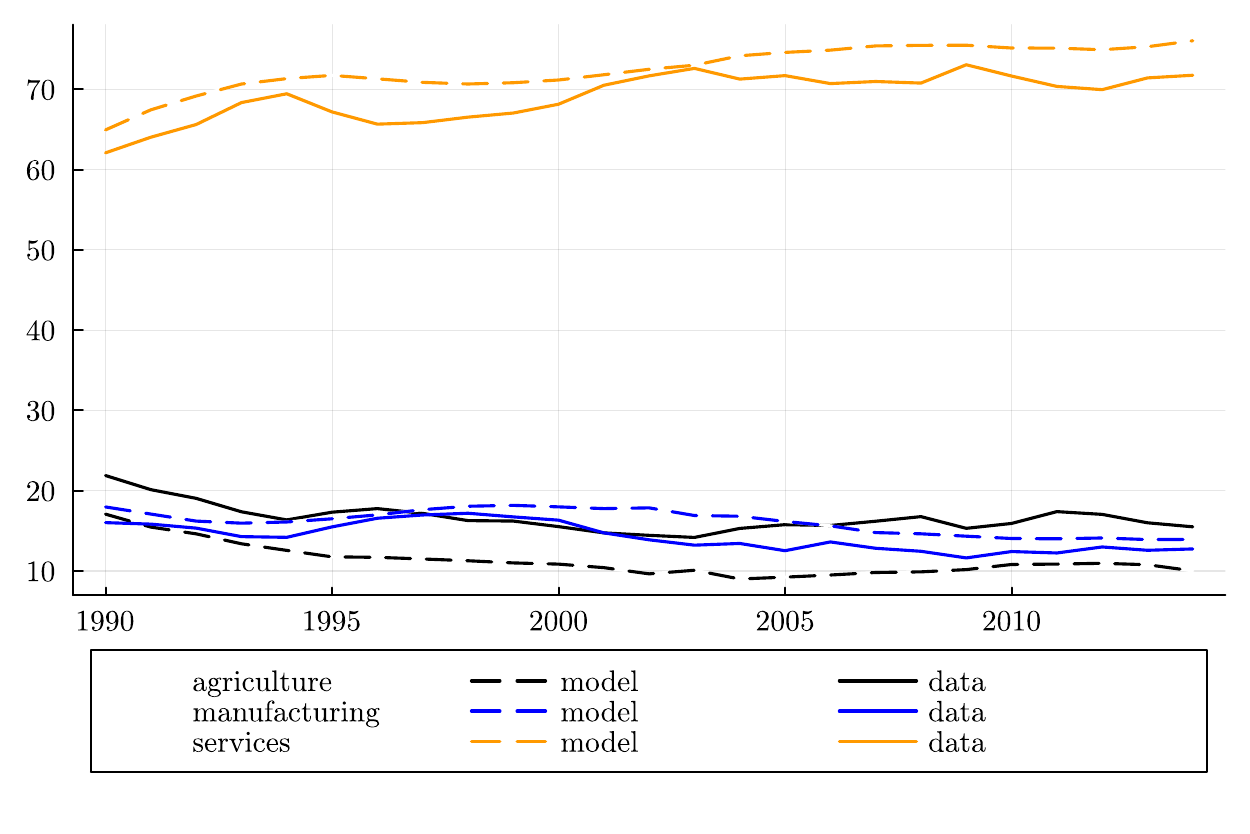} }}
    \subfloat[\centering U.S. ]{{\includegraphics[width=0.33\textwidth]{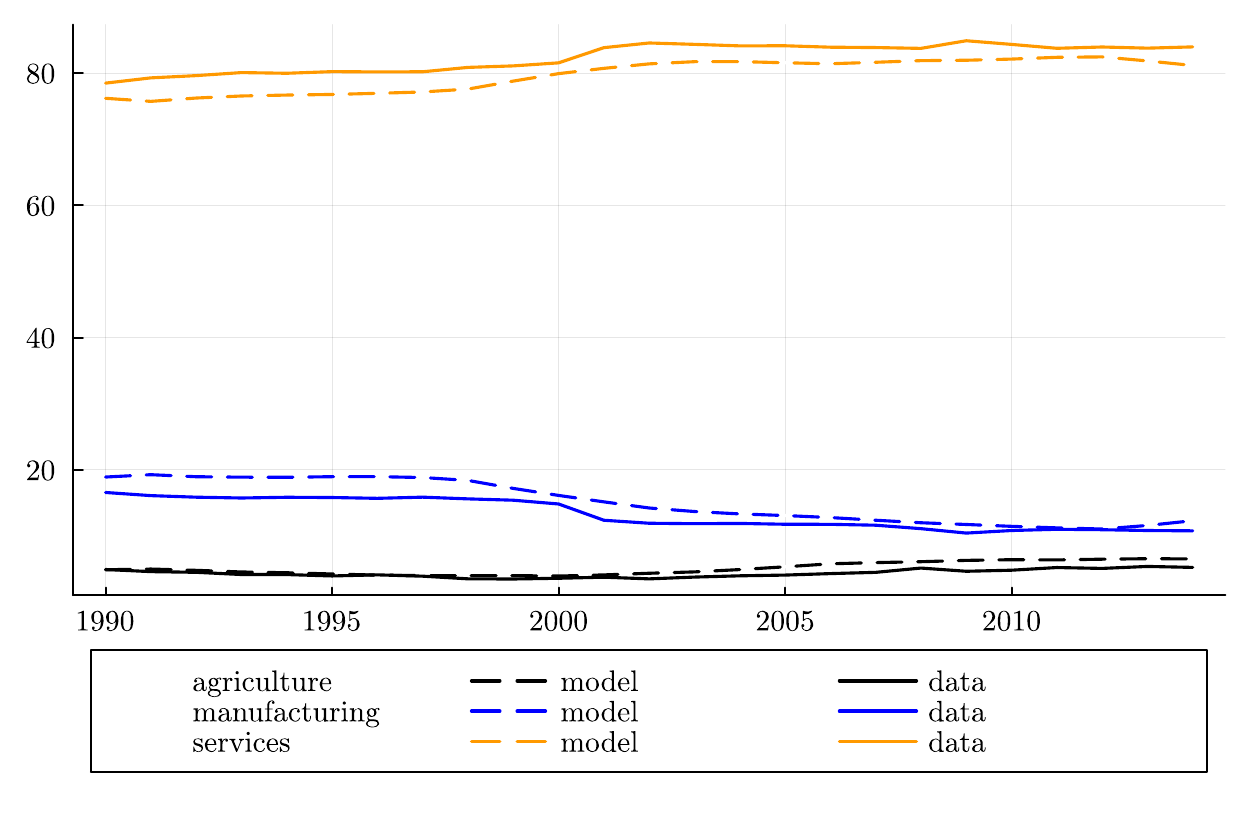} }}
    \centering
    \caption{Model fit: Sectoral value-added share}\label{fig: bl_vashare}%
    \begin{figurenotes}
       Each panel shows the sectoral value-added shares in the baseline equilibrium (dashed lines) and in the data (solid lines).
    \end{figurenotes}
\end{figure}

To examine the model's ability to match the data, Figure~\ref{fig: bl_vashare} compares the model-implied (dashed lines) value-added shares in three sectors (black for agriculture, blue for manufacturing, and orange for services), $va_{n,t}^j$ in \eqref{eq: va_nt^j}, with the data counterparts (solid lines) for six selected countries, Canada, China, Germany, Japan, Mexico, and the U.S. In the six countries, the model captures the overall trend of falling manufacturing and rising services over time. The fit of the model is notably better for Germany, Japan, and the U.S., while in the other three countries, the model overestimates the service share and underestimates the agricultural share, despite the overall trend being well captured.

The model fit of sectoral expenditure shares in final consumption, $\omega_{n,t}^j = P_{n,t}^j C_{n,t}^j/E_{n,t}$ and saving rates, $\rho_{n,t}$ in \eqref{eq: rho} is included in Supplemental Appendix~\ref{ap: fit}. The model captures the shift of final expenditure from agriculture to manufacturing, and then to services over time, while across all countries, it overpredicts service expenditure and underestimates agricultural expenditure.

\section{Counterfactuals}\label{sec: counterfactuals}

We now use the model to examine the impacts of tariffs on the sectoral allocation of economic activities. We first consider a permanent rise in the U.S. import tariffs on manufacturing by 20 percentage points, effective starting in 2001 (U.S. unilateral tariffs). The year 2001 is chosen given the significant decline in U.S. manufacturing, which has never recovered since then. We keep all the fundamentals, such as non-tariff barriers and productivity, unchanged from the baseline to isolate the pure effect of the tariff. We compare the results to those under different preferences. We then consider the trade war scenario in which U.S. manufacturing exports are also subject to a 20-percentage-point higher tariff by all other countries, effective starting in 2001 and lasting indefinitely.
Finally, in order to find the U.S. optimal manufacturing tariffs, we compute U.S. welfare for various time-invariant manufacturing tariffs starting in 2001. In this exercise, the U.S. unilaterally sets manufacturing tariffs, and other countries do not retaliate.
In what follows, we treat the increase in tariffs as a surprise shock; everyone realizes the tariff hike in 2001, while people have perfect foresight regarding the tariff schedule from then onward.

\subsection{U.S. Unilateral Tariffs}\label{sec: unilateral tariffs}

\begin{figure}[tb!]%
    \subfloat[\centering Real per capita consumption]{{\includegraphics[width=0.45\textwidth]{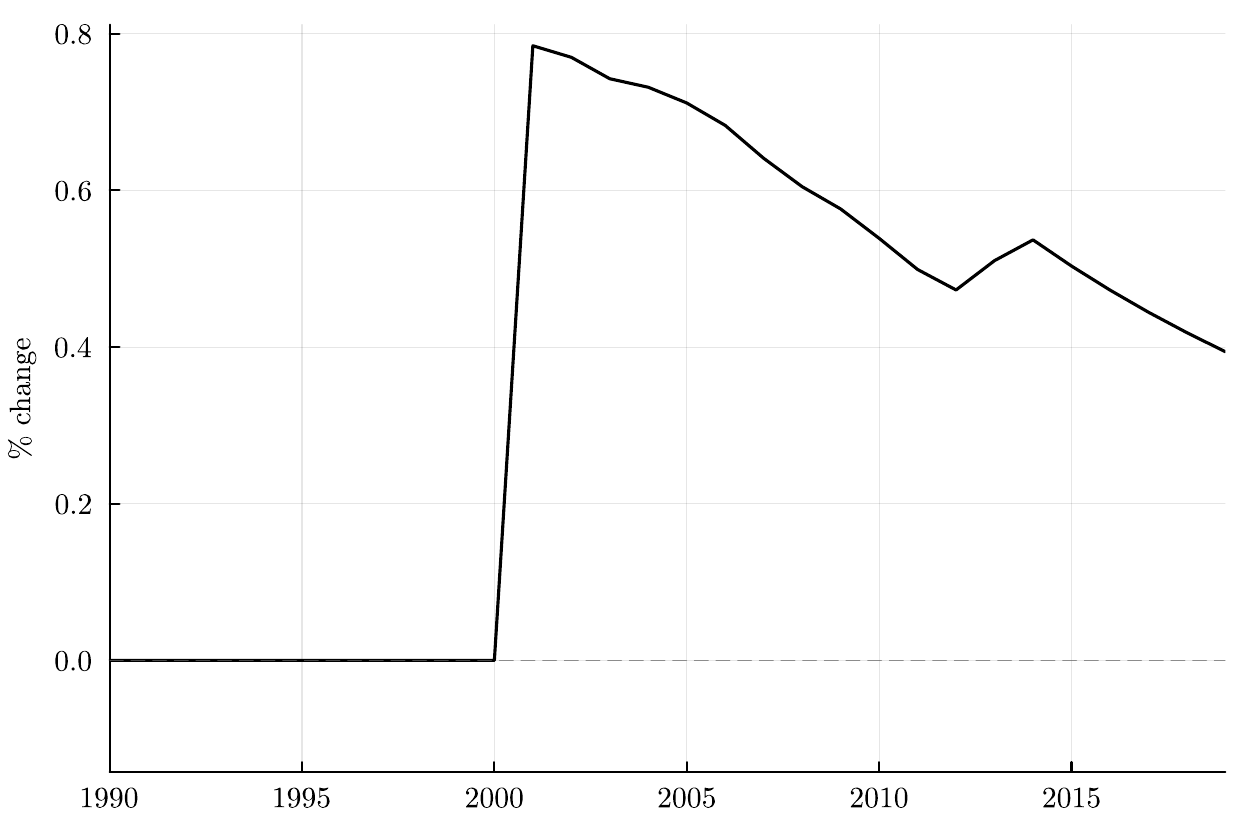} }}%
    \subfloat[\centering Saving rate]{{\includegraphics[width=0.45\textwidth]{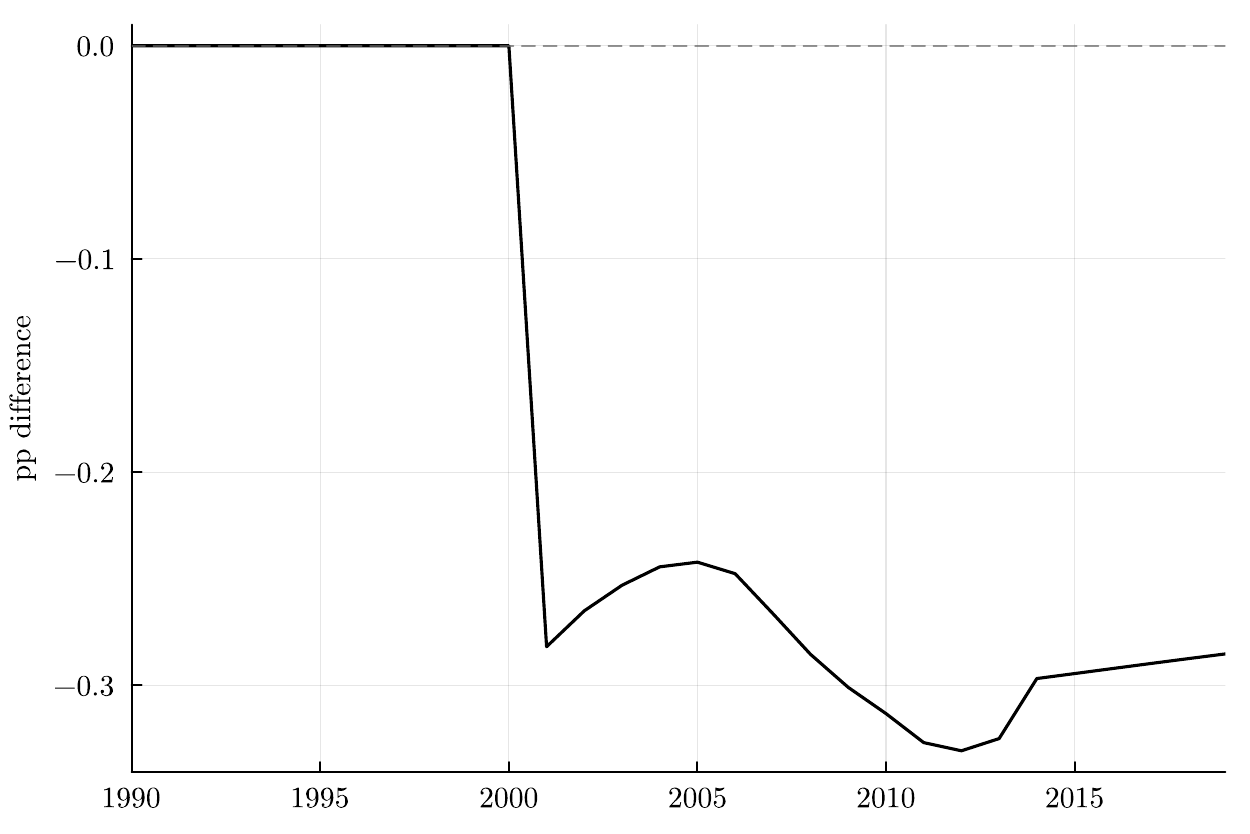} }}%
    \quad
    \subfloat[\centering Sectoral expenditure share]{{\includegraphics[width=0.45\textwidth]{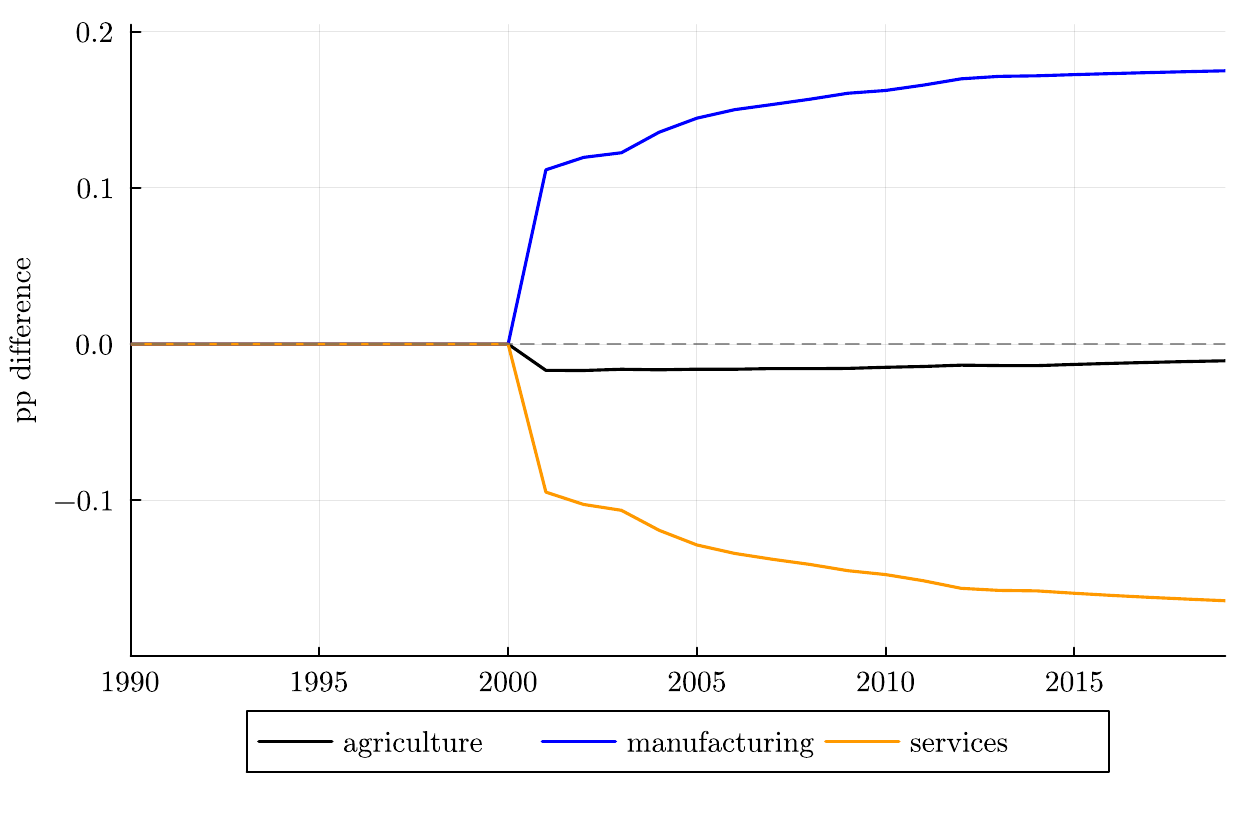} }}%
    \subfloat[\centering Sectoral value-added share]{{\includegraphics[width=0.45\textwidth]{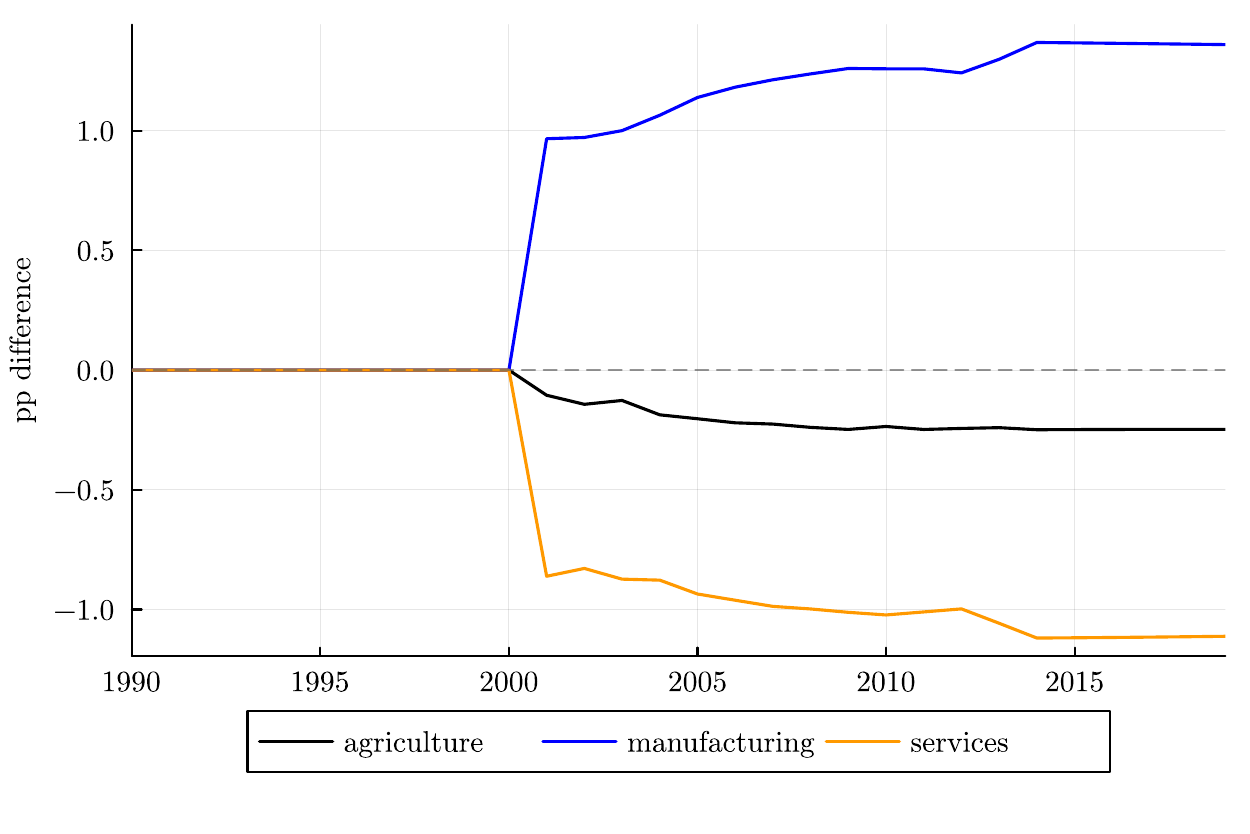} }}%
    \centering
    \caption{Impacts of a U.S. manufacturing tariff increase}\label{fig: newtradepolicy_4_nh}%
    \begin{figurenotes}%
        Each panel shows the impacts of a counterfactual 20-percentage-point increase in the U.S. manufacturing tariff since 2001 on the transition paths in the U.S. For real per capita consumption, the vertical axis represents the percent change from the baseline to the counterfactual equilibrium. For the other variables, the vertical axis measures the percentage point change from the baseline.
    \end{figurenotes}
\end{figure}

Figure~\ref{fig: newtradepolicy_4_nh} summarizes the impacts of a 20-percentage-point increase in U.S. unilateral manufacturing tariffs on the four key variables in the U.S., (A) real per capita consumption, (B) saving rate, (C) sectoral expenditure share, and (D) sectoral value-added share.\footnote{The choice of a 20-percentage-point increase is unrelated to the optimal tariff, which is slightly above 20 percent. The similarity between the two numbers is purely coincidental.}  As in Figure~\ref{fig: tradecost_tau} in Supplemental Appendix~\ref{ap: baseline parameters}, the average U.S. manufacturing tariff is about 3--4 percent. In this exercise, the U.S. manufacturing tariffs increase to 23--24 percent on average. All results are shown relative to the baseline transition path.

\vspace{0.1cm}

\textbf{Real Per Capita Consumption and Saving:} \ \
Panel (A) of Figure~\ref{fig: newtradepolicy_4_nh} shows the percent changes in real per capita consumption $C_{US,t}/L_{US,t}$ from the baseline to this counterfactual equilibrium. U.S. real per capita consumption is 0.78 percent higher than in the baseline equilibrium in 2001. In relation to Proposition~\ref{prop: proposition 1}, this result confirms that a 20-percentage-point increase in the tariff is small enough to raise the real consumption. The figure also shows that the difference between the baseline and counterfactual is largest in the first year after the shock, and it diminishes over time. This is because the household has an incentive to front-load consumption.

Consistent with the intuition discussed in Section~\ref{sec: two-period}, the household anticipating the permanent lump-sum transfer of tariff revenue is disincentivized to save, leading to a lower-than-baseline saving rate in the counterfactual.
Panel (B) shows that the saving rate, $\rho_{US,t}$ in \eqref{eq: rho}, is approximately 0.3 percentage points lower than the baseline equilibrium. A lower saving rate implies a lower capital stock, resulting in lower real consumption in the long run. While the real per capita consumption remains above the baseline level for the first 55 years after the shock, it falls below the baseline level afterward.

\vspace{0.1cm}

\textbf{Sectoral Expenditure Share:} \ \ Panel (C) of Figure~\ref{fig: newtradepolicy_4_nh} shows the impact on sectoral consumption expenditure shares, $\omega_{US,t}^j = P_{US,t}^jC_{US,t}^j/E_{US,t}$, relative to the baseline equilibrium. As we discussed in Proposition~\ref{prop: proposition 2}, under nonhomothetic CES preferences, there are two competing effects of tariffs. On the one hand, a higher relative price of manufacturing shifts the expenditure share to manufacturing due to the relative price effect. On the other hand, if tariffs raise U.S. real per capita consumption ($d(C_{US,t}/L_{US,t}) > 0$) and its expenditure-weighted income elasticity is sufficiently high ($\overline{\epsilon}_{US,t}=\sum_j \epsilon^j \omega_{US,t}^j > 1$), the income effect counteracts the first effect. In our counterfactual scenario where both $d(C_{US,t}/L_{US,t})>0$ (panel (A)) and $\overline{\epsilon}_{US,t}>1$ hold, the two effects indeed move in opposite directions, but the relative price effect dominates; the manufacturing expenditure share in the U.S. rises by approximately 0.15 percentage points compared to the baseline equilibrium.

\vspace{0.1cm}

\textbf{Sectoral Value-added Share:} \ \ Panel (D) of Figure~\ref{fig: newtradepolicy_4_nh}  represents the impacts on sectoral value-added shares, $va_{US,t}^j$ in \eqref{eq: va_nt^j}. A 20-percentage-point additional tariff on U.S. manufacturing imports leads to approximately a one-percentage-point increase in the manufacturing value-added share, which primarily comes at the expense of a lower share in the service sector. This is in line with our intuition that manufacturing tariffs have qualitatively similar effects on both sectoral expenditure shares and value-added shares  (Proposition~\ref{prop: proposition 3}).

In terms of magnitude, the sectoral value-added shares respond more to the tariff shock than the sectoral expenditure shares. The manufacturing expenditure share under the counterfactual is approximately 0.15 percentage points higher than the baseline, whereas its value-added share is 0.9--1.4 percentage points higher. This discrepancy is due to our quantitative model featuring international trade as well as input-output linkages through intermediate inputs.\footnote{In other words, if we abstract from these two features, the sectoral expenditure shares coincide with the value-added shares as in \citet{CLM}.}

\vspace{0.1cm}

\textbf{Welfare:} \ \ We evaluate the overall welfare implications of the tariff shock measured by consumption equivalent from the viewpoint of 2001 (see Supplemental Appendix~\ref{ap: welfare formula} for the formula). The second column of Table~\ref{tab: welfare main} shows that the U.S. welfare rises by 0.43 percent while all other countries are worse off.\footnote{Real per capita consumption of the U.S. will fall below the baseline level in six decades after the shock. Yet, our welfare measures discount the future real consumption, resulting in a positive welfare implication.}
The imposed tariff directly lowers U.S. imports from other countries, thereby reducing their factor prices and welfare due to the terms-of-trade effect. Canada and Mexico experience the largest welfare losses (1.58 and 0.84 percent, respectively), largely due to the U.S. being their primary trading partner and a major importer of their manufacturing goods.

\begin{table}[!t]
\small
\begin{center}
\caption{Welfare impacts of the U.S. tariffs}
\label{tab: welfare main}
\begin{tabular}{crrcrr} \toprule
 & \multicolumn{2}{c}{Unilateral U.S.\ tariffs} & \multicolumn{1}{c}{} & \multicolumn{2}{c}{Retaliatory tariffs} \\ \cmidrule{2-3} \cmidrule{5-6}
 & \multicolumn{1}{c}{nh CES} & \multicolumn{1}{c}{h CES} & \multicolumn{1}{c}{} & \multicolumn{1}{c}{nh CES} & \multicolumn{1}{c}{h CES} \\ \midrule
Australia & -0.057 & -0.063 &  & -0.073 & -0.083 \\
Austria & -0.148 & -0.169 &  & -0.106 & -0.120 \\
Belgium & -0.136 & -0.154 &  & -0.102 & -0.116 \\
Brazil & -0.219 & -0.240 &  & -0.161 & -0.177 \\
Canada & -1.579 & -1.884 &  & -1.330 & -1.595 \\
China & -0.377 & -0.448 &  & -0.225 & -0.264 \\
Germany & -0.099 & -0.109 &  & -0.072 & -0.080 \\
Denmark & -0.059 & -0.069 &  & -0.050 & -0.059 \\
Spain & -0.118 & -0.133 &  & -0.090 & -0.102 \\
Finland & -0.161 & -0.187 &  & -0.116 & -0.134 \\
France & -0.086 & -0.097 &  & -0.064 & -0.072 \\
U.K. & -0.080 & -0.090 &  & -0.056 & -0.064 \\
Greece & -0.191 & -0.207 &  & -0.157 & -0.170 \\
India & -0.751 & -0.826 &  & -0.438 & -0.476 \\
Ireland & -0.088 & -0.101 &  & -0.080 & -0.094 \\
Italy & -0.122 & -0.136 &  & -0.090 & -0.100 \\
Japan & -0.077 & -0.087 &  & -0.050 & -0.056 \\
Korea & -0.171 & -0.193 &  & -0.139 & -0.156 \\
Mexico & -0.835 & -0.906 &  & -0.926 & -1.016 \\
Netherlands & -0.123 & -0.139 &  & -0.094 & -0.107 \\
Portugal & -0.165 & -0.184 &  & -0.134 & -0.150 \\
Rest of World & -0.127 & -0.146 &  & -0.092 & -0.106 \\
Sweden & -0.212 & -0.241 &  & -0.150 & -0.170 \\
U.S. & 0.433 & 0.487 &  & -0.165 & -0.205 \\
Turkey & -0.174 & -0.191 &  & -0.105 & -0.114 \\ \bottomrule
\end{tabular}

\end{center}\begin{flushleft}
{\small \textit{Note}: The table shows percent changes in welfare (consumption equivalent, see Supplemental Appendix~\ref{ap: welfare formula}) under nonhomothetic CES (nh CES) and homothetic CES preferences (h CES) in the two different counterfactual scenarios, U.S. unilateral tariffs in Section~\ref{sec: unilateral tariffs} and trade war in Section~\ref{sec: trade war}.}
\end{flushleft}
\end{table}

\vspace{0.1cm}

\begin{figure}[tb!]%
    \subfloat[\centering Manufacturing expenditure share]{{\includegraphics[width=0.5\textwidth]{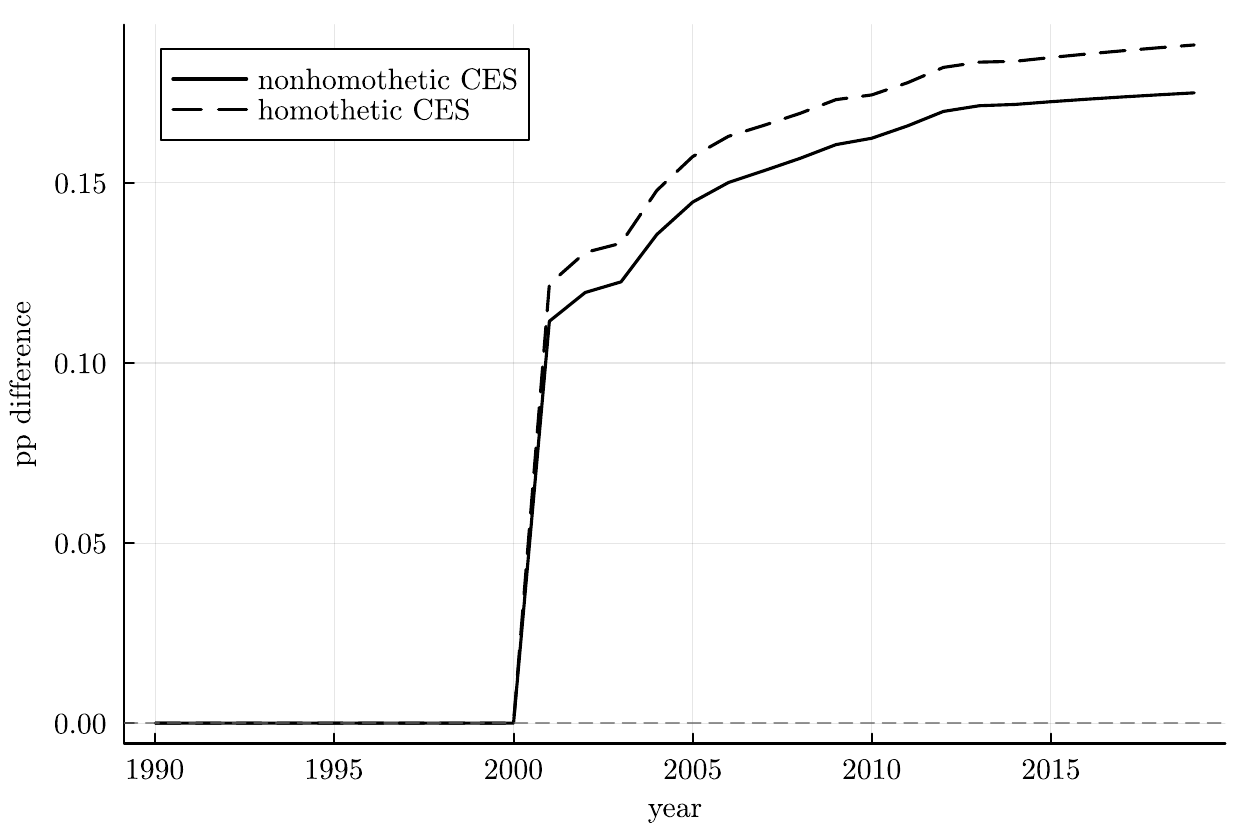} }}%
    \subfloat[\centering Manufacturing value-added share]{{\includegraphics[width=0.5\textwidth]{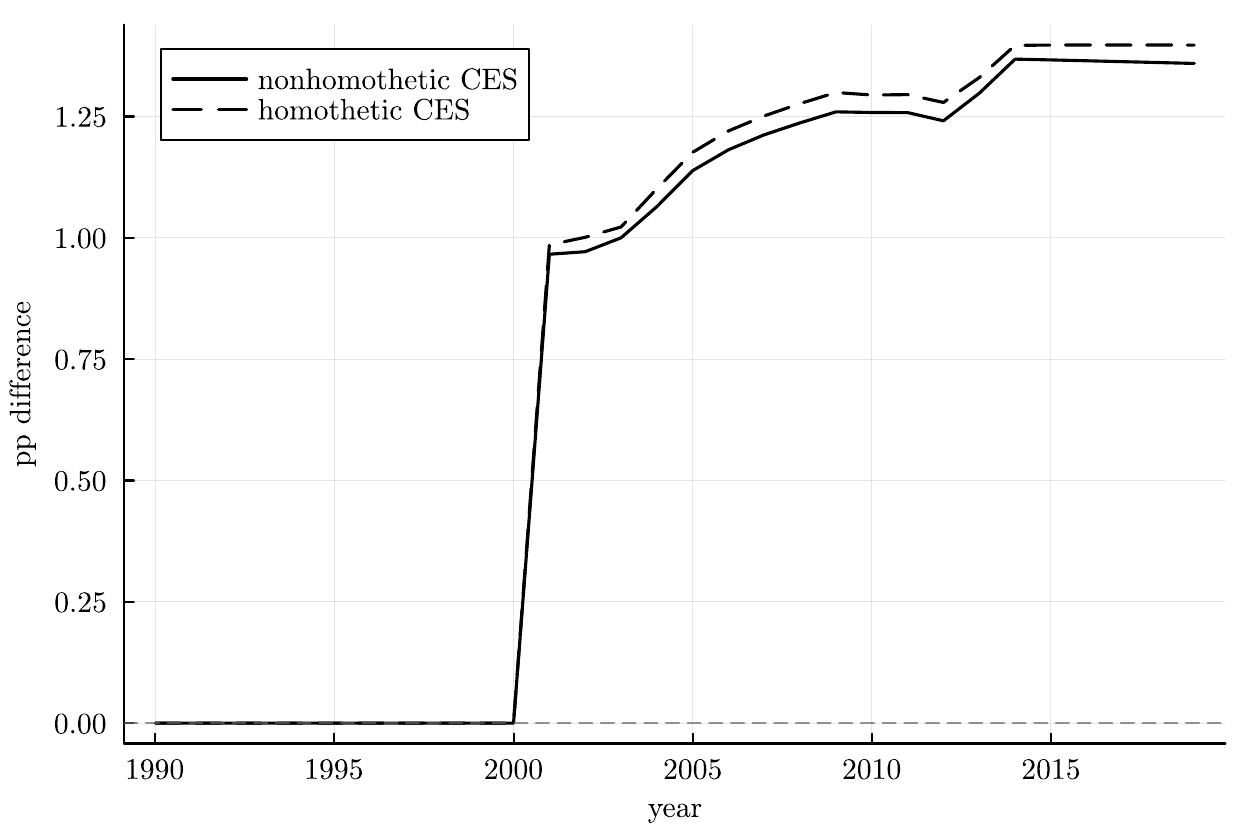} }}%
    \centering
    \caption{Comparison of impacts under different preferences}\label{fig: newtradepolicy_4_nh_h_cd}%
    \begin{figurenotes}
        Each panel shows the impacts of a counterfactual 20-percentage-point increase in the U.S. manufacturing tariff since 2001 on the transition paths in the U.S. under the two different preferences, nonhomothetic CES (\verb|nh CES|, solid lines) and homothetic CES (\verb|h CES|, dashed lines). The vertical axes represent percentage point change from the baseline. The thick lines in panels (A) and (B) correspond, respectively, to the manufacturing lines in panels (C) and (D) of Figure~\ref{fig: newtradepolicy_4_nh}.
    \end{figurenotes}
\end{figure}

\textbf{Impacts under Different Preferences:} \ \ To highlight the role of nonhomotheticity in accounting for the impacts of tariffs on the sectoral allocation, Figure~\ref{fig: newtradepolicy_4_nh_h_cd} compares the results for manufacturing expenditure shares in panel (A) and value-added shares in panel (B) across nonhomothetic and homothetic CES preferences.

Under homothetic CES preferences, both the expenditure share and value-added share respond more in magnitude to the tariff shock than under nonhomothetic CES preferences. This is because homothetic CES preferences shut down the negative income effect on manufacturing shares, which counteracts the positive relative price effect. For the manufacturing expenditure shares presented in panel (A), the gap between the dashed and solid lines, 0.01 percentage point, measures the mitigating effect of the income channel. Panel (B) shows that the manufacturing value-added share is 0.02--0.04 percentage point higher under homothetic CES than under nonhomothetic CES preferences, a magnitude similar to that observed for expenditure shares.

Regarding the welfare implications, the second column of Table~\ref{tab: welfare main} shows that the welfare impacts are larger in magnitude under homothetic CES preferences than under nonhomothetic CES preferences for all countries except Australia. According to Proposition~\ref{prop: proposition 1}, homothetic CES preferences overestimate welfare impacts if the expenditure-weighted income elasticity is sufficiently high, $\overline{\epsilon}_{n,t} > 1$. Consistent with this qualitative result, we confirm that the inequality holds for all countries from 2001 onward.

\subsection{Trade War}\label{sec: trade war}

Next, we consider a trade war scenario: the U.S. imposes an additional 20-percentage-point tariff on all the manufacturing imports, and every other country retaliates against the U.S. by imposing tariffs on U.S. exports with the same magnitude. For example, in 2001, the other 24 countries' average manufacturing tariff against the U.S. is 7.0 percent in the data, so this increases to 27 percent.\footnote{These 24 countries include the RoW.} As before, the trade war is assumed to come as a surprise shock. Results are displayed in Figure~\ref{fig: newtradepolicy_7_nh} in Supplemental Appendix~\ref{ap: trade war}. Even with retaliatory tariffs from around the world, U.S. real per capita consumption rises for the first 10 years relative to the baseline, but then turns negative. The saving rate is lower than the baseline equilibrium, which follows the same intuition as in the unilateral tariffs. Implications for sectoral expenditure shares and value-added shares remain the same qualitatively, but are smaller in magnitude. The manufacturing value-added share rises by 0.7--0.9 percentage points relative to the baseline, as compared to 0.9--1.4 percentage points in the case of the unilateral tariff. As the retaliatory scenario is more plausible in the real world, our result suggests that the sectoral implications of tariffs on manufacturing are quantitatively smaller.

Welfare implications of the trade war are summarized in the fourth and fifth columns of Table~\ref{tab: welfare main}. We confirm that under the trade war scenario, all countries are worse off compared to the baseline equilibrium. The U.S. welfare loss is 0.17 percent, which is larger than the welfare losses of most European countries and Japan. Canada has the largest welfare loss of 1.30 percent.

In large manufacturers such as China, Germany, and Japan, the welfare losses are larger under U.S. unilateral tariffs than under the trade war. When the U.S. unilaterally imposes tariffs, trade values from China, Germany, and Japan to the U.S. decline. Due to the terms-of-trade effect, factor prices and thus welfare in these countries also fall. If they retaliate against the U.S., however, they will receive tariff revenues, and more importantly, lower imports from the U.S. mitigate the negative terms-of-trade effect. In contrast to those three countries, Mexico, the second-worst-off country under the trade war, experiences welfare losses (0.93 percent) that are greater than those under the U.S. unilateral tariff (0.84 percent). Mexico is heavily reliant on the U.S. economy as both the largest export destination and the primary import source; the economy is worse off if all the other countries retaliate against the U.S. The last column of Table~\ref{tab: welfare main} confirms that, as in the case of U.S. unilateral tariffs, the welfare impacts are larger in magnitude under homothetic CES than under nonhomothetic CES preferences.\footnote{In the trade war scenario, we can check $\overline{\epsilon}_{n,t}>1$ for all the countries $n$ and $t= 2001$ onward, as implied by Proposition~\ref{prop: proposition 1}.}

\subsection{Optimal Time-invariant Tariffs}

\begin{figure}[tb!]%
    \centering
        \includegraphics[width=0.8\textwidth]{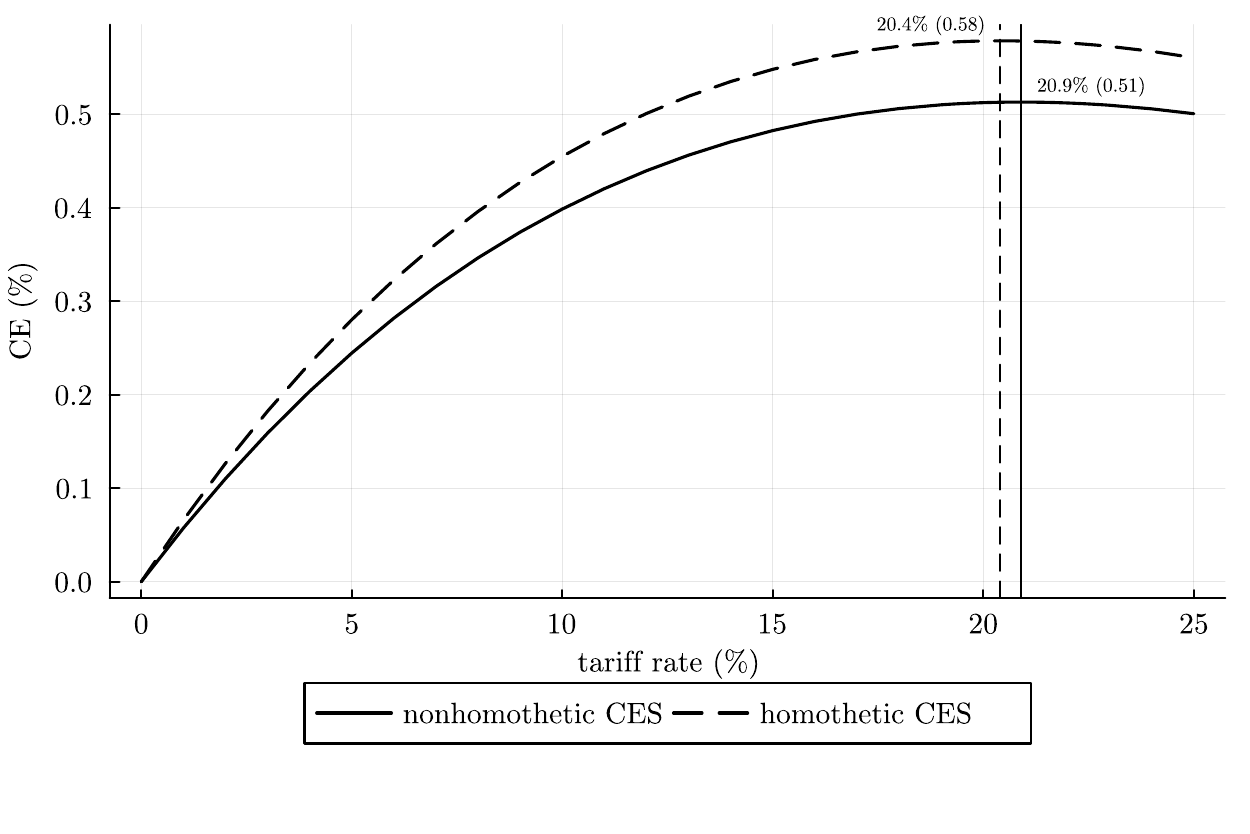}%
    \caption{Optimal tariffs under different preferences}\label{fig: optimal tariffs}%
    \begin{figurenotes}%
        The figure plots welfare, measured as the consumption-equivalent change relative to a zero time-invariant manufacturing tariff, against alternative time-invariant manufacturing tariff rates for the United States. It uses $\{\Omega_{n,t}^{j}\}$ and $\{\zeta_{n,t}\}$ calibrated from the data.
    \end{figurenotes}
\end{figure}

As the last quantitative exercise, we compute the U.S. optimal time-invariant manufacturing tariffs under different preferences. We assume that in 2001, the U.S. unilaterally sets a manufacturing tariff that comes as a surprise. After the manufacturing tariff is set, everyone in the world economy knows that this manufacturing tariff lasts forever. We also assume that no other country retaliates against this tariff.

Figure~\ref{fig: optimal tariffs} plots U.S. welfare against time-invariant manufacturing tariff rates under homothetic and nonhomothetic CES preferences. The welfare is measured by consumption equivalent relative to that in the zero-tariff case. As we detail in Supplemental Appendix~\ref{ap: welfare formula}, the consumption equivalent is the percent change in per capita consumption to equate the discounted sum of per capita consumption to the baseline level. The initial period to compute the discounted sum is 2001. Thus, we are comparing welfare under different tariffs from the viewpoint of 2001. A marginal increase in manufacturing tariffs from zero generates a larger welfare gain under homothetic CES preferences than under nonhomothetic CES preferences. Since the U.S. has a sufficiently high expenditure-weighted income elasticity ($\overline\epsilon_{US,t}>1$), this pattern is consistent with Proposition~\ref{prop: proposition 1}. This overestimation under homothetic CES preferences, driven by an aggregate price index independent of real income, appears not only near zero tariffs but also across the wide range of tariff rates considered. At the respective optima, the welfare gain is 0.58 percent under homothetic CES preferences, 14 percent larger than the 0.51 percent gain under nonhomothetic CES preferences.

The optimal tariff is higher under nonhomothetic CES preferences (20.9 percent) than under homothetic CES preferences (20.4 percent). As the optimal-tariff formula in \eqref{eq: optimal tariff} shows, this ordering reflects the inverse ordering of the U.S. manufacturing import share ($s_{US,RoW}^m$), which is largely shaped by the U.S. manufacturing expenditure share ($\omega_{US}^m$). This relationship is confirmed in Figure~\ref{fig: expshare_optimum} in Supplemental Appendix~\ref{ap: counterfactual additional}: evaluated at the optimal tariff, the U.S. manufacturing expenditure share under nonhomothetic CES preferences is consistently lower than its homothetic-CES counterpart since 2001.

\section{Conclusion}\label{sec: concl}
\addtocontents{toc}{}
\let\addcontentsline\oldaddcontentsline %

This paper examines the effects of tariffs on sectoral composition and welfare in the presence of two key drivers of structural change---sectoral complementarity and nonhomothetic preferences---as well as the standard protective role of tariffs. We ask whether manufacturing tariffs can help manufacturing regain its share in sectoral consumption expenditure and value added, and how their welfare effects differ between homothetic and nonhomothetic CES preferences.

Using a two-country model, we show qualitatively that an increase in tariffs from a low level raises income and shifts demand from manufacturing to services through the income effect driven by nonhomothetic preferences. However, tariffs targeting manufacturing shift demand in the opposite direction via the relative price effect, given an elasticity of substitution less than one. In addition to these structural-change-driven effects, tariffs replace foreign manufacturers with domestic producers, thereby increasing the manufacturing value-added share.

To quantitatively assess these mechanisms, we extend the model to a multi-country dynamic framework with capital accumulation and input-output linkages. Using data from 24 countries over the period from 1990 to 2014 and calibrating fundamentals such as sectoral productivity and non-tariff trade barriers, we compute transition paths in terms of levels rather than relative changes.

We simulate a 20-percentage-point increase in U.S. manufacturing tariffs on all trading partners beginning in 2001. We find that the relative-price effect and the trade-protection effect dominate the income effect. Specifically, the U.S. manufacturing value-added share rises by 6.4--11.2 percent (0.9--1.4 percentage points) between 2001 and 2014. Despite these shifts in sectoral composition, the welfare gain is quantitatively modest: U.S. welfare rises by only 0.43 percent, at the expense of all other countries. Canada experiences the largest welfare loss, at 1.58 percent. If other countries retaliate with equally large tariff increases on U.S. manufacturing exports, all countries are worse off, and U.S. welfare falls by 0.17 percent.

Our analysis also shows that nonhomothetic preferences are important for evaluating the welfare effects of tariffs. Under homothetic CES preferences, the same unilateral tariff experiment implies a U.S. welfare gain of 0.49 percent, 14 percent (0.06 percentage point) higher than under nonhomothetic CES preferences. This difference reflects the fact that, in countries with a higher expenditure-weighted income elasticity of sectoral demand, tariff-induced income gains shift expenditure toward more income-elastic services and raise the aggregate price index. Homothetic CES preferences shut down this channel and therefore overstate the welfare gains from unilateral tariffs. This pattern also appears in our optimal-tariff exercise: the unilateral optimal time-invariant U.S. manufacturing tariff is 20.4 percent and 20.9 percent under homothetic and nonhomothetic CES preferences, respectively. At these optima, the welfare gain is 14 percent (0.07 percentage point) higher under homothetic CES preferences. These findings suggest that widely used homothetic CES preferences may overestimate the gains from unilateral tariff policy in economies at advanced stages of structural change.

\newpage

\begin{flushleft}
    \textbf{Declaration of generative AI and AI-assisted technologies in the manuscript preparation process}
\end{flushleft}

During the preparation of this work, the authors used ChatGPT and Claude to assist with language editing and proofreading, checking analytical derivations and computational calculations, writing and debugging programming code, and collecting and organizing data. After using these tools, the authors reviewed and edited the content and take full responsibility for the submitted article.

\setstretch{1.2}
\bibliographystyle{apalike}
\bibliography{main}

\clearpage 
\setstretch{1.2}
\appendix %
\counterwithin*{equation}{section} %
\renewcommand\theequation{\thesection\arabic{equation}} %
\setcounter{page}{1} %

\setcounter{table}{0}
\renewcommand{\thetable}{A\arabic{table}}
\setcounter{figure}{0}
\renewcommand{\thefigure}{A\arabic{figure}}

\renewcommand*{\thefootnote}{\fnsymbol{footnote}} %

\begin{center}
{\LARGE Supplemental Appendix of \\ ``Trade Policy and Structural Change''} \\
\vspace{0.5cm} 
{\large Hayato Kato\footnotemark[1] \hspace{1cm} Kensuke Suzuki\footnotemark[2] \hspace{1cm}  Motoaki Takahashi\footnotemark[3]} \\
\vspace{0.5cm} 
August, 2026
\end{center}

\footnotetext[1]{Institute of Developing Economies (IDE-JETRO); CESifo Research Network. Email: \href{mailto:hayato_kato@ide.go.jp}{hayato\_kato@ide.go.jp}}

\footnotetext[2]{Department of Economics, Clark University; Economic Research Center, Nagoya University. Email: \href{mailto:ksuzuki@clarku.edu}{ksuzuki@clarku.edu}}

\footnotetext[3]{Graduate School of Economics, the University of Osaka. Email: \href{motoaki.takahashi@econ.osaka-u.ac.jp}{motoaki.takahashi@econ.osaka-u.ac.jp}}

\tableofcontents

\

\setcounter{footnote}{0} %
\renewcommand*{\thefootnote}{\arabic{footnote}} %

\section{Proof of Proposition 1} \label{ap: proof of prop 1}

In a two-country, $J$ sector model, we consider an increase in Home's sector $j$ tariff in a tariff-free world. Sectors are asymmetric unless otherwise noted. For notational convenience, we define the following:
{\small
\begin{align*}
&\omega_n^j = \frac{P_n^j C_n^j}{\sum_{h=1}^J P_n^h C_n^h} =
\frac{
(C_n/L_n)^{\epsilon^j(1-\sigma)}(P_n^j)^{1-\sigma}
}{
\sum_{h=1}^J
(C_n/L_n)^{\epsilon^h(1-\sigma)}(P_n^h)^{1-\sigma}
},
\qquad
\overline\epsilon_n =
\sum_{k=1}^J \omega_n^k\epsilon^k,
&& n\in\{H,F\} \\[0.5em]
&IM_H^k = \frac{\pi_{HF}^k \omega_H^k E_H}{1+\tau^k}, \qquad
EX_H^k = \pi_{FH}^k \omega_F^k E_F,
&& k \in \{1,\dots, J\}
\\[0.5em]
&s_{HF}^k = \frac{IM_H^k}{\sum_{h=1}^J IM_H^h} = \frac{\pi_{HF}^k \omega_H^k/(1+\tau^k)}{\sum_{h=1}^J \pi_{HF}^h \omega_H^h/(1+\tau^h)},
&& k \in \{1,\dots, J\}
\\[0.5em]
&s_{FH}^k = \frac{EX_H^k}{\sum_{h=1}^J EX_H^h} = \frac{\pi_{FH}^k \omega_F^k}{\sum_{h=1}^J \pi_{FH}^h \omega_F^h},
&& k \in \{1,\dots, J\}
\\[0.5em]
&\overline\pi_{HF} = \sum_{k=1}^J s_{HF}^k \pi_{HF}^k, \qquad \overline\pi_{FH} = \sum_{k=1}^J s_{FH}^k \pi_{FH}^k, \qquad \overline\epsilon_{HF} = \sum_{k=1}^J s_{HF}^k \epsilon^k, \qquad \overline\epsilon_{FH} = \sum_{k=1}^J s_{FH}^k \epsilon^k,
\\[0.5em]
&\Lambda_{HF} = \sum_{k=1}^J \frac{\pi_{HF}^k \omega_H^k}{1+\tau^k}, \qquad \Lambda_{FH} = \sum_{k=1}^J \pi_{FH}^k \omega_F^k, \qquad \Lambda_{FF} = \sum_{k=1}^J \pi_{FF}^k \omega_H^k = 1- \Lambda_{FH},
\end{align*}}
noting $\tau_{HF}^j=\tau^j$ for some $j \in \{ 1,\dots,J\}$, $\tau^k=0$ for $k \neq j$, and $\tau_{FH}^j=0$ for all $j$. We proceed in seven steps.

\medskip

\noindent \textit{Step 1: Derivation of the welfare formula}

The per-capita expenditure function in country $n \in \{ H,F\}$ is
\begin{align*}
&\frac{E_n}{L_n}
= \left[ \sum_{k=1}^J \left( \frac{C_n}{L_n}\right)^{\epsilon^k(1-\sigma)}\left( P_n^k \right)^{1-\sigma} \right]^{\frac{1}{1-\sigma}}, &&n \in \{ H,F\}
\end{align*}
where the sectoral price indices are
\begin{align*}
    &P_H^k = \left[(1/A_H^k)^{-\theta} + (w_F d_{HF}^k(1+\tau^k)/A_F^k)^{-\theta}\right]^{-\frac{1}{\theta}}, &&k \in \{ 1,\dots,J\} \\
    &P_F^k = \left[(d_{FH}^k(1+\tau^k)/A_H^k)^{-\theta} + (w_F/A_F^k)^{-\theta}\right]^{-\frac{1}{\theta}}, &&k \in \{ 1,\dots,J\}
\end{align*}
noting that Home labor is chosen as the num{\'e}raire so that $w_H=1$. We define the aggregate price index as $P_n = E_n/C_n$:
\begin{align*}
&P_n = \left[ \sum_{k=1}^J \left( \frac{C_n}{L_n}\right)^{(\epsilon^k-1)(1-\sigma)}\left( P_n^k \right)^{1-\sigma} \right]^{\frac{1}{1-\sigma}}. &&n \in \{ H,F\}
\end{align*}

We differentiate Home's real per capita consumption with respect to Home's sector $j$ tariff to obtain
\begin{align*}
&\frac{d\ln (C_H/L_H)}{d\ln (1+\tau^j)} &&= \frac{d\ln (E_H/L_H)}{d\ln (1+\tau^j)} - \frac{d\ln P_H}{d\ln (1+\tau^j)} \\
& &&= \frac{d\ln (E_H/L_H)}{d\ln (1+\tau^j)} - \left[ \sum_{k=1}^J \omega_H^k \frac{d\ln P_H^k}{d\ln (1+\tau^j)} + (\overline{\epsilon}_H - 1)\frac{d\ln (C_H/L_H)}{d\ln (1+\tau^j)}\right] \\
& &&=
\frac{d\ln (E_H/L_H)}{d\ln (1+\tau^j)} - \left[ \omega_H^j \pi_{HF}^j + \Lambda_{HF} \varepsilon_{w,F} + (\overline{\epsilon}_H - 1)\frac{d\ln (C_H/L_H)}{d\ln (1+\tau^j)}\right] \\
&\Leftrightarrow \frac{d\ln (C_H/L_H)}{d\ln (1+\tau^j)} &&= \frac{1}{\overline{\epsilon}_H}\left[\frac{d\ln (E_H/L_H)}{d\ln (1+\tau^j)} - \left( \omega_H^j \pi_{HF}^j + \Lambda_{HF} \varepsilon_{w,F}\right) \right],
\end{align*}
From the second to the third line, we use
\begin{align*}
    \frac{d\ln P_H^k}{d\ln (1+\tau^j)} = \pi_{HF}^k \left[ \mathbf{1}_{\{k=j\}} + \varepsilon_{w,F}\right], \qquad \varepsilon_{w,F} = \frac{d\ln w_F}{d\ln (1+\tau^j)},
\end{align*}
where $\mathbf{1}_{\{k=j\}}$ is a dummy taking one if $k=j \in \{ 1,\dots,J\}$ and zero otherwise.

To see the derivative of Home's aggregate expenditure, we turn to Home's budget constraint:
\begin{align*}
E_H
&= L_H + \tau^j IM_H^j \\
&= L_H + \frac{\tau^j \pi_{HF}^j \omega_H^j E_H}{1+\tau^j}.
\end{align*}
We differentiate this to obtain
\begin{align*}
\frac{d\ln E_H}{d\ln (1+\tau^j)}
&= \frac{(1+\tau^j)IM_H^j}{E_H}
+ \frac{\tau^j IM_H^j}{E_H}\frac{d\ln IM_H}{d\ln (1+\tau^j)} \\
&= \pi_{HF}^j \omega_H^j
+ \frac{\tau^j \pi_{HF}^j \omega_H^j}{1+\tau^j} \frac{d\ln IM_H}{d\ln (1+\tau^j)},
\end{align*}
We substitute this back into the previous equation and evaluate it at zero tariff:
\begin{align*}
    \left.\frac{d\ln (C_H/L_H)}{d\ln (1+\tau^j)}\right|_{\tau^j=0} &= \frac{1}{\overline{\epsilon}_H}\left[ \pi_{HF}^j \omega_H^j - \left( \omega_H^j \pi_{HF}^j + \Lambda_{HF} \varepsilon_{w,F}\right) \right] \\
    &= -\frac{\Lambda_{HF}\varepsilon_{w,F}}{\overline\epsilon_H}.
\end{align*}

\noindent \textit{Step 2: Derivation of the Foreign wage elasticity $\varepsilon_{w,F}$}

The aggregate trade balance requires
\begin{align*}
&\sum_{k=1}^J IM_H^k
= \sum_{k=1}^J EX_H^k, \\
&\Leftrightarrow \sum_{k=1}^J \frac{\pi_{HF}^k \omega_H^k E_H}{1+\tau^k} = \sum_{k=1}^J \pi_{FH}^k \omega_F^k E_F, \\
&\Leftrightarrow \Lambda_{HF}E_H = \Lambda_{FH}E_F,
\end{align*}
where total expenditure in Home and Foreign are respectively $E_H=L_H + \tau^j IM_H^j$ and $E_F = w_F L_F$.
We differentiate this to obtain
\begin{align*}
    \frac{d\ln \Lambda_{HF}}{d\ln (1+\tau^j)} + \frac{d\ln E_H}{d\ln (1+\tau^j)} = \frac{d\ln \Lambda_{FH}}{d\ln (1+\tau^j)} + \varepsilon_{w,F}.
\end{align*}
Each term above is given by
\begin{align*}
    &\frac{d\ln \Lambda_{HF}}{d\ln (1+\tau^j)} &&= \frac{\partial \ln \Lambda_{HF}}{\partial \ln w_F} \varepsilon_{w,F} +  \frac{\partial \ln \Lambda_{HF}}{\partial \ln (1+\tau^j)}, \\
    &\frac{d\ln E_H}{d\ln (1+\tau^j)} &&= \pi_{HF}^j \omega_H^j + \frac{\tau^j \pi_{HF}^j \omega_H^j}{1+\tau^j} \frac{d\ln IM_H}{d\ln (1+\tau^j)}, \\
    &\frac{d\ln \Lambda_{FH}}{d\ln (1+\tau^j)} &&= \frac{\partial \ln \Lambda_{FH}}{\partial \ln w_F} \varepsilon_{w,F}.
\end{align*}
We substitute these expressions back into the previous equation and evaluate it at zero tariff:
\begin{align*}
    &\frac{\partial \ln \Lambda_{HF}}{\partial \ln w_F} \varepsilon_{w,F} +  \frac{\partial \ln \Lambda_{HF}}{\partial \ln (1+\tau^j)} +  \pi_{HF}^j \omega_H^j \bigg|_{\tau^j=0}= \frac{\partial \ln \Lambda_{FH}}{\partial \ln w_F} \varepsilon_{w,F} +\varepsilon_{w,F}  \bigg|_{\tau^j=0}, \\
    &\Leftrightarrow \varepsilon_{w,F} \big|_{\tau^j=0} = \dfrac{\dfrac{\partial \ln \Lambda_{HF}}{\partial \ln (1+\tau^j)} + \pi_{HF}^j \omega_H^j}{\dfrac{\partial \ln \Lambda_{FH}}{\partial \ln w_F} - \dfrac{\partial \ln \Lambda_{HF}}{\partial \ln w_F} + 1}.
\end{align*}
In the following steps, we derive the derivatives in the right hand side of the above equation.

\medskip

\noindent \textit{Step 3: Derivation of $\partial \ln \Lambda_{HF}/\partial \ln (1+\tau^j)$}

Holding the Foreign wage fixed, we differentiate $\pi_{HF}^k \omega_H^k/(1+\tau^k)$ for sector $k \in \{ 1,\dots,J\}$ with respect to Home's sector $j$ tariff to obtain
\begin{align*}
\frac{\partial \ln (\pi_{HF}^k \omega_H^k/(1+\tau^k))}{\partial \ln (1+\tau^j)}
&=
\frac{\partial \ln \pi_{HF}^k}{\partial \ln (1+\tau^j)}\mathbf{1}_{\{k=j\}} + \frac{\partial \ln \omega_H^k}{\partial \ln (1+\tau^j)} - \mathbf{1}_{\{k=j\}}.
\end{align*}
Each term in the right hand side is
\begin{align*}
    \frac{\partial \ln \pi_{HF}^k}{\partial \ln (1+\tau^j)} &= -\theta \pi_{HH}^k \mathbf{1}_{\{k=j\}}, \\
    \frac{\partial \ln \omega_H^k}{\partial \ln (1+\tau^j)}
    &= (1-\sigma) \left[\frac{\partial \ln P_H^k}{\partial \ln (1+\tau^j)}\mathbf{1}_{\{k=j\}}
    -\sum_{h=1}^J \omega_H^h \frac{\partial \ln P_H^h}{\partial \ln (1+\tau^j)}
    + \left( \epsilon^k-\overline\epsilon_H \right) \frac{\partial \ln (C_H/L_H)}{\partial \ln (1+\tau^j)} \right] \\
    &= (1-\sigma) \left[\pi_{HF}^k \mathbf{1}_{\{k=j\}} - \pi_{HF}^j \omega_H^j
    + \left( \epsilon^k-\overline\epsilon_H \right) \frac{\partial \ln (C_H/L_H)}{\partial \ln (1+\tau^j)} \right].
\end{align*}
We substitute these into the previous equation and aggregate it across sectors to obtain {\small
\begin{align*}
    \frac{\partial \ln \Lambda_{HF}}{\partial \ln (1+\tau^j)} &= \sum_{k=1}^J \frac{\pi_{HF}^k \omega_H^k/(1+\tau^k)}{\sum_{h=1}^J \pi_{HF}^h \omega_H^h/(1+\tau^h)} \frac{\partial \ln (\pi_{HF}^k \omega_H^k/(1+\tau^k))}{\partial \ln (1+\tau^j)} \\
    &= \sum_{k=1}^J s_{HF}^k \left[ -\theta \pi_{HH}^k \mathbf{1}_{\{ k=j\}} + (1-\sigma) \left\{\pi_{HF}^j \left( \mathbf{1}_{\{k=j\}} - \omega_H^j\right)
    + \left( \epsilon^k-\overline\epsilon_H \right) \frac{\partial \ln (C_H/L_H)}{\partial \ln (1+\tau^j)} \right\}
    - \mathbf{1}_{\{ k=j\}} \right].
\end{align*} }

Applying the welfare formula in Step 1 to the case of $\varepsilon_{w,F}=0$, the expression above simplifies to
\begin{align*}
    \frac{\partial \ln \Lambda_{HF}}{\partial \ln (1+\tau^j)}\bigg|_{\tau^j=0}
    &= s_{HF}^j \left[ (1-\sigma) \left( \pi_{HF}^j - \Lambda_{HF}\right) -\left( \theta \pi_{HH}^j + 1 \right) \right].
\end{align*}

\medskip

\noindent \textit{Step 4: Derivation of $\partial \ln \Lambda_{HF}/\partial \ln w_F$}

Holding Home's sector $j$ tariff fixed, we differentiate $\pi_{HF}^k \omega_H^k/(1+\tau^k)$ for sector $k \in \{ 1,\dots,J\}$ with respect to Foreign wage to obtain
\begin{align*}
\frac{\partial \ln (\pi_{HF}^k \omega_H^k/(1+\tau^k))}{\partial \ln w_F}
&=
\frac{\partial \ln \pi_{HF}^k}{\partial \ln w_F} + \frac{\partial \ln \omega_H^k}{\partial \ln w_F}.
\end{align*}
Each term in the right hand side of the above equation is
\begin{align*}
    &\frac{\partial \ln \pi_{HF}^k}{\partial \ln w_F} &&= -\theta \pi_{HH}^k, \\
    &\frac{\partial \ln \omega_H^k}{\partial \ln w_F} &&= (1-\sigma) \left[\frac{\partial \ln P_H^k}{\partial \ln w_F}
    -\sum_{h=1}^J \omega_H^h \frac{\partial \ln P_H^h}{\partial \ln w_F}
    + \left( \epsilon^k-\overline\epsilon_H \right) \frac{\partial \ln (C_H/L_H)}{\partial \ln w_F} \right] \\
    & &&= (1-\sigma) \left[\pi_{HF}^k
    -\sum_{h=1}^J \omega_H^h \pi_{HF}^h
    + \left( \epsilon^k-\overline\epsilon_H \right) \frac{\partial \ln (C_H/L_H)}{\partial \ln w_F} \right].
\end{align*}
We substitute these into the previous equation and aggregate it across sectors to obtain
\begin{align*}
    \frac{\partial \ln \Lambda_{HF}}{\partial \ln w_F} &= \sum_{k=1}^J \frac{\pi_{HF}^k \omega_H^k/(1+\tau^k)}{\sum_{h=1}^J \pi_{HF}^h \omega_H^h/(1+\tau^h)} \frac{\partial \ln (\pi_{HF}^k \omega_H^k/(1+\tau^k))}{\partial \ln w_F} \\
    &= \sum_{k=1}^J s_{HF}^k \left[-\theta \pi_{HH}^k + (1-\sigma) \left\{ \pi_{HF}^k
    -\sum_{h=1}^J \omega_H^h \pi_{HF}^h
    + \left( \epsilon^k-\overline\epsilon_H \right) \frac{\partial \ln (C_H/L_H)}{\partial \ln w_F}  \right\} \right].
\end{align*}

What remains to see is the derivative of Home welfare with respect to the Foreign wage, $\partial \ln (C_H/L_H)/\partial \ln w_F$. From the per capita expenditure function in Home, we have
\begin{align*}
    \frac{\partial \ln (E_H/L_H)}{\partial \ln w_F}\bigg|_{\tau^j=0} =
    \overline\epsilon_H \frac{\partial \ln (C_H/L_H)}{\partial \ln w_F} + \sum_{k=1}^J \omega_H^k \frac{\partial \ln P_H^k}{\partial \ln w_F}\bigg|_{\tau^j=0},
\end{align*}
holding Home's sector $j$ tariff fixed. Each term in the equation above is given by
\begin{align*}
    &\frac{\partial \ln (E_H/L_H)}{\partial \ln w_F}\bigg|_{\tau^j=0}
    &&= 0, \\
    &\frac{\partial \ln P_H^k}{\partial \ln w_F}\bigg|_{\tau^j=0} &&= \pi_{HF}^k.
\end{align*}
Substituting these back into the previous equation yields
\begin{align*}
    &0
    = \overline\epsilon_H \frac{\partial \ln (C_H/L_H)}{\partial \ln w_F} + \sum_{k=1}^J \omega_H^k \pi_{HF}^k \bigg|_{\tau^j=0}, \\
    &\Leftrightarrow \frac{\partial \ln (C_H/L_H)}{\partial \ln w_F}\bigg|_{\tau^j=0} = -\frac{\Lambda_{HF}}{\overline\epsilon_H}.
\end{align*}
Using these, we have
\begin{align*}
    \frac{\partial \ln \Lambda_{HF}}{\partial \ln w_F}\bigg|_{\tau^j=0}
    &= -\theta \left( 1-\overline{\pi}_{HF} \right) + (1-\sigma) \left(\overline{\pi}_{HF} - \frac{\Lambda_{HF}\overline{\epsilon}_{HF}}{\overline{\epsilon}_H} \right).
\end{align*}

\medskip
\noindent \textit{Step 5: Derivation of $\partial \ln \Lambda_{FH}/\partial \ln w_F$}

Holding Home's sector $j$ tariff fixed, we differentiate $\pi_{FH}^k \omega_H^k$ for sector $k \in \{ 1,\dots,J\}$ with respect to Foreign wage to obtain \begin{align*} \frac{\partial \ln (\pi_{FH}^k \omega_F^k)}{\partial \ln w_F} &= \frac{\partial \ln \pi_{FH}^k}{\partial \ln w_F} + \frac{\partial \ln \omega_F^k}{\partial \ln w_F}. \end{align*} Each term in the right hand side of the above equation is
\begin{align*}
&\frac{\partial \ln \pi_{FH}^k}{\partial \ln w_F} &&= \theta \pi_{FF}^k, \\
&\frac{\partial \ln \omega_F^k}{\partial \ln w_F} &&= (1-\sigma) \left[\frac{\partial \ln P_F^k}{\partial \ln w_F} -\sum_{h=1}^J \omega_F^h \frac{\partial \ln P_F^h}{\partial \ln w_F} + \left( \epsilon^k-\overline\epsilon_F \right) \frac{\partial \ln (C_F/L_F)}{\partial \ln w_F} \right] \\
& &&= (1-\sigma) \left[\pi_{FF}^k -\sum_{h=1}^J \omega_F^h \pi_{FF}^h + \left( \epsilon^k-\overline\epsilon_F \right) \frac{\partial \ln (C_F/L_F)}{\partial \ln w_F} \right].
\end{align*}
We substitute these into the previous equation and aggregate it across sectors to obtain
\begin{align*}
\frac{\partial \ln \Lambda_{FH}}{\partial \ln w_F} &= \sum_{k=1}^J \frac{\pi_{FH}^k \omega_F^k}{\sum_{h=1}^J \pi_{FH}^h \omega_F^h} \frac{\partial \ln (\pi_{FH}^k \omega_F^k)}{\partial \ln w_F} \\
&= \sum_{k=1}^J s_{FH}^k \left[\theta \pi_{FF}^k + (1-\sigma) \left\{ \pi_{FF}^k -\sum_{h=1}^J \omega_F^h \pi_{FF}^h + \left( \epsilon^k-\overline\epsilon_F\right) \frac{\partial \ln (C_F/L_F)}{\partial \ln w_F} \right\} \right].
\end{align*}
What remains to see is $\partial \ln (C_F/L_F)/\partial \ln w_F$. From the per capita expenditure function in Foreign, we have \begin{align*} \frac{\partial \ln (E_F/L_F)}{\partial \ln w_F}\bigg|_{\tau^j=0} = \overline\epsilon_F \frac{\partial \ln (C_F/L_F)}{\partial \ln w_F} + \sum_{k=1}^J \omega_F^k \frac{\partial \ln P_F^k}{\partial \ln w_F}\bigg|_{\tau^j=0}, \end{align*} holding sector $j$ tariff fixed. Each term in the equation above is given by \begin{align*} &\frac{\partial \ln (E_F/L_F)}{\partial \ln w_F}\bigg|_{\tau^j=0} &&= 1, \\ &\frac{\partial \ln P_F^k}{\partial \ln w_F}\bigg|_{\tau^j=0} &&= \pi_{FF}^k. \end{align*} Substituting these back into the previous equation yields
\begin{align*}
&1 = \overline\epsilon_F \frac{\partial \ln (C_F/L_F)}{\partial \ln w_F} + \sum_{k=1}^J \omega_F^k \pi_{FF}^k \bigg|_{\tau^j=0}, \\
&\Leftrightarrow \frac{\partial \ln (C_F/L_F)}{\partial \ln w_F}\bigg|_{\tau^j=0} = \frac{1-\sum_{k=1}^J \omega_F^k \pi_{FF}^k}{\overline\epsilon_F}= \frac{\Lambda_{FH}}{\overline\epsilon_F}.
\end{align*}
Using this, we have \begin{align*} \frac{\partial \ln \Lambda_{FH}}{\partial \ln w_F}\bigg|_{\tau^j=0} &= \theta \left( 1-\overline{\pi}_{FH} \right) + (1-\sigma) \left(-\overline{\pi}_{FH} + \frac{\Lambda_{FH}\overline{\epsilon}_{FH}}{\overline{\epsilon}_F} \right). \end{align*}

\medskip

\noindent \textit{Step 6: Explicit expression of the welfare formula}

We use the results in the previous steps to rewrite the effect of Home's tariff on its own welfare
\begin{align} \label{ap: welfare_shosoku}
\left.\frac{d\ln (C_H/L_H)}{d\ln(1+\tau^j)}\right|_{\tau^j=0} &= -\frac{\Lambda_{HF}\varepsilon_{w,F}}{\overline\epsilon_H},
\end{align}
where
\begin{align*}
    \varepsilon_{w,F}\big|_{\tau^j=0} = - \frac{s_{HF}^j \left[1 + \theta \pi_{HH}^j - \Lambda_{HF} - (1-\sigma)( \pi_{HF}^j - \Lambda_{HF}) \right]}{ 1 + \theta(2 - \overline{\pi}_{HF} - \overline{\pi}_{FH}) - (1-\sigma)\left( \overline{\pi}_{HF} + \overline{\pi}_{FH} - \Lambda_{HF}\overline{\epsilon}_{HF}/\overline{\epsilon}_H - \Lambda_{FH}\overline{\epsilon}_{FH}/\overline{\epsilon}_F\right) }.
\end{align*}
An analogous expression holds for the the effect of Home's tariff on Foreign welfare:
\begin{align} \label{ap: welfare_shosoku_F}
\left.\frac{d\ln (C_F/L_F)}{d\ln(1+\tau^j)}\right|_{\tau^j=0} &= \frac{\Lambda_{FH}\varepsilon_{w,F}}{\overline\epsilon_F}.
\end{align}

Under symmetric sectoral productivities and non-tariff barriers, $A_H^j = A_H$, $A_F^j=A_F$ and $d_{ni}^j=d_{ni}$ for all $j$, the following relations hold in the tariff-free world:
\begin{align*}
    &\pi_{HF}^k = \pi_{HF}, \qquad \pi_{FH}^k = \pi_{FH}, && k \in \{1,\dots, J\} \\[0.5em]
    &s_{HF}^k = \frac{IM_H^k}{\sum_{h=1}^J IM_H^h} = \frac{\pi_{HF} \omega_H^k}{\sum_{h=1}^J \pi_{HF} \omega_H^h} = \omega_H^k,
    && k \in \{1,\dots, J\} \\[0.5em]
    &s_{FH}^k = \frac{EX_H^k}{\sum_{h=1}^J EX_H^h} = \frac{\pi_{FH} \omega_F^k}{\sum_{h=1}^J \pi_{FH} \omega_F^h} = \omega_F^k,
    && k \in \{1,\dots, J\} \\[0.5em]
    &\overline\pi_{HF} = \sum_{k=1}^J s_{HF}^k \pi_{HF}^k = \left( \sum_{k=1}^J \omega_H^k \right)\pi_{HF} = \Lambda_{HF} = \pi_{HF}, \\
    &\overline\pi_{FH} = \sum_{k=1}^J s_{FH}^k \pi_{FH}^k = \left(\sum_{k=1}^J \omega_F^k \right)\pi_{FH} = \Lambda_{FH} = \pi_{FH}, \\
    &\overline\epsilon_{HF} = \sum_{k=1}^J s_{HF}^k \epsilon^k = \sum_{k=1}^J \omega_H^k \epsilon^k = \overline{\epsilon}_H, \\
    &\overline\epsilon_{FH} = \sum_{k=1}^J s_{FH}^k \epsilon^k = \sum_{k=1}^J \omega_F^k \epsilon^k = \overline{\epsilon}_F.
\end{align*}
In this case, the Foreign wage elasticity is negative:
\begin{align} \label{ap: foreign_wage_elasticity}
    \varepsilon_{w,F}\big|_{\tau^j=0} = -\frac{\omega_H^j \pi_{HH}(1+\theta)}{ 1 + \theta(2 - \pi_{HF} - \pi_{FH}) } < 0.
\end{align}
since $\pi_{HF}, \pi_{FH} \in (0,1)$. Therefore, we can conclude that Home's sector $j$ tariff has a positive effect on its own welfare and a negative effect on Foreign welfare, $d\ln (C_H/L_H)/d\ln (1+\tau^j) >0$ and $d\ln (C_F/L_F)/d\ln (1+\tau^j) <0$ evaluated at $\tau^j=0$.

\medskip

\noindent \textit{Step 7: Comparing the welfare effect under different preferences}

Assuming symmetric productivities and non-tariff barriers, the levels of Foreign wage under both homothetic and nonhomothetic CES preferences coincide because they are determined by the aggregate trade balance condition independent of sectoral expenditure shares:
\begin{align} \label{ap: trade_balance}
    &\sum_{k=1}^J IM_H^k\bigg|_{\tau^j=0} = \sum_{k=1}^J EX_H^k\bigg|_{\tau^j=0}, \\
    &\Leftrightarrow \left( \sum_{k=1}^J \omega_H^k\right) \pi_{HF}L_H = \left( \sum_{k=1}^J \omega_F^k\right) \pi_{FH}w_F L_F, \\
    &\Leftrightarrow \frac{(w_F d_{HF}/A_F)^{-\theta}}{(1/A_H)^{-\theta}+(w_F d_{HF}/A_F)^{-\theta}} L_H = \frac{(d_{FH}/A_H)^{-\theta}}{(d_{FH}/A_H)^{-\theta}+(w_F/A_F)^{-\theta}} w_F L_F,
\end{align}
where the equation does not include the nonhomotheticity parameters $\epsilon^j$. For later reference, it is worth noting that $w_F$ decreases with $A_H$ and $L_F$, and it increases with $A_F$ and $L_H$.

We take the difference between the welfare effect under nonhomothetic and homothetic CES preferences to obtain
\begin{align}
\frac{d\ln (C_H/L_H)^{\text{nh}}}{d\ln(1+\tau^j)} - \frac{d\ln (C_H/L_H)^{\text{h}}}{d\ln(1+\tau^j)}\bigg|_{\tau^j=0} &= \left(-\frac{\Lambda_{HF}\varepsilon_{w,F}}{\overline\epsilon_H}\right)^{\text{nh}} - \left(-\Lambda_{HF}\varepsilon_{w,F}\right)^{\text{h}} \\
&= \frac{\pi_{HF}\pi_{HH}(1+\theta)}{ 1 + \theta(2 - \pi_{HF} - \pi_{FH}) }\left( \frac{\omega_H^{j, \text{nh}}}{\overline{\epsilon}_H} - \frac{1}{J} \right),
\end{align}
where the superscripts $\mathrm{nh}$ and $\mathrm{h}$ hereafter denote nonhomothetic and homothetic CES preferences, respectively. We note that $\epsilon^j=1$ and $\omega_H^{j,\text{h}}=1/J$ for all $j$ under homothetic CES preferences. The difference is negative if $\omega_H^{j,\text{nh}} < \overline\epsilon_H/J$. A sufficient condition for this inequality is
\begin{align*}
    &\overline\epsilon_H = \sum_{j=1}^J \epsilon^j \omega_H^{j,\text{nh}} = \sum_{j=1}^J\frac{\epsilon^j(C_H/L_H)^{\epsilon^j(1-\sigma)}}{\sum_{k=1}^J \epsilon^k(C_H/L_H)^{\epsilon^k(1-\sigma)}} > 1 \quad \text{and} \quad \omega_H^{j,\text{nh}} = \frac{(C_H/L_H)^{\epsilon^j(1-\sigma)}}{\sum_{k=1}^J (C_H/L_H)^{\epsilon^k(1-\sigma)}} < \frac{1}{J}.
\end{align*}
An analogous expression holds for Foreign:
\begin{align}
\frac{d\ln (C_F/L_F)^{\text{nh}}}{d\ln(1+\tau^j)} - \frac{d\ln (C_F/L_F)^{\text{h}}}{d\ln(1+\tau^j)}\bigg|_{\tau^j=0} &= \left(\frac{\Lambda_{FH}\varepsilon_{w,F}}{\overline\epsilon_F}\right)^{\text{nh}} - \left(\Lambda_{FH}\varepsilon_{w,F}\right)^{\text{h}}\\
&= \frac{\pi_{FH}\pi_{HH}(1+\theta)}{ 1 + \theta(2 - \pi_{HF} - \pi_{FH}) }\left( \frac{1}{J} - \frac{\omega_H^{j, \text{nh}}}{\overline{\epsilon}_F} \right),
\end{align}
which is positive if $\omega_H^{j,\text{nh}} <  \overline{\epsilon}_F/J$. A sufficient condition for this is $\omega_H^{j,\text{nh}} < 1/J$ and
\begin{align*}
    &\overline\epsilon_F = \sum_{j=1}^J \epsilon^j \omega_F^{j,\text{nh}} = \sum_{j=1}^J\frac{\epsilon^j(C_F/L_F)^{\epsilon^j(1-\sigma)}}{\sum_{k=1}^J \epsilon^k(C_F/L_F)^{\epsilon^k(1-\sigma)}} > 1.
\end{align*}

In the three-sector model considered in the text, we can further break down the sufficient condition ensuring that the welfare effect under nonhomothetic CES preferences is smaller in magnitude than under homothetic CES preferences. In the tariff-free world, the real per capita consumption in Home is pinned down by
\begin{align} \label{ap: P_H^o}
   &\left[\sum_{j=a,m,s} \left\{ \left( \frac{C_H}{L_H}\right)^{\epsilon^j} P_H^j \right\}^{1-\sigma} \right]^{\frac{1}{1-\sigma}}  = \frac{E_H}{L_H} = 1, \notag \\
   &\Leftrightarrow \sum_{j=a,m,s} \left( \frac{C_H}{L_H}\right)^{\epsilon^j(1-\sigma)} = \left(P_H^o \right)^{\sigma-1} = \left[ \left( \frac{1}{A_H}\right)^{-\theta} + \left( \frac{w_F d_{FH}}{A_F}\right)^{-\theta}\right]^{\frac{1-\sigma}{\theta}},
\end{align}
where $P_H^j = P_H^o$ for all $j$ is the sectoral price index common to all sectors, and the Foreign wage is in turn determined by the trade balance condition \eqref{ap: trade_balance}. In the following, we focus on the range of parameters in which $C_H/L_H>1$. Letting $F(C_H/L_H) = \sum_j (C_H/L_H)^{\epsilon^j(1-\sigma)}$, we observe $F'(C_H/L_H)>0$ and $F(1) = 3$. The condition for $C_H/L_H > 1$ is thus $F(C_H/L_H) = (P_H^o)^{\sigma-1} > F(1)$, or equivalently
\begin{align} \label{ap: assumption_C/L>1}
    A_H^{\theta} +\left(\frac{A_F}{w_F d_{FH}} \right)^{\theta} > 3^{\frac{\theta}{1-\sigma}}.
\end{align}
Since we know $dw_F/dA_H<0$ and $dw_F/dL_F<0$ from the trade balance condition, this inequality is more likely to hold when $A_H$ and $L_F$ are larger. As long as $C_H/L_H>1$ or equivalently the inequality above holds, we can see $d(C_H/L_H)/d\epsilon^j < 0$ for $j \in \{a,s \}$ from \eqref{ap: P_H^o}, $d\omega_H^{m,\text{nh}}/d\epsilon^j < 0$ for $j \in \{a,s \}$, and $d\overline\epsilon_H/d\epsilon^s > 0$. In sum, assuming the range of parameters ensuring \eqref{ap: assumption_C/L>1}, Home's welfare gain from its manufacturing tariff under nonhomothetic CES preferences is more likely to be smaller than under homothetic CES preferences when $\epsilon^s$ is larger. \qed

\

\section{Proof of Proposition 2} \label{ap: proof of prop 2}

In a two-country, $J$ sector model, we consider an increase in Home's sector $j$ tariff in a tariff-free world under sectoral symmetry.
An analogous expression for \eqref{eq: dln(exp share)} holds as follows:
\begin{align}
    \frac{d\ln \omega_H^k}{d\ln (1+\tau^j)} &= (1-\sigma)\biggr[\frac{d\ln P_H^j}{d\ln (1+\tau^j)} - \sum_{h=1}^J\omega_H^h \frac{d\ln P_H^h}{d\ln (1+\tau^j)} \biggl] + (1-\sigma)\bigl( \epsilon^k - \overline{\epsilon}_H \bigr) \frac{d\ln (C_H/L_H)}{d\ln (1+\tau^j)}. \label{ap: dln(exp share)}
\end{align}
Using the results in Supplemental Appendix~\ref{ap: proof of prop 1}, we can simplify the expression above as follows:
\begin{align*}
    \frac{d\ln \omega_H^k}{d\ln (1+\tau^j)} \bigg|_{\tau^j=0} = (1-\sigma) \left[ \pi_{HF}\left( \mathbf{1}_{\{k=j\}}-\omega_H^j \right) + \left( \epsilon^k - \overline{\epsilon}_H \right)\frac{d\ln (C_H/L_H)}{d\ln (1+\tau^j)}\right],
\end{align*}
where the welfare formula $d\ln (C_H/L_H)/d\ln (1+\tau^j)$ is given by \eqref{ap: welfare_shosoku} and $\mathbf{1}_{\{k=j\}}$ is a dummy taking one if $k=j$ and zero otherwise.
Under homothetic CES preferences, there is no income effect and thus the welfare formula takes zero. Since $1-\sigma>0$, the sign of the derivative above is determined solely by the sign of $\mathbf{1}_{\{k=j\}}-\omega_H^j$. Clearly, this term capturing the relative price effect is positive for the protected sector $k=j$ and negative for the non-protected sectors $k \neq j$.

Under nonhomothetic CES preferences, whether the income effect dampens or amplifies the relative price effect depends on the sign of $\epsilon^k - \overline{\epsilon}_H$. In the three sector model considered in the text, the direction of this change is easy to see:
\begin{align*}
    \frac{d\ln \omega_H^k}{d\ln (1+\tau^m)} \bigg|_{\tau^m=0} = \begin{cases}
        (1-\sigma) \biggl[ \underbrace{-\pi_{HF}\omega_H^m}_{<0} + \underbrace{\left( \epsilon^a - \overline{\epsilon}_H \right)\frac{d\ln (C_H/L_H)}{d\ln (1+\tau^m)}}_{<0}\biggr] &k = a \\
        (1-\sigma) \biggl[\underbrace{\pi_{HF}(1-\omega_H^m)}_{>0} + \underbrace{\left( 1 - \overline{\epsilon}_H \right)\frac{d\ln (C_H/L_H)}{d\ln (1+\tau^m)}}_{<0}\biggr] &k = m \\
        (1-\sigma) \biggl[ \underbrace{-\pi_{HF}\omega_H^m}_{<0} + \underbrace{\left( \epsilon^s - \overline{\epsilon}_H \right)\frac{d\ln (C_H/L_H)}{d\ln (1+\tau^m)}}_{>0}\biggr] &k = s
    \end{cases}.
\end{align*}
Assuming the range of parameters in which $\overline{\epsilon}_H > 1$, relative to the case of homothetic CES, the agricultural expenditure share falls more, the manufacturing share rises less, and the service share falls less.

Similarly, the effect on the sectoral expenditure shares in Foreign
\begin{align*}
    \frac{d\ln \omega_F^k}{d\ln (1+\tau^m)} \bigg|_{\tau^m=0} = \begin{cases}
        (1-\sigma) \left( \epsilon^a - \overline{\epsilon}_F \right)\dfrac{d\ln (C_F/L_F)}{d\ln (1+\tau^m)}>0 &k = a \\
        (1-\sigma) \left( 1 - \overline{\epsilon}_F \right)\dfrac{d\ln (C_F/L_F)}{d\ln (1+\tau^m)}>0 &k = m \\
        (1-\sigma) \left( \epsilon^s - \overline{\epsilon}_F \right)\dfrac{d\ln (C_F/L_F)}{d\ln (1+\tau^m)} < 0 &k = s
    \end{cases}.
\end{align*}
The relative price effect is zero in Foreign since a decrease in its wage due to the terms-of-trade effect lowers the price indices in all sectors proportionately. Therefore, under homothetic CES preferences where only the relative effect operates, Home's tariff does not change the sectoral expenditure shares in Foreign. Under nonhomothetic CES preferences, for the range of parameters satisfying $\overline{\epsilon}_F > 1$, the income effect operates in exactly the opposite direction from that in Home. Both agricultural and manufacturing expenditure shares rise, while the service share falls. \qed

\

\section{Proof of Proposition 3} \label{ap: proof of prop 3}

In a two-country, $J$ sector model, we consider a tariff increase only in sector $j$ in a tariff-free world under sectoral symmetry. Noting $w_H=1$, the value-added share of sector $k$ in Home is given by
\begin{align*}
    va_H^k &= \frac{VA_H^k}{\sum_{h=1}^J VA_H^h} = \frac{L_H^k}{\sum_{h=1}^J L_H^h} \\
    &= \frac{P_H^k C_H^k}{L_H} + \frac{NX_H^k}{L_H} - \frac{\widetilde{T}_H}{L_H} = \frac{P_H^k C_H^k}{E_H} \frac{E_H}{L_H} + \frac{NX_H^k}{ L_H} - \frac{\widetilde{T}_H^k}{ L_H} \\
    &= \omega_H^k \left( 1+\frac{\widetilde{T}_H^j}{L_H} \right) + \frac{NX_H^k}{L_H} - \mathbf{1}_{\{ k=j\}}\frac{\widetilde{T}_H^k}{L_H},
\end{align*}
where $\widetilde{T}_H^j = \tau^j IM_H^j = \tau^j \pi_{HF} \omega_H^j E_H/(1+\tau^j)$ is tariff revenue in Home.
As in the text, the effect of sector $j$ tariff is given by
\begin{align}
    \frac{d\ln va_H^k}{d\ln(1+\tau^j)}\bigg|_{\tau^j=0}
    &=
    \underbrace{
    \frac{P_H^k C_H^k}{L_H^k}
    \left[
    \frac{d\ln \omega_H^k}{d\ln(1+\tau^j)}
    +
    \frac{d(\widetilde{T}_H^j/L_H)}{d(1+\tau^j)}
    \right]
    }_{\text{(a) Expenditure adjusted by tariff revenue}}  \notag \\
    &\quad
    +
    \underbrace{
    \frac{NX_H^k}{L_H^k}
    \frac{
    d\ln \left( NX_H^k/L_H\right)
    }{
    d\ln(1+\tau^j)
    }
    }_{\text{(b) Net exports}}
    -
    \underbrace{
    \frac{\mathbf{1}_{\{ k=j\}}}{va_H^k}
    \frac{
    d(\widetilde{T}_H^k/L_H)
    }{
    d(1+\tau^j)
    }
    }_{\text{(c) Tariff revenue}}.
    \label{ap: decomposition_va}
\end{align}

Term (a) in \eqref{ap: decomposition_va} is largely governed by the effect on sectoral expenditure shares discussed in Supplemental Appendix~\ref{ap: proof of prop 2}, but is pushed in the positive direction by the additional tariff-revenue-per-income term, which we will see shortly.

The change in net exports, term (b) in \eqref{ap: decomposition_va}, is further decomposed as
\begin{align*}
    \frac{NX_H^k}{L_H^k}\frac{d\ln (NX_H^k/L_H)}{d\ln (1+\tau^j)}\bigg|_{\tau^j=0} &=
    \frac{EX_H^k}{L_H^k}\frac{d\ln (EX_H^k/L_H)}{d\ln (1+\tau^j)}
    - \frac{IM_H^k}{L_H^k}\frac{d\ln (IM_H^k/L_H)}{d\ln (1+\tau^j)}\\
    \frac{EX_H^k}{L_H^k}\frac{d\ln (EX_H^k/L_H)}{d\ln (1+\tau^j)}\bigg|_{\tau^j=0}&= \frac{1}{va_H^k}  \frac{\pi_{FH}^k \omega_F^k w_F}{L_H/L_F}\left[\frac{d\ln \pi_{FH}^k}{d\ln (1+\tau^j)} + \frac{d\ln \omega_F^k}{d\ln (1+\tau^j)} + \frac{d\ln w_F}{d\ln (1+\tau^j)} \right], \\
    \frac{IM_H^k}{L_H^k}\frac{d\ln (IM_H^k/L_H)}{d\ln (1+\tau^j)}\bigg|_{\tau^j=0}&= \frac{\pi_{HF}^k \omega_H^k}{va_H^k} \left[ \frac{d\ln \pi_{HF}^k}{d\ln (1+\tau^j)} +\frac{d\ln \omega_H^k}{d\ln (1+\tau^j)} + \frac{d (\widetilde{T}_H^j/L_H)}{d\ln (1+\tau^j)} -\mathbf{1}_{\{ k=j\}} \right],
\end{align*}
noting $\widetilde{T}_H^j$ is zero when it is evaluated at $\tau^j=0$. An analogous expression holds for Foreign. Changes in tariffs by Home affect their net exports to Foreign in three channels. First, tariffs directly make Foreign's varieties more expensive, and hence reduce Home's imports and raise their net exports, which is captured by the very last term in the third equation above, $\mathbf{1}_{\{ k=j\}}$. Second, tariffs affect the demand for sector $j$ composite good in both countries by
changing their sectoral expenditure shares and Home's tariff revenue. Finally, tariffs affect the trade shares, $d\ln \pi_{ni}^k$, by changing the relative cost of production, i.e., the comparative advantage of the two countries.

Similarly, the change in tariff revenue, term (c) in \eqref{ap: decomposition_va}, is rewritten as
\begin{align*}
    -\frac{1}{va_H^k} \frac{d(\widetilde{T}_H^k/L_H)}{d(1+\tau^j)}\bigg|_{\tau^j=0} = -\frac{\pi_{HF}^k \omega_H^k}{va_H^k} < 0,
\end{align*}
which is always negative.

In the three sector model considered in the text, term (a) evaluated at $\tau^m=0$ is further simplified as
\begin{align*}
    \text{(a) Expenditure}
    =
    \begin{cases}
    \displaystyle
    \frac{P_H^a C_H^a}{L_H^a}
    \biggl[
    \sigma\pi_{HF}\omega_H^m
    +
    \underbrace{(1-\sigma)(\epsilon^a-\overline{\epsilon}_H)\Theta_H}_{<0}
    \biggr]
    & k=a
    \\[1.2em]
    \displaystyle
    \frac{P_H^m C_H^m}{L_H^m}
    \biggl[
    \pi_{HF}\{(1-\sigma)+\sigma\omega_H^m\}
    +
    \underbrace{(1-\sigma)(1-\overline{\epsilon}_H)\Theta_H}_{<0}
    \biggr]
    & k=m
    \\[1.2em]
    \displaystyle
    \frac{P_H^s C_H^s}{L_H^s}
    \biggl[
    \sigma\pi_{HF}\omega_H^m
    +
    \underbrace{(1-\sigma)(\epsilon^s-\overline{\epsilon}_H)\Theta_H}_{>0}
    \biggr]
    & k=s
    \end{cases},
\end{align*}
where $\Theta_H = d\ln(C_H/L_H)/d\ln(1+\tau^m) < 0$ at $\tau^m=0$.
Under homothetic CES preferences, term (a) is positive in every sector. Under nonhomothetic CES preferences with $\overline\epsilon_H>1$, the income effect weakens this positive effect in agriculture and manufacturing, while strengthening it in services.

Similarly, assuming $\overline\epsilon_H>1$ and $\overline\epsilon_F>1$, term (b) evaluated at $\tau^m=0$ is simplified as
\begin{align*}
    \text{(b) Net exports}
    =
    \begin{cases}
    \displaystyle
    \Upsilon^{a}
    \underbrace{
    -(1-\sigma)\Gamma^a\omega_H^a
    (\epsilon^a-\overline{\epsilon}_H)\Theta_H
    }_{>0}
    +
    \underbrace{
    (1-\sigma)\Gamma^a\omega_F^a
    (\epsilon^a-\overline{\epsilon}_F)\Theta_F
    }_{>0}
    & k=a
    \\[1.4em]
    \displaystyle
    \Upsilon^{m}
    \underbrace{
    -(1-\sigma)\Gamma^m\omega_H^m
    (1-\overline{\epsilon}_H)\Theta_H
    }_{>0}
    +
    \underbrace{
    (1-\sigma)\Gamma^m\omega_F^m
    (1-\overline{\epsilon}_F)\Theta_F
    }_{>0}
    & k=m
    \\[1.4em]
    \displaystyle
    \Upsilon^{s}
    \underbrace{
    -(1-\sigma)\Gamma^s\omega_H^s
    (\epsilon^s-\overline{\epsilon}_H)\Theta_H
    }_{<0}
    +
    \underbrace{
    (1-\sigma)\Gamma^s\omega_F^s
    (\epsilon^s-\overline{\epsilon}_F)\Theta_F
    }_{<0}
    & k=s
    \end{cases},
\end{align*}
where $\Theta_F = d\ln (C_F/L_F)/d\ln (1+\tau^m)$ is the effect of Home's manufacturing tariff on Foreign's welfare and is negative at $\tau^m=0$. The non-income-effect terms are summarized as {\small
\begin{align*}
    \Upsilon^k
    =
    \begin{cases}
    \displaystyle
    -
    \Gamma^a\omega_H^m
    \biggl[
    \sigma\pi_{HF}\omega_H^a
    +
    \frac{\pi_{HH}(1+\theta)}{\vartheta}
    \left\{
    \omega_F^a(1+\theta\pi_{FF})
    +
    \omega_H^a\theta\pi_{HH}
    \right\}
    \biggr] < 0
    & k=a
    \\[1.4em]
    \displaystyle
    \Gamma^m\omega_H^m
    \biggl[
    \pi_{HH}(1+\theta)
    +
    \sigma\pi_{HF}(1-\omega_H^m)
    -
    \frac{\pi_{HH}(1+\theta)}{\vartheta}
    \left\{
    \omega_F^m(1+\theta\pi_{FF})
    +
    \omega_H^m\theta\pi_{HH}
    \right\}
    \biggr] > 0
    & k=m
    \\[1.4em]
    \displaystyle
    -
    \Gamma^s\omega_H^m
    \biggl[
    \sigma\pi_{HF}\omega_H^s
    +
    \frac{\pi_{HH}(1+\theta)}{\vartheta}
    \left\{
    \omega_F^s(1+\theta\pi_{FF})
    +
    \omega_H^s\theta\pi_{HH}
    \right\}
    \biggr] < 0
    & k=s
    \end{cases},
\end{align*} }
\noindent where
\begin{align*}
    &\vartheta = 1 + \theta(2-\pi_{HF}-\pi_{FH})=1+\theta(\pi_{HH}+\pi_{FF}), && \\
    &\Gamma^k = \frac{\pi_{HF}}{\pi_{HH}\omega_H^k + \pi_{HF} \omega_F^k}. &&k \in \{ a,m,s\}
\end{align*}
Under homothetic CES preferences, the sign of term (b) is determined by that of $\Upsilon^k$, so that it is negative in the non-protected sectors, agriculture and services, and positive in the protected sector, manufacturing. Under nonhomothetic CES preferences with $\overline\epsilon_H>1$ and $\overline\epsilon_F>1$, term (b) for manufacturing is unambiguously positive. In Home, the tariff raises real income, which shifts expenditure away from manufacturing and reduces manufacturing imports from Foreign. In Foreign, the tariff lowers real income, which shifts expenditure toward manufacturing and expands Home's manufacturing exports to Foreign. Both channels increase Home's net exports in the manufacturing sector.

Summing the three terms yields {\small \begin{align*}
    \frac{d\ln va_H^k}{d\ln(1+\tau^m)}\bigg|_{\tau^m=0}
    =
    \begin{cases}
    \displaystyle
    \Gamma^a\pi_{HH}\omega_H^m
    \left(
    \sigma\omega_H^a
    -
    \frac{1+\theta}{\vartheta}X^a
    \right)
    +
    \underbrace{\Gamma_H^a
    (\epsilon^a-\overline{\epsilon}_H)\Theta_H}_{<0}
    +
    \underbrace{\Gamma_F^a
    (\epsilon^a-\overline{\epsilon}_F)\Theta_F}_{>0}
    & k=a
    \\[1.4em]
    \displaystyle
    \Gamma^m\pi_{HH}\omega_H^m
    \left(
    \theta+1-\sigma+\sigma\omega_H^m
    -
    \frac{1+\theta}{\vartheta}X^m
    \right)
    +
    \underbrace{\Gamma_H^m
    (1-\overline{\epsilon}_H)\Theta_H}_{<0}
    +
    \underbrace{\Gamma_F^m
    (1-\overline{\epsilon}_F)\Theta_F}_{>0}
    & k=m
    \\[1.4em]
    \displaystyle
    \Gamma^s\pi_{HH}\omega_H^m
    \left(
    \sigma\omega_H^s
    -
    \frac{1+\theta}{\vartheta}X^s
    \right)
    +
    \underbrace{\Gamma_H^s
    (\epsilon^s-\overline{\epsilon}_H)\Theta_H}_{<0}
    +
    \underbrace{\Gamma_F^s
    (\epsilon^s-\overline{\epsilon}_F)\Theta_F}_{>0}
    & k=s
    \end{cases}
\end{align*} }
where
\begin{align*}
    &X^k = \omega_F^k(1+\theta\pi_{FF})
    +
    \omega_H^k\theta\pi_{HH}, &&k \in \{a,m,s \}\\
    &\Gamma_H^k = \frac{(1-\sigma)\pi_{HH}\omega_H^k}{\pi_{HH}\omega_H^k + \pi_{HF}\omega_F^k}, \qquad \Gamma_F^k = \frac{(1-\sigma)\pi_{HF}\omega_F^k}{\pi_{HH}\omega_H^k + \pi_{HF}\omega_F^k}. &&k \in \{a,m,s \}
\end{align*}
Under homothetic CES preferences, we use $\omega_n^k = 1/3$ and $\Theta_n=0$ for all $n$ and $k$ to simplify the equation above as
\begin{align*}
\frac{d\ln va_H^k}{d\ln(1+\tau^m)}\bigg|_{\tau^m=0} = \begin{cases} \displaystyle -\pi_{HF}\pi_{HH}(1+\theta-\sigma)/3<0 & k=a \\[1em]
\displaystyle 2\pi_{HF}\pi_{HH}(1+\theta-\sigma)>0 & k=m \\[1em]
\displaystyle -\pi_{HF}\pi_{HH}(1+\theta-\sigma)/3<0 & k=s \end{cases}.
\end{align*}
Sectoral value-added shares respond in the same direction as sectoral expenditure shares as we checked in Supplemental Appendix~\ref{ap: proof of prop 2}.

Under nonhomothetic CES preferences, the sign of the effect cannot generally be determined without further assumptions. Imposing symmetry between the two countries ($L_H=L_F$, $A_H=A_F$, and $d_{HF}=d_{FH}=d$) and assuming $\overline{\epsilon}_H = \overline{\epsilon}_F = \overline{\epsilon}>1$, the effect on sectoral value-added reduces to
\begin{align*}
\frac{d\ln va_H^k}{d\ln(1+\tau^m)}\bigg|_{\tau^m=0} = \begin{cases} \displaystyle \underbrace{-\pi\pi_D\omega^m(1+\theta-\sigma)}_{<0} + \underbrace{(1-\sigma)(\pi_D-\pi)(\epsilon^a-\overline{\epsilon})\Theta_H}_{<0} & k=a \\[1.2em] \displaystyle \underbrace{\pi\pi_D(1+\theta-\sigma)(1-\omega^m)}_{>0} + \underbrace{(1-\sigma)(\pi_D-\pi)(1-\overline{\epsilon})\Theta_H}_{<0} & k=m \\[1.2em] \displaystyle \underbrace{-\pi\pi_D\omega^m(1+\theta-\sigma)}_{<0} + \underbrace{(1-\sigma)(\pi_D-\pi)(\epsilon^s-\overline{\epsilon})\Theta_H}_{>0} & k=s \end{cases},
\end{align*}
noting $w_F=w_H=1$, $\omega_H^k = \omega_F^k = \omega^k$, and
\begin{align*}
    &\pi_{HH}=\pi_{FF}= \pi_D = \frac{(1/A)^{-\theta}}{(1/A)^{-\theta} + (d/A)^{-\theta}} = \frac{d^{\theta}}{1+d^{\theta}} > \frac{1}{2}, \qquad \pi_{HF}=\pi_{FH}=\pi=1-\pi_D < \frac{1}{2}, \\
    &\vartheta = 1 + 2\theta \pi_D, \qquad \Theta_H = - \Theta_F = \frac{\pi \pi_D(1+\theta)\omega^m}{\overline{\epsilon} \vartheta}.
\end{align*}

\

\section{Optimal Tariff} \label{ap: optimal tariff}

\noindent \textbf{Proposition D:} \qquad We consider an economy with two countries and $J$ sectors. Suppose Home imposes a tariff in sector $j$ to maximize its welfare ($\tau_{HF}^j=\tau^j>0$ and $\tau^k=0$ for $k \neq j \in \{ 1,\dots,J\}$), measured by real per capita consumption $u_H = C_H/L_H$, while Foreign does not impose tariffs ($\tau_{FH}^k=0$ for all $k$). If an interior optimal tariff exists, it must satisfy
\begin{align}
    \tau^j
    &= \frac{1}{s_{HF}^j \varepsilon_{X,F}}.
    \label{ap_eq: optimal tariff}
\end{align}
Here $\varepsilon_{X,F}$ is the Foreign export supply elasticity with respect to the net-of-tariff world price:
\begin{align*}
    \varepsilon_{X,F} = \frac{d\ln M_H^j}{d\ln p_{HF}^j},
\end{align*}
where $p_{HF}^j$ is the tariff-exclusive price of Foreign-produced sector $j$ varieties in Home, and $M_H^j$ is the quantity index of Foreign-produced sector-$j$ varieties imported by Home:
\begin{align*}
    p_{HF}^j = \frac{w_F d_{HF}^j}{A_F^j}, \qquad
    M_H^j = \frac{IM_H^j}{p_{HF}^j}
    =
    \frac{1}{p_{HF}^j}
    \frac{\pi_{HF}^j \omega_H^j E_H}{1+\tau^j},
\end{align*}
and $s_{HF}^j$ is sector $j$'s share of Home's total tariff-exclusive imports:
\begin{align*}
    s_{HF}^j
    = \frac{IM_H^j}{\sum_{h=1}^J IM_H^h}
    =
    \frac{\pi_{HF}^j\omega_H^j/(1+\tau^j)}
    {
    \sum_{h=1}^J \pi_{HF}^h\omega_H^h/(1+\tau^h)
    }.
\end{align*}

\medskip

\noindent \textbf{Proof of Proposition D:} \qquad We use the same notation as in Supplemental Appendix~\ref{ap: proof of prop 1} and proceed in three steps. We note that Home labor is chosen as the numeraire, so that $w_H=1$, while the Foreign wage is denoted by $w_F$.

\medskip

\noindent
\textit{Step 1: Optimal-tariff formula}

As in Step 1 in Supplemental Appendix~\ref{ap: proof of prop 1}, differentiating Home's per-capita expenditure function with respect to $\ln(1+\tau^j)$ gives
\begin{align*}
    \frac{d\ln(E_H/L_H)}{d\ln(1+\tau^j)}
    &=
    \overline\epsilon_H
    \frac{d\ln (C_H/L_H)}{d\ln(1+\tau^j)}
    +
    \sum_{k=1}^J
    \omega_H^k
    \frac{d\ln P_H^k}{d\ln(1+\tau^j)}
    \\
    &=
    \overline\epsilon_H
    \frac{d\ln (C_H/L_H)}{d\ln(1+\tau^j)}
    +
    \omega_H^j\pi_{HF}^j
    +
    \left(
    \Lambda_{HF}
    +
    \frac{\tau^j\pi_{HF}^j\omega_H^j}{1+\tau^j}
    \right)
    \varepsilon_{w,F},
\end{align*}
where
\begin{align*}
    \frac{d\ln P_H^k}{d\ln(1+\tau^j)}
    =
    \pi_{HF}^k
    \left[
    \mathbf{1}_{\{k=j\}}
    +
    \varepsilon_{w,F}
    \right],
    \qquad
    \varepsilon_{w,F}
    =
    \frac{d\ln w_F}{d\ln(1+\tau^j)}.
\end{align*}
Since the derivative of real per capita consumption is zero at an interior optimum, $d\ln (C_H/L_H)/d\ln (1+\tau^j)=0$, we can simplify the equation above as
\begin{align}
    \frac{d\ln(E_H/L_H)}{d\ln(1+\tau^j)}
    =
    \omega_H^j\pi_{HF}^j
    +
    \left(
    \Lambda_{HF}
    +
    \frac{\tau^j\pi_{HF}^j\omega_H^j}{1+\tau^j}
    \right)
    \varepsilon_{w,F}.
    \label{ap: exp_derivative_prices}
\end{align}

We now turn to Home's budget constraint:
\begin{align*}
    E_H
    =
    L_H+\tau^j IM_H^j.
\end{align*}
Differentiating this with respect to $\ln(1+\tau^j)$ gives
\begin{align}
    \frac{d\ln E_H}{d\ln(1+\tau^j)}
    &=
    \frac{(1+\tau^j)IM_H^j}{E_H}
    +
    \frac{\tau^jIM_H^j}{E_H}
    \frac{d\ln IM_H^j}{d\ln(1+\tau^j)}
    \notag
    \\
    &=
    \pi_{HF}^j\omega_H^j
    -
    \frac{\tau^j}{1+\tau^j}
    \pi_{HF}^j\omega_H^j
    \varepsilon_{M,H},
    \label{ap: exp_derivative_budget}
\end{align}
where $\varepsilon_{M,H}$ is the elasticity of Home's tariff-exclusive imports in sector $j$ with respect to $\ln (1+\tau^j)$:
\begin{align*}
    \varepsilon_{M,H}
    =
    -\frac{d\ln IM_H^j}{d\ln (1+\tau^j)}.
\end{align*}

Noting $d\ln(E_H/L_H)/d\ln(1+\tau^j)=d\ln E_H/d\ln(1+\tau^j)$ since $L_H$ is fixed, we equate \eqref{ap: exp_derivative_prices} and \eqref{ap: exp_derivative_budget} to obtain
\begin{align*}
    \pi_{HF}^j\omega_H^j
    -
    \frac{\tau^j}{1+\tau^j}
    \pi_{HF}^j\omega_H^j
    \varepsilon_{M,H}
    &=
    \omega_H^j\pi_{HF}^j
    +
    \left(
    \Lambda_{HF}
    +
    \frac{\tau^j\pi_{HF}^j\omega_H^j}{1+\tau^j}
    \right)
    \varepsilon_{w,F}.
\end{align*}
After canceling $\pi_{HF}^j\omega_H^j$ from both sides, we have
\begin{align*}
    &-\frac{\tau^j}{1+\tau^j}
    \pi_{HF}^j\omega_H^j
    \left(
    \varepsilon_{M,H}
    +
    \varepsilon_{w,F}
    \right)
    =
    \Lambda_{HF}\varepsilon_{w,F},
    \\
    &\Leftrightarrow
    \frac{\tau^j}{1+\tau^j}
    =
    -\frac{
    \Lambda_{HF}\varepsilon_{w,F}
    }{
    \pi_{HF}^j\omega_H^j
    \left(
    \varepsilon_{M,H}
    +
    \varepsilon_{w,F}
    \right)
    },
    \\
    &\Leftrightarrow
    \tau^j
    =
    - \frac{\varepsilon_{w,F}}
    {s_{HF}^j
    \left(
    \varepsilon_{M,H}
    +
    \varepsilon_{w,F}
    \right)
    }.
\end{align*}
From the second to the third line, we use
\begin{align*}
    s_{HF}^j
    &=
    \frac{\pi_{HF}^j \omega_H^j/(1+\tau^j)}
    {\sum_{h=1}^J \pi_{HF}^h \omega_H^h/(1+\tau^h)}
    =
    \frac{\pi_{HF}^j \omega_H^j/(1+\tau^j)}
    {\Lambda_{HF}},
    \\
    &\Leftrightarrow
    \frac{\Lambda_{HF}}{\pi_{HF}^j \omega_H^j}
    =
    \frac{1}{s_{HF}^j(1+\tau^j)}.
\end{align*}

Furthermore, since $IM_H^j=p_{HF}^jM_H^j$, we differentiate $\ln IM_H^j=\ln p_{HF}^j+\ln M_H^j$ with respect to $\ln(1+\tau^j)$ to obtain
\begin{align*}
    \frac{d\ln IM_H^j}{d\ln(1+\tau^j)}
    &=
    \frac{d\ln p_{HF}^j}{d\ln(1+\tau^j)}
    +
    \frac{d\ln M_H^j}{d\ln(1+\tau^j)}.
\end{align*}
Noting $p_{HF}^j=w_Fd_{HF}^j/A_F^j$, and $d_{HF}^j$ and $A_F^j$ are fixed, we have
\begin{align*}
    \frac{d\ln p_{HF}^j}{d\ln(1+\tau^j)}
    =
    \frac{d\ln w_F}{d\ln(1+\tau^j)}
    =
    \varepsilon_{w,F}.
\end{align*}
Using $d\ln IM_H^j/d\ln(1+\tau^j)=-\varepsilon_{M,H}$, we obtain
\begin{align*}
    \frac{d\ln M_H^j}{d\ln(1+\tau^j)}
    =
    -(\varepsilon_{M,H}+\varepsilon_{w,F}).
\end{align*}
Therefore, the Foreign export supply elasticity can be written as
\begin{align*}
    &\varepsilon_{X,F}
    =
    \frac{d\ln M_H^j}{d\ln p_{HF}^j}
    =
    \frac{
    d\ln M_H^j/d\ln(1+\tau^j)
    }{
    d\ln p_{HF}^j/d\ln(1+\tau^j)
    }
    =
    -\frac{\varepsilon_{M,H}+\varepsilon_{w,F}}{\varepsilon_{w,F}}, \\
    &\Leftrightarrow -\frac{\varepsilon_{w,F}}
    {\varepsilon_{M,H}+\varepsilon_{w,F}}
    =
    \frac{1}{\varepsilon_{X,F}}.
\end{align*}
Using this relationship, we have
\begin{align*}
    \tau^j
    =
    \frac{1}{s_{HF}^j\varepsilon_{X,F}}.
\end{align*}
In the remaining steps, we derive the explicit expressions for $\varepsilon_{M,H}$ and $\varepsilon_{w,F}$.

\medskip

\noindent
\textit{Step 2: Derivation of the elasticity of Home imports $\varepsilon_{M,H}$}

We differentiate Home's sector $j$ tariff-exclusive imports, $IM_H^j =\pi_{HF}^j \omega_H^j E_H/(1+\tau^j)$, with respect to $\ln(1+\tau^j)$:
\begin{align*}
    \frac{d\ln IM_H^j}{d\ln(1+\tau^j)}
    =
    \frac{d\ln\pi_{HF}^j}{d\ln(1+\tau^j)}
    +
    \frac{d\ln\omega_H^j}{d\ln(1+\tau^j)}
    +
    \frac{d\ln E_H}{d\ln(1+\tau^j)}
    -
    1.
\end{align*}
Evalauated at an interior optimal tariff, each term is given by
\begin{align*}
    \frac{d\ln\pi_{HF}^j}{d\ln(1+\tau^j)}
    &=
    -\theta\pi_{HH}^j
    \left(
    1+\varepsilon_{w,F}
    \right),
    \\
    \frac{d\ln\omega_H^j}{d\ln(1+\tau^j)}
    &=
    (1-\sigma)
    \left[
    (1-\omega_H^j)\pi_{HF}^j
    +
    (\Lambda_{HH}-\pi_{HH}^j)\varepsilon_{w,F}
    \right],
    \\
    \frac{d\ln E_H}{d\ln(1+\tau^j)}
    &=
    \omega_H^j\pi_{HF}^j
    +
    (1-\Lambda_{HH})\varepsilon_{w,F}.
\end{align*}
Substituting these expressions into $\varepsilon_{M,H}=-d\ln IM_H^j/d\ln (1+\tau^j)$ yields
\begin{align}
    \varepsilon_{M,H}
    &=
    (\theta+1)\pi_{HH}^j
    +
    \sigma(1-\omega_H^j)\pi_{HF}^j
    +
    \left[
    \sigma\Lambda_{HH}
    +
    (\theta+1-\sigma)\pi_{HH}^j
    -
    1
    \right]
    \varepsilon_{w,F}.
    \label{app: import_elasticity_home_numeraire}
\end{align}

\medskip

\noindent
\textit{Step 3: Derivation of the Foreign wage elasticity $\varepsilon_{w,F}$}

The aggregate trade-balance condition is
\begin{align*}
    \sum_{k=1}^J IM_H^k
    =
    \sum_{k=1}^J EX_H^k.
\end{align*}
Differentiating both sides with respect to $\ln(1+\tau^j)$ and dividing by the common total trade value gives
\begin{align*}
    \sum_{k=1}^J
    s_{HF}^k
    \frac{d\ln IM_H^k}{d\ln(1+\tau^j)}
    =
    \sum_{k=1}^J
    s_{FH}^k
    \frac{d\ln EX_H^k}{d\ln(1+\tau^j)}.
\end{align*}

We first compute the left-hand side. For each sector $k$, we have
\begin{align*}
    \frac{d\ln IM_H^k}{d\ln(1+\tau^j)}
    &=
    \frac{d\ln\pi_{HF}^k}{d\ln(1+\tau^j)}
    +
    \frac{d\ln\omega_H^k}{d\ln(1+\tau^j)}
    +
    \frac{d\ln E_H}{d\ln(1+\tau^j)}
    -
    \mathbf{1}_{\{k=j\}}.
\end{align*}
Using the results from Step 1, we obtain
\begin{align*}
    \sum_{k=1}^J
    s_{HF}^k
    \frac{d\ln IM_H^k}{d\ln(1+\tau^j)}
    &=
    -s_{HF}^j
    \left[
    \sigma+(\theta+1-\sigma)\pi_{HH}^j
    \right]
    +
    \sigma\omega_H^j\pi_{HF}^j
    \\
    &\quad
    +
    \left[
    1
    -
    (\theta+1-\sigma)\overline\pi_{HH}
    -
    \sigma\Lambda_{HH}
    \right]
    \varepsilon_{w,F}.
\end{align*}

We next compute the right-hand side. Noting that Foreign does not impose tariffs, its budget constraint is $E_F=w_F L_F$. Differentiating this with repect to $\ln (1+\tau^j)$ yields
\begin{align*}
    \frac{d\ln E_F}{d\ln(1+\tau^j)}
    =
    \frac{d\ln(E_F/L_F)}{d\ln(1+\tau^j)}
    =
    \frac{d\ln w_F}{d\ln(1+\tau^j)}
    =
    \varepsilon_{w,F},
\end{align*}
where $L_F$ is fixed.

We now derive the response of Foreign real per capita consumption. Foreign's per-capita expenditure function is
\begin{align*}
    \frac{E_F}{L_F}
    =
    \left[
    \sum_{k=1}^J
    \left( \frac{C_F}{L_F} \right)^{\epsilon^k(1-\sigma)}
    (P_F^k)^{1-\sigma}
    \right]^{\frac{1}{1-\sigma}}.
\end{align*}
Differentiating this expression with respect to $\ln(1+\tau^j)$ gives
\begin{align*}
    \frac{d\ln(E_F/L_F)}{d\ln(1+\tau^j)}
    &=
    \overline\epsilon_F
    \frac{d\ln(C_F/L_F)}{d\ln(1+\tau^j)}
    +
    \sum_{k=1}^J
    \omega_F^k
    \frac{d\ln P_F^k}{d\ln(1+\tau^j)}.
\end{align*}
Home's tariff affects Foreign sectoral price indices only through the Foreign wage $w_F$:
\begin{align*}
    \frac{d\ln P_F^k}{d\ln(1+\tau^j)}
    =
    \pi_{FF}^k\varepsilon_{w,F}.
\end{align*}
Averaging across sectors with weights $\omega_F^k$, we obtain
\begin{align*}
    \sum_{k=1}^J
    \omega_F^k
    \frac{d\ln P_F^k}{d\ln(1+\tau^j)}
    =
    \sum_{k=1}^J
    \omega_F^k\pi_{FF}^k
    \varepsilon_{w,F}
    =
    \Lambda_{FF}\varepsilon_{w,F},
\end{align*}
where
\[
    \Lambda_{FF}
    =
    \sum_{k=1}^J
    \pi_{FF}^k\omega_F^k
    =
    1-\Lambda_{FH}.
\]
Since $E_F = w_F L_F$ and $d\ln(E_F/L_F)/d\ln(1+\tau^j)=\varepsilon_{w,F}$, we have
\begin{align*}
    \varepsilon_{w,F}
    &=
    \overline\epsilon_F
    \frac{d\ln(C_F/L_F)}{d\ln(1+\tau^j)}
    +
    \Lambda_{FF}\varepsilon_{w,F}
    \\
    &=
    \overline\epsilon_F
    \frac{d\ln(C_F/L_F)}{d\ln(1+\tau^j)}
    +
    (1-\Lambda_{FH})\varepsilon_{w,F}.
\end{align*}
Rearranging this yields
\begin{align}
    \frac{d\ln(C_F/L_F)}{d\ln(1+\tau^j)}
    =
    \frac{\Lambda_{FH} \varepsilon_{w,F}}{\overline\epsilon_F}.
    \label{app: foreign_real_income_derivative}
\end{align}

This real-income response is needed because Foreign's sectoral expenditure shares depend on real per capita consumption under nonhomothetic CES preferences. Differentiating $\omega_F^k$ gives
\begin{align*}
    \frac{d\ln\omega_F^k}{d\ln(1+\tau^j)}
    &=
    (1-\sigma)
    \left[
    \frac{d\ln P_F^k}{d\ln(1+\tau^j)}
    -
    \sum_{h=1}^J
    \omega_F^h
    \frac{d\ln P_F^h}{d\ln(1+\tau^j)}
    \right. \left.
    +
    (\epsilon^k-\overline\epsilon_F)
    \frac{d\ln(C_F/L_F)}{d\ln(1+\tau^j)}
    \right]
    \\
    &=
    (1-\sigma)
    \left[
    \pi_{FF}^k
    -
    \Lambda_{FF}
    +
    \frac{\Lambda_{FH}}{\overline\epsilon_F}
    (\epsilon^k-\overline\epsilon_F)
    \right]
    \varepsilon_{w,F}
    \\
    &=
    (1-\sigma)
    \left[
    \pi_{FF}^k
    -
    (1-\Lambda_{FH})
    +
    \frac{\Lambda_{FH}}{\overline\epsilon_F}
    (\epsilon^k-\overline\epsilon_F)
    \right]
    \varepsilon_{w,F}.
\end{align*}

We can now compute the response of Home's exports in sector $k$. Noting $EX_H^k=\pi_{FH}^k\omega_F^kE_F$, we have
\begin{align*}
    \frac{d\ln EX_H^k}{d\ln(1+\tau^j)}
    =
    \frac{d\ln\pi_{FH}^k}{d\ln(1+\tau^j)}
    +
    \frac{d\ln\omega_F^k}{d\ln(1+\tau^j)}
    +
    \frac{d\ln E_F}{d\ln(1+\tau^j)}.
\end{align*}
The response of Foreign's expenditure share on Home-produced sector $k$ varieties is
\begin{align*}
    \frac{d\ln\pi_{FH}^k}{d\ln(1+\tau^j)}
    =
    \theta\pi_{FF}^k\varepsilon_{w,F},
\end{align*}
and, as shown above, $d\ln E_F/d\ln(1+\tau^j)=\varepsilon_{w,F}$. Therefore,
\begin{align*}
    \frac{d\ln EX_H^k}{d\ln(1+\tau^j)}
    &=
    \theta\pi_{FF}^k\varepsilon_{w,F}    +
    (1-\sigma)
    \left[
    \pi_{FF}^k
    -
    (1-\Lambda_{FH})
    +
    \frac{\Lambda_{FH}}{\overline\epsilon_F}
    (\epsilon^k-\overline\epsilon_F)
    \right]
    \varepsilon_{w,F}
    +
    \varepsilon_{w,F}
    \\
    &=
    \biggl[
    1
    +
    \theta\pi_{FF}^k
    +
    (1-\sigma)
    \left\{
    \pi_{FF}^k
    -
    (1-\Lambda_{FH})
    +
    \frac{\Lambda_{FH}}{\overline\epsilon_F}
    (\epsilon^k-\overline\epsilon_F)
    \right\}
    \biggr]
    \varepsilon_{w,F}.
\end{align*}
Averaging this expression with weights $s_{FH}^k$ gives
\begin{align*}
    \sum_{k=1}^J
    s_{FH}^k
    \frac{d\ln EX_H^k}{d\ln(1+\tau^j)}
    &=
    \biggl[
    1
    +
    \theta\overline\pi_{FF}
    +
    (1-\sigma)
    \left\{
    \Lambda_{FH}
    -
    \overline\pi_{FH}
    +
    \frac{\Lambda_{FH}}{\overline\epsilon_F}
    (\overline\epsilon_{FH}-\overline\epsilon_F)
    \right\}
    \biggr]
    \varepsilon_{w,F}.
\end{align*}

Equating the import and export sides gives
\begin{align*}
&-s_{HF}^j
\left[
\sigma+(\theta+1-\sigma)\pi_{HH}^j
\right]
+
\sigma\omega_H^j\pi_{HF}^j
+
\left[
1
-
(\theta+1-\sigma)\overline\pi_{HH}
-
\sigma\Lambda_{HH}
\right]
\varepsilon_{w,F}
\\
&\qquad =
\biggl[
1
+
\theta\overline\pi_{FF}
+
(1-\sigma)
\left\{
\Lambda_{FH}
-
\overline\pi_{FH}
+
\frac{\Lambda_{FH}}{\overline\epsilon_F}
(\overline\epsilon_{FH}-\overline\epsilon_F)
\right\}
\biggr]
\varepsilon_{w,F}.
\end{align*}
Solving for $\varepsilon_{w,F}$ yields
\begin{align}
    \varepsilon_{w,F}
    =
    -
    \frac{
    s_{HF}^j
    \left[
    \sigma+(\theta+1-\sigma)\pi_{HH}^j
    \right]
    -
    \sigma\omega_H^j\pi_{HF}^j
    }{
    (\theta+1-\sigma)\overline\pi_{HH}
    +
    \sigma\Lambda_{HH}
    +
    \theta\overline\pi_{FF}
    -
    (1-\sigma)
    \left[
    \overline\pi_{FH}
    -
    \Lambda_{FH}
    -
    (\Lambda_{FH}/\overline\epsilon_F)
    (\overline\epsilon_{FH}-\overline\epsilon_F)
    \right]
    }.
    \label{ap: foreign_wage_elasticity_home_numeraire}
\end{align}
This completes the proof. \qed

\medskip

\noindent \textit{Relationship between $s_{HF}^j$ and $\omega_H^j$ under sectoral symmetry}

If productivities and trade costs are symmetric across sectors, $A_H^k=A_H$, $A_F^k=A_F$, $d_{HF}^k=d_{HF}>1$, and $d_{FH}^k=d_{FH}>1$ for all $k$, then Home's sectoral tariff-exclusive import share $s_{HF}^j$ increases with its sectoral expenditure share $\omega_H^j$. Letting $\rho_{HF}=A_F/(w_F d_{HF} A_H)$, we can write Foreign's sectoral trade share in Home as
\begin{align*}
    \pi_{HF}^k
    =
    \frac{(w_F d_{HF}/A_F)^{-\theta}}
    {(1/A_H)^{-\theta}+(w_F d_{HF}/A_F)^{-\theta}}
    =
    \begin{cases}
    \dfrac{\rho_{HF}^{\theta}(1+\tau^j)^{-\theta}}
    {1+\rho_{HF}^{\theta}(1+\tau^j)^{-\theta}}
    & k=j
    \\[1.2em]
    \dfrac{\rho_{HF}^{\theta}}
    {1+\rho_{HF}^{\theta}}
    & k\neq j
    \end{cases}.
\end{align*}
Using these expressions, the sectoral import share can be written as
\begin{align*}
    s_{HF}^j
    &=
    \frac{\pi_{HF}^j\omega_H^j/(1+\tau^j)}
    {\sum_{h=1}^J \pi_{HF}^h\omega_H^h/(1+\tau^h)}
    \\
    &=
    \frac{
    \dfrac{\rho_{HF}^{\theta}(1+\tau^j)^{-(1+\theta)}}
    {1+\rho_{HF}^{\theta}(1+\tau^j)^{-\theta}}
    \omega_H^j
    }
    {
    \dfrac{\rho_{HF}^{\theta}}{1+\rho_{HF}^{\theta}}
    \sum_{h\neq j}\omega_H^h
    +
    \dfrac{\rho_{HF}^{\theta}(1+\tau^j)^{-(1+\theta)}}
    {1+\rho_{HF}^{\theta}(1+\tau^j)^{-\theta}}
    \omega_H^j
    }
    \\
    &=
    \frac{
    \dfrac{(1+\tau^j)^{-(1+\theta)}}
    {1+\rho_{HF}^{\theta}(1+\tau^j)^{-\theta}}
    \omega_H^j
    }
    {
    \dfrac{1}{1+\rho_{HF}^{\theta}}
    (1-\omega_H^j)
    +
    \dfrac{(1+\tau^j)^{-(1+\theta)}}
    {1+\rho_{HF}^{\theta}(1+\tau^j)^{-\theta}}
    \omega_H^j
    }.
\end{align*}
which increases with $\omega_H^j$.

\

\section{Welfare} \label{ap: welfare formula}

To measure changes in welfare moving from the baseline to a counterfactual situation, we calculate a constant fraction $\lambda_n$ (different from $\lambda$, the parameter governing the capital adjustment cost in \eqref{eq: capital dynamics}) of per capita consumption that would be paid to the country $n$'s representative consumer in each year in the baseline to achieve the same utility in the counterfactual.
Letting $\{ C_{n,t} \}_t$ and $\{ C^*_{n,t} \}_t$ be the consumption streams in country $n$ in the baseline and in the counterfactual respectively, this fraction $\lambda_n$ is given by
\begin{align*}
    &\sum_{t=0}^{\infty} \beta^t \zeta_{n,t} L_{n,t} \frac{(C_{n,t}^*/L_{n,t})^{1-\psi}}{1-\psi}
    = \sum_{t=0}^{\infty} \beta^t \zeta_{n,t} L_{n,t} \frac{\left( \left( 1+\frac{\lambda_n}{100} \right)C_{n,t}/L_{n,t} \right)^{1-\psi}}{1-\psi}, \\
                        &\Leftrightarrow \lambda_n = 100 \times \left[\left\{ \frac{\sum_{t=0}^{\infty} \beta^t \zeta_{n,t} L_{n,t}(C^*_{n,t}/L_{n,t})^{1-\psi}}{\sum_{t=0}^{\infty} \beta^t \zeta_{n,t} L_{n,t}(C_{n,t}/L_{n,t})^{1-\psi} } \right\}^{\frac{1}{1-\psi}}  - 1 \right].
\end{align*}

In addition, our economy reaches the steady state at $t=T$. Letting the variables without the time subscript represent the steady state values, the formula can be rewritten as
\begin{align*}
    \lambda_n = 100 \times \left[\left\{ \frac{ \sum_{t=0}^{T} \beta^t \zeta_{n,t} L_{n,t} \left( \frac{C^*_{n,t}}{L_{n,t}} \right)^{1-\psi} + \frac{\beta^{T+1}}{1-\beta} \zeta_n L_n \left( \frac{C_n^*}{L_n} \right)^{1-\psi} }{ \sum_{t=0}^{T} \beta^t \zeta_{n,t} L_{n,t} \left( \frac{C_{n,t}}{L_{n,t}} \right)^{1-\psi} + \frac{\beta^{T+1}}{1-\beta} \zeta_n L_n \left( \frac{C_n}{L_n} \right)^{1-\psi} } \right\}^{\frac{1}{1-\psi}}  - 1 \right],
\end{align*}
where $L_{n,t}=L_n$ and $\zeta_{n,t}=\zeta_n$ are time invariant after $T$ and are common in both the baseline and the counterfactual scenarios. We also note
\begin{align*}
    \sum_{t=T+1}^{\infty} \beta^t \zeta_{n} L_n \left( \dfrac{C^*_{n}}{L_{n}} \right)^{1-\psi}
    = \beta^{T+1} \zeta_{n} L_n \left( \dfrac{C^*_{n}}{L_{n}} \right)^{1-\psi} \sum_{t=0}^{\infty} \beta^t
    = \beta^{T+1} \zeta_{n} L_n \left( \dfrac{C^*_{n}}{L_{n}} \right)^{1-\psi} \frac{1}{1-\beta}.
\end{align*}

\

\newpage

\section{List of ISIC3 Industries Included in the Three Sectors}\label{ap: sector}

\begin{table}[h]
    \centering
    \small
    \caption{Three sectors and corresponding ISIC3 codes}\label{Tab: sector_list}
    \begin{tabular}{ccl} \toprule 
    Sector        & ISIC3     &   Description                                            \\ \midrule 
                  & A to B    & Agriculture, Hunting, Forestry and Fishing   \\
    Agriculture   & C         & Mining and Quarrying                         \\
                  & D15 to 16 & Food, Beverages and Tobacco                  \\ \midrule 
                  & D17 to 19 & Textiles, Textile, Leather and Footwear      \\
                  & D21 to 22 & Pulp, Paper, Printing and Publishing  \\
                  & D23       & Coke, Refined Petroleum and Nuclear Fuel     \\
                  & D24       & Chemicals and Chemical Products              \\
    Manufacturing & D25       & Rubber and Plastics                          \\
                  & D26       & Other Non-Metallic Mineral                   \\
                  & D27 to 28 & Basic Metals and Fabricated Metal            \\
                  & D29       & Machinery, Nec                               \\
                  & D30 to 33 & Electrical and Optical Equipment             \\
                  & D34 to 35 & Transport Equipment                          \\
                  & D n.e.c.  & Manufacturing, Nec; Recycling                \\ \midrule 
                  & E         & Electricity, Gas and Water Supply            \\
                  & F         & Construction                                 \\
                  & G         & Wholesale and Retail Trade                   \\
    Service       & H         & Hotels and Restaurants                       \\
                  & I60 to 63 & Transport and Storage                        \\
                  & I64       & Post and Telecommunications                  \\
                  & J         & Financial Intermediation                     \\
                  & K         & Real Estate, Renting and Business Activities \\
                  & L to Q    & Community Social and Personal Services      \\ \bottomrule 
    \end{tabular}
    \end{table}

\section{Calibration of Fundamentals}\label{ap: shock}

We calibrate the iceberg trade costs (including tariffs and non-tariff barriers), $b_{n,t}^j$, and average productivity, $A_{n,t}^j$, following Levchenko and Zhang (2016). Apart from Levchenko and Zhang (2016), we use information from price indices on WIOD to make sequences of productivity $A_{n,t}^j$ comparable across countries and over time within each sector. Since the initial year in the Long-Run WIOD is 1965, we calibrate the relevant parameters since 1965. We compute dynamic equilibria since 1990. We feed the capital stocks in 1990 from the data as the initial capital stocks in the model computation. As shown below, the reference year (1965) appears only in the calibration of productivity $A_{n,t}^j$, but the choice of this reference year does not affect our results.

To begin with, we express the trade share normalized by its own trade share as follows:
\begin{align*}     \frac{\pi_{ni,t}^{j}}{\pi_{nn,t}^{j}}&=\frac{\left(\frac{\widetilde{c}_{i,t}^{j}{b_{ni,t}^{j}}}{A_{i,t}^{j}} \right)^{-\theta^j}}{\left(\frac{\widetilde{c}_{n,t}^{j}}{A_{n,t}^{j}} \right)^{-\theta^j}} = \prn{{{\widetilde{c}}_{i,t}^{j}}/{{A}_{i,t}^{j}}}^{-\theta^j} \times \prn{{{\widetilde{c}}_{n,t}^{j}}/{{A}_{n,t}^{j}}}^{\theta^j} \times \prn{b_{ni,t}^{j}}^{-\theta^j}.
\end{align*}

\noindent Taking the log of both sides yields
\begin{align*}     \ln\prn{\frac{\pi_{ni,t}^{j}}{\pi_{nn,t}^{j}}}&=\ln\prn{{{\widetilde{c}}_{i,t}^{j}}/{{A}_{i,t}^{j}}}^{-\theta^j} + \ln\prn{{{\widetilde{c}}_{n,t}^{j}}/{{A}_{n,t}^{j}}}^{\theta^j}  -\theta^j \ln\prn{b_{ni,t}^{j}}.
\end{align*}

\noindent The total trade costs $b_{ni,t}^j$ can be decomposed into tariffs and non-tariff barriers:
\begin{align*}     \ln\prn{\frac{\pi_{ni,t}^{j}}{\pi_{nn,t}^{j}}}&=\ln\prn{{{\widetilde{c}}_{i,t}^{j}}/{{A}_{i,t}^{j}}}^{-\theta^j} + \ln\prn{{{\widetilde{c}}_{n,t}^{j}}/{{A}_{n,t}^{j}}}^{\theta^j}  -\theta^j \ln\prn{d_{ni,t}^{j}} - \theta^j \ln\prn{\widetilde{\tau}_{ni,t}^{j}},
\end{align*}
where $\widetilde{\tau}_{ni,t}^{j} = 1+{\tau}_{ni,t}^{j}$. We express the log of non-tariff barriers $d_{ni,t}^j$ with the set of bilateral observables commonly used in the gravity estimation:
\begin{align*}
    \ln\prn{d_{ni,t}^{j}} = \text{dist}_{k(ni)}^j + \text{CB}_{ni,t}^j + \text{CU}_{ni,t}^j + \text{RTA}_{ni,t}^j + ex_{i,t}^j + \nu_{ni,t}^j,
\end{align*}

\noindent where $\text{dist}_{k(ni),t}^j$ is the contribution to trade costs of the distance between $n$ and $i$ being in a certain interval,\footnote{We follow Eaton and Kortum (2002) and intervals are defined, in miles, $[0,350]$, $[350,750]$, $[750,1500]$, $[1500,3000]$, $[3000,6000]$, $[6000,\text{max}]$.}  $\text{CB}_{ni,t}^j$ is the indicator if the two countries $n$ and $i$ share the border, $\text{CU}_{ni,t}^j$ indicates if they are in a currency union, $\text{RTA}_{ni,t}^j$ indicates if they are in a regional trade agreement (WTO definition), $ex_{it}^j$ is the exporter fixed effects, and $\nu_{ni,t}^j$ is the bilateral error term. Note that each component in the bilateral trade cost is indexed by $t$, and we estimate them as the fixed effects interacted with years. This implies that, for instance, the contribution of distance to trade costs can vary over time due to the technological progress of transportation. Exporter fixed effects are included to allow asymmetry in trade costs in the spirit of Waugh (2010). We plug this into the trade share equation~\eqref{eq: trade share} in the text and estimate the following using the Pseudo Poisson Maximum Likelihood (PPML) for each sector $j$ while pooling all sampled countries and years:
\begin{align*}     \ln\prn{\frac{\widetilde{\pi}_{ni,t}^{j}}{{\pi}_{nn,t}^{j}}}
      = & \underbrace{ \prn{\ln\prn{{{\widetilde{c}}_{i,t}^{j}}/{{A}_{i,t}^{j}}}^{-\theta^j} - \theta^j ex_{it}^j}} _{\text{exporter-year F.E.}} + \underbrace{\ln\prn{{{\widetilde{c}}_{n,t}^{j}}/{{A}_{n,t}^{j}}}^{\theta^j}}_\text{importer-year F.E.}  \\
    &\underbrace{-\theta^j \prn{ \text{dist}_{k(ni),t}^j + \text{CB}_{ni,t}^j + \text{CU}_{ni,t}^j + \text{RTA}_{ni,t}^j }}_{\text{bilateral observables}} - \theta^j \nu_{ni,t}^j.
\end{align*}

\noindent where $\ln\prn{\frac{\widetilde{\pi}_{ni,t}^{j}}{{\pi}_{nn,t}^{j}}} = \ln\prn{\frac{{\pi}_{ni,t}^{j}}{{\pi}_{nn,t}^{j}}} + \theta^j \ln\prn{\widetilde{\tau}_{ni,t}^{j}}$. Estimating the gravity equation above allows us to identify the technology-cum-unit-cost term, $\ln\prn{{{\widetilde{c}}_{n,t}^{j}}/{{A}_{i,t}^{j}}}^{\theta^j}$, for each country and year as an importer-year fixed effect, relative to the reference country and year (U.S. in 1965), which we denote by $S_{n,t}^j = { \prn{{{\widetilde{c}}_{n,t}^{j}}/{{A}_{n,t}^{j}}}^{\theta^j}}/{ \prn{{{\widetilde{c}}_{US,1965}^{j}}/{{A}_{US,1965}^{j}}}^{\theta^j}}$. We can then tease out the term $(-\theta^jex_{it}^j)$ from the exporter-year fixed effects. By combining all the terms in the bilateral trade costs, we can recover the asymmetric bilateral trade costs.

To back out productivity, we need a few preliminary steps. First, following Shikher (2013), we recover the sectoral price indices as follows. We define the own trade share relative to that of the reference country and year:
\begin{align*}     \frac{\pi_{nn,t}^{j}}{\pi_{US,US,1965}^{j}}&= \frac{\prn{{{\widetilde{c}}_{n,t}^{j}}/{{A}_{n,t}^{j}}}^{-\theta^j}}{\prn{{{\widetilde{c}}_{US,1965}^{j}}/{{A}_{US,1965}^{j}}}^{-\theta^j}}\prn{\frac{P_{n,t}^j}{P_{US,1965}^j}}^{\theta^j} = \frac{1}{S_{n,t}^j} \prn{\frac{P_{n,t}^j}{P_{US,1965}^j}}^{\theta^j}.
\end{align*}
\noindent Hence, for given trade elasticity $\theta^j$, we have\footnote{Note that the price indices are recovered relative to the U.S. in 1965 for each sector, implying that the U.S. price index is 1 for all sectors in 1965.}
\begin{equation} \label{eq:rel_price_wo_norm}
    \frac{P_{n,t}^j}{P_{US,1965}^j} = \prn{ \frac{\pi_{nn,t}^{j}}{\pi_{US,US,1965}^{j}} {S_{n,t}^j}}^{1/\theta^j}.
\end{equation}

\noindent It is important to note that, for each sector and year, $S_{n,t}^j$ is only identified up to normalization. This implies that the sequence of prices given by equation \eqref{eq:rel_price_wo_norm} is only comparable across countries but not over time. To see this point clearly, consider a sequence of any positive scalar $\mprn{a_{t}^j}$. It is easy to show that the two sequences, $\mprn{S_{n,t}^j}$ and $\mprn{a_{t}^jS_{n,t}^j}$, generate the same trade share $\mprn{\pi_{ni}^j}$. Therefore, we need to rescale the sequence of $S_{n,t}^j$ by identifying $\mprn{a_{t}^j}$ to measure the productivity growth over time.

To identify the shifters $\{a_{t}^{j}\}$, we take advantage of the gross output price index provided by the WIOD Socio Economic Accounts. Let $\{P_{US,t,Data}^{j} \}$ be the gross output price index of sector $j$ in the U.S. and year $t$. Note that $\{P_{US,t,Data}^{j} \}_{t}$ are comparable over time within sector $j$. Since the sequence $\mprn{a_{t}^j}_{t}$ is defined for each sector, we will use the gross price index for the three sectors in the U.S. and back out $\mprn{a_{t}^j}$ according to
\begin{equation}
\begin{split}
    P_{US,t}^{j} &=(a_{t}^{j})^{1/\theta^{j}}\left( \sum_{n=1}S_{n,t}^{-1}b_{US,n,t}^{-\theta^{j}} \right)^{-1/\theta^{j}} \\
    \Leftrightarrow a_{t}^{j} &= \prn{P_{US,t,Data}^{j}}^{\theta^j} \left( \sum_{n=1}S_{n,t}^{-1}b_{US,n,t}^{-\theta^{j}} \right)
\end{split}
\end{equation}
Now redefine $S_{n,t}^{j}$ by $a_{t}^{j}S_{n,t}^{j}$. Such redefined $\{S_{n,t}^{j} \}_{n,t}$ are comparable over time and across countries within sector $j$.

Being armed with the sectoral price indices after rescaling the sequence of $\mprn{S_{n,t}^j}$, we next back out the exogenous demand shifters for intermediate inputs, $\kappa_{n,t}^{jh}$, by solving the system of equations for each $j$, $n$, and $t$:
\begin{equation}
    g_{n,t}^{j,h}=\frac{\kappa_{n,t}^{j,h}(P_{n,t}^{h})^{1-\sigma^{j}}}{\sum_{h'=a,m,s}\kappa_{n,t}^{j,h'}(P_{n,t}^{h'})^{1-\sigma^{j}}}.
\end{equation}
by restricting $\sum_{h'}\kappa_{n,t}^{j,h'}=1$ for each $j$, $n$, and $t$. The left-hand side of the equation, $g_{n,t}^{j,h}$, is the share of expenditure spent on input from sector $h$ in total input costs of $j$, which is directly observed in the IO table. After obtaining $\kappa_{n,t}^{j,h}$, we can recover the CES price index for the composite intermediate good $\xi_{n,t}^j$ defined in the text.

We analogously back out the exogenous demand shifter in the capital goods production function, $\kappa_{n,t}^{Kh}$, by solving the system of equations for each $n$ and $t$:
\begin{equation}
    g_{n,t}^{K,h}=\frac{\kappa_{n,t}^{K,h}(P_{n,t}^{h})^{1-\sigma^{K}}}{\sum_{h'=a,m,s}\kappa_{n,t}^{K,h'}(P_{n,t}^{h'})^{1-\sigma^{K}}},
\end{equation}
\noindent by restricting $\sum_{h'}\kappa_{n,t}^{K,h'}=1$. This gives the price index of investment good $P_{n,t}^K$ defined in the text.

In order to obtain the factor prices, we construct the sequence of capital stock over time for each country. Starting from the initial capital stock in 1965 for each country provided by the PWT, we use the gross fixed capital formation from the WIOD and follow~\eqref{eq: capital dynamics} to construct the nationwide capital stock. Since capital stock is measured as the real variable in the model, we need to obtain the initial period capital stock in the current U.S. Dollars\footnote{We use the capital stock at current PPP multiplied by the price level of capital stock to obtain the initial capital stock.} and then divide the nominal value by the price index of the investment good obtained in the previous step.\footnote{The underlying assumption is that the capital stock in period $t$ is priced at $P_{n,t}^K$.} We then compute the real investment in each year by dividing the gross fixed capital formation (in current U.S. Dollars) by the investment good price index and accumulate the capital stock as implied by the model.

Using the value added from the WIOD, we apply the labor share from the PWT to obtain the wage bill and the return to capital. The wage bill and the total number of employment give the wage, $w_{n,t}$, and the return to capital and the capital stock give the rental price of capital, $r_{n,t}$.

Together with the composite intermediate input price index, $\xi_{n,t}^j$, and factor prices, $r_{n,t},w_{n,t}$, we can compute the cost of the input bundle according to~\eqref{eq: cost}. Finally, we can recover the productivity $A_{n,t}^j$ by\footnote{By construction, sectoral productivity takes 1 for the U.S. in 1965 in all sectors.}
\begin{equation*}
    \frac{A_{n,t}^j}{A_{US,1965}^j} =  (S_{n,t}^j)^{1/\theta^j}\prn{\frac{\widetilde{c}_{n,t}^{j}}{-\widetilde{c}_{US,1965}^{j}}}.
\end{equation*}

Using the sectoral price indices computed above, we calibrated the sectoral demand shifter $\Omega_{n,t}^j$ as follows. First, we guess the vector of $\mprn{ \Omega_{n,t}^j }$. Given the data on consumption expenditure $E_{n,t}$ from the WIOD, population $L_{n,t}$ from the PWT, sectoral prices $P_{n,t}^j$, and guessed values of $\Omega_{n,t}^j$, solve the consumption index $C_{n,t}^j$ according to~\eqref{eq: sector cons}. Using the computed consumption index, we can find the unique vector of $\Omega_{n,t}^j$ (up to normalization for each $n$ and $t$) by applying the Perron-Frobenius theorem to~\eqref{eq: expenditure share}. We then use the value of $\Omega_{n,t}^j$ as the new guess and repeat the steps until we find the fixed points. We do this process separately for both homothetic and nonhomothetic CES preferences. Therefore, We use different $\mprn{ \Omega_{n,t}^j }$ for the quantification of these two utility functions.

The intertemporal demand shifter $\zeta_{n,t}$ is backed out sequentially according to~\eqref{eq: Euler}. Using the consumption index $C_{n,t}$ obtained above, we can construct the series of $\zeta_{n,t}$ for each country by normalizing the one in the last sample year $\zeta_{n,2014}$ to be unity.

We also calibrate the sectoral demand shifter $\Omega_{n,t}^j$ and the intertemporal demand shifter $\zeta_{n,t}$ under the homothetic CES preference (i.e., $\epsilon^j=1$ for all $j$). See Supplemental Appendix~\ref{ap: baseline parameters} for such calibrated productivity as well as average tariff rates.

We take the five-year moving averages of all time-varying parameters except tariffs $\{\tau_{ni,t}^{j} \}$ and populations $\{L_{n,t}\}$ so that the model-generated sequences do not abruptly swing over time.

\section{Computation of Steady States}\label{ap: steady state}

This section outlines the algorithm to solve the steady states of the model. In what follows, we drop the time subscript from the variables. The following summarizes the solution algorithm.

\begin{enumerate}
    \item Guess wages across countries, $\{ w_n\}_n \in \mathbb{R}^N$, normalized such that $w_{US} = 1$.
  \begin{enumerate}
      \item Compute $r_n$ as follows.
    \begin{enumerate}
        \item Guess rental rates across countries, $\{ r_n \}_n \in \mathbb{R}^N$.
        \begin{enumerate}
            \item Compute $P_n^j$ as follows.
            \begin{itemize}
                \item Guess sectoral price indices across countries, $\{ P_n^j\}_{n,j} \in \mathbb{R}^{NJ}$.
                \begin{itemize}
                    \item Compute $\xi_n^j$ using (F1).
                    \item Compute $\widetilde{c}_n^j$ using (F2).
                    \item Compute $P_n^j$ using (F3).
                \end{itemize}
                \item Check if $P_n^j$ obtained in the last step is close to $P_n^j$ initially guessed. If it does, stop. Otherwise, update $\{ P_n^j \}_{n,j}$ and return to the first step.
            \end{itemize}
            \item Compute $P_n^K$ using (F4).
            \item Compute $r_n$ using (F5).
        \end{enumerate}
        \item Check if $r_n$ obtained in the last step is close to $r_n$ initially guessed. If it does, stop. Otherwise, update $\{ r_n \}_{n}$ and return to step i.
    \end{enumerate}
    \item Compute $\pi_{ni}^j$ using (F6).
    \item Compute $K_n$ using (F7).
    \item Compute $g_n^{j,j'}$ using (F8).
    \item Compute $g_n^{K,j}$ using (F9).
    \item Compute $X_n^j$ as follows.
        \begin{itemize}
                \item Guess sectoral spending across countries, $\{ X_n^j\}_{n,j} \in \mathbb{R}^{NJ}$.
                \begin{itemize}
                    \item Compute $\widetilde{T}_n$ using (M1).
                    \item Compute $T^P$ using (M2).
                    \item Compute national income $NI_n$ using (H1).
                    \item Compute $E_n$ using (H2).
                    \item Compute $C_n$ applying the Newton method to (H3).
                    \item Compute $\omega_n^j$ using (H4).
                    \item Compute $F_n^j$ using (H5).
                    \item Compute $Y_n^j$ using (M3).
                    \item Compute $X_n^j$ using (H6).
                \end{itemize}
                \item Check if $X_n^j$ obtained in the last step is close to $X_n^j$ initially guessed. If it does, stop. Otherwise, update $\{ X_n^j \}_{n,j}$ and return to the first step.
        \end{itemize}
        \item Compute $w_n$ using (M4).
  \end{enumerate}
  \item Check if $w_n$ obtained in the last step is close to $w_n$ initially guessed. If it does, stop and normalize $w_{US}$ to one. Otherwise, update $\{ w_n \}_{n}$ and return to step 1.
\end{enumerate}

\begin{table}[H]
\caption{Equilibrium conditions at steady state} \label{tb: algorithm steady state}
    \centering
\begin{tabular}{clc} \hline \hline
   (F1) & $\xi_{n}^j = \left[ \sum_{j'} \kappa_{n}^{j,j'} (P_{n}^{j'})^{1-\sigma^j} \right]^{\frac{1}{1-\sigma^j}}$ & $\forall (n,j)$ \\
   (F2) & $\widetilde{c}_{n,t}^j = (r_{n})^{\gamma_{n}^j \alpha_{n}} (w_{n})^{\gamma_{n}^j (1-\alpha_{n})} (\xi_{n}^j)^{1-\gamma_{n}^j}$ & $\forall (n,j)$ \\
   (F3) & $P_{n}^j = \left[ \sum_i^N \left( \dfrac{\widetilde{c}_{i}^j b_{n,i}^j}{A_{i}^j} \right)^{-\theta^j} \right]^{-\frac{1}{\theta^j}}$ & $\forall (n,j)$ \\
   (F4) & $P_{n}^K = \dfrac{1}{\kappa_{n}^K} \left[ \sum_j \kappa_{n}^{K,j} (P_{n}^j)^{1-\sigma^K} \right]^{\frac{1}{1-\sigma^K}}$ & $\forall (n)$ \\
   (F5) & $r_n = \dfrac{1-\alpha_n(1-\lambda \delta_n)}{\alpha_n(1-\phi_n)\lambda} P_n^K$ & $\forall (n)$ \\
   (F6) & $\pi_{ni}^j = \left( \dfrac{\widetilde{c}_{i}^j b_{ni}^j}{A_{i}^j P_{n}^j} \right)^{-\theta^j}$ & $\forall (n,i,j)$ \\
   (F7) & $K_{n} = \dfrac{\alpha_{n}}{1-\alpha_{n}} \dfrac{w_{n} L_{n}}{r_{n}}$ & $\forall (n)$ \\
   (F8) & $g_{n}^{j,j'} = \dfrac{\kappa_{n}^{j,j'}(P_{n}^{j'})^{1-\sigma^j} }{ \sum_{j''} \kappa_{n}^{j,j''}(P_{n}^{j''})^{1-\sigma^j} }$ & $\forall (n,j,j')$ \\
   (F9) & $g_{n}^{K,j} = \dfrac{\kappa_{n}^{K,j}(P_{n}^{K,j})^{1-\sigma^K} }{ \sum_{j'} \kappa_{n}^{K,j'}(P_{n}^{K,j'})^{1-\sigma^K} }$ & $\forall (n,j)$ \\
   (H1) & $NI_{n} = (1-\phi_n)(w_{n} L_{n} + r_{n} K_{n} + \widetilde{T}_{n}) + L_{n} T^P$ & $\forall (n)$ \\
   (H2) & $E_{n} = NI_{n} - P_{n}^K \delta_{n}K_{n}$ & $\forall (n)$ \\
   (H3) & $E_{n} = L_{n}\left[\sum_{j}\Omega_{n}^{j}\left\{ \left(\dfrac{C_{n}}{L_{n}}\right)^{\epsilon^{j}}P_{n}^{j}\right\}^{1-\sigma} \right]^{\frac{1}{1-\sigma}}$ & $\forall (n)$ \\
   (H4) & $\omega_{n}^j = \dfrac{\Omega_{n}^{j}\left\{ \left(\frac{C_{n}}{L_{n}}\right)^{\epsilon^{j}}P_{n}^{j}\right\}^{1-\sigma}}{ \sum_{j'} \Omega_{n}^{j'}\left\{ \left(\frac{C_{n}}{L_{n}}\right)^{\epsilon^{j'}}P_{n}^{j'}\right\}^{1-\sigma} } \ \left( = \dfrac{P_{n}^j C_{n}^j}{E_{n}} \right)$ & $\forall (n,j)$ \\
   (H5) & $F_{n}^j = \omega_{n}^j E_{n} + g_{n}^{K,j} P_{n}^K \delta_n K_{n}$ & $\forall (n,j)$ \\
   (H6) & $X_{n}^{j}=F_{n}^j+
        \sum_{j'}(1-\gamma_{n}^{j'})g_{n}^{j',j}Y_{n}^{j'}$ & $\forall (n,j)$ \\
   (M1) & $\widetilde{T}_{n} = \sum_j \sum_i^N \tau_{ni}^j X_{n,t}^j \dfrac{\pi_{ni}^j}{\widetilde{\tau}_{ni}^j}$ & $\forall (n)$ \\
   (M2) & $T^{P}=\sum_{i}^{N}\phi_{i}(w_{i}L_{i}+r_{i}K_{i}+\wtilde{T}_{i})\big/ \sum_{i}^{N}L_{i}$ & \\
   (M3) & $Y_{n}^j = \sum_{i}^N X_{i}^j\dfrac{\pi_{in}^j}{\widetilde{\tau}_{in}^j}$ & $\forall (n,j)$ \\
   (M4) & $w_{n} = (1-\alpha_n)\sum_j \gamma_{n}^j Y_{n}^j/L_n$ & $\forall (n,j)$ \\ \hline \hline
\end{tabular} \\
{\footnotesize
\begin{flushleft}
    \textit{Note}: $b_{ni}^j = d_{ni}^j \widetilde{\tau}_{ni}^j$ and $\widetilde{\tau}_{ni}^j = 1+\tau_{ni}^j$. Roughly, (F*) is a condition for firms/production; (H*) for household; (M*) for market clearing.
\end{flushleft} }
\end{table}

\section{Computation of Transition Paths} \label{ap: transition paths}

We compute transition paths in the following way. Let $T+1$ be the terminal period, $N$ the number of countries, and $J$ the number of sectors.

\begin{enumerate}
\item Give pre-determined values (data in our case) to the initial capital stock, $\{ K_{n,1}\}_n$, and arbitrary values to the other variables at the initial period $1$.
\item Give the steady-state values to variables at the terminal period $T+1$ including $\{ K_{n,T+1}, K_{n,T+2}\}_n$, $\{ C_{n,T+1} \}_n$, $\{ \overline{\epsilon}_{n,T+1} \}_n$, $\{ E_{n,T+1} \}_n$, $\{ r_{n,T+1} \}_n$, and $\{ P_{n,T+1}^K \}_n$. Note that $K_{n,t}$ is a pre-determined variable at period $t$. Therefore, $K_{n,T+2}$ is determined at period $T+1$ given the steady-state value of $K_{n,T+1}$ at the same period.
\item Guess nominal investment rates across countries and time, $\{ \rho_{n,t}\}_{n,t} \in \mathbb{R}^{N(T+1)}$.
\item Compute variables forward in time from $t=1$ including $\{ w_{n,t}\}_{n,t} \in \mathbb{R}^{N(T+1)}$, $\{ I_{n,t}\}_{n,t} \in \mathbb{R}^{N(T+1)}$, $\{ K_{n,t}\}_{n,t} \in \mathbb{R}^{N(T+2)}$, $\{ \overline{\epsilon}_{n,t}\}_{n,t} \in \mathbb{R}^{N(T+1)}$, and others in the sub-steps below.
\begin{enumerate}
\item Compute $w_{n,t}$ in each period $t$ as follows, noting that in period $t$ capital stock across countries, $\{ K_{n,t}\}_{n} \in \mathbb{R}^N$ are predetermined.
 \begin{enumerate}
    \item Guess wages across countries in period $t$, $\{ w_{n,t}\}_{n} \in \mathbb{R}^N$, normalized such that $w_{US, t} = 1$.
  \begin{enumerate}
      \item Compute $r_{n,t}$ using (F5).
      \item Compute $P_{n,t}^j$ as follows.
        \begin{itemize}
            \item Guess sectoral price indices across countries in period $t$, $\{ P_{n,t}^j\}_{n,j} \in \mathbb{R}^{NJ}$.
              \begin{itemize}
                \item Compute $\xi_{n,t}^j$ using (F1).
                \item Compute $\widetilde{c}_{n,t}^j$ using (F2).
                \item Compute $P_{n,t}^j$ using (F3).
              \end{itemize}
            \item Check if $P_{n,t}^j$ obtained in the last step is close to $P_{n,t}^j$ initially guessed. If it does, stop. Otherwise, update $\{ P_{n,t}^j \}_{n,j}$ and return to the first step.
        \end{itemize}
    \item Compute $P_{n,t}^K$ using (F4).
    \item Compute $\pi_{ni,t}^j$ using (F6).
        \item Compute $g_{n,t}^{j,j'}$ using (F7).
    \item Compute $g_{n,t}^{K,j}$ using (F8).
    \item Compute $X_{n,t}^j$ as follows.
        \begin{itemize}
                \item Guess sectoral spending across countries in period $t$, $\{ X_{n,t}^j\}_{n,j} \in \mathbb{R}^{NJ}$.
                \begin{itemize}
                    \item Compute $\widetilde{T}_{n,t}$ using (M1).
                    \item Compute $T^P$ using (M2).
                    \item Compute national income $NI_{n,t}$ using (H1).
                    \item Compute $E_{n,t}$ using (H2).
                    \item Compute $C_{n,t}$ applying the Newton method to (H3).
                    \item Compute $\omega_{n,t}^j$ using (H4).
                    \item Compute $Y_{n,t}^j$ using (M3).
                                        \item Compute $X_{n,t}^j$ using (H5).
                \end{itemize}
                \item Check if $X_{n,t}^j$ obtained in the last step is close to $X_{n,t}^j$ initially guessed. If it does, stop. Otherwise, update $\{ X_{n,t}^j \}_{n,j}$ and return to the first step.
        \end{itemize}
        \item Compute $w_{n,t}$ using (M4).
  \end{enumerate}
  \item Check if $w_{n,t}$ obtained in the last step is close to $w_{n,t}$ initially guessed. If so, stop and normalize $w_{US,t}$ to one. Otherwise, update $\{ w_{n,t} \}_{n}$ and return to step i.
 \end{enumerate}
 \item Compute $I_{n,t} = \rho_{n,t} NI_{n,t}/P_{n,t}^K$.
 \item Compute $K_{n,t+1}=(1-\delta_{n,t})K_{n,t}+I_{n,t}^{\lambda}(\delta_{n,t}K_{n,t})^{1-\lambda}$.
 \item Compute $\overline{\epsilon}_{n,t} = \sum_j \omega_{n,t}^j \epsilon^j$.
\end{enumerate}
\item Update $\rho_{n,t}$ backward in time from period $t=T$ as $\rho_{n,t}(1+ \eta Z_{n,t})$ using (H6) ``Euler equation residual''$Z_{n,t}$ with a dampening parameter $\eta$ and associated functions (H7) and (H8). Note that we restrict the updated $\rho_{n,t}$ to be in $(0,1)$.
\item Check if $\rho_{n,t}$ obtained in step 5 is close to $\rho_{n,t}$ initially guessed. If it does, stop. Otherwise, update $\{ \rho_{n,t} \}_{n,t}$ and return to step 3.
\end{enumerate}

\begin{table}[H]
\caption{Equilibrium conditions} \label{tb: algorithm transition path}
    \centering
\begin{tabular}{clc} \hline \hline
   (F1) & $\xi_{n,t}^j = \left[ \sum_{j'} \kappa_{n,t}^{j,j'} (P_{n,t}^{j'})^{1-\sigma^j} \right]^{\frac{1}{1-\sigma^j}}$ & $\forall (n,j,t)$ \\
   (F2) & $\widetilde{c}_{n,t}^j = (r_{n,t})^{\gamma_{n,t}^j \alpha_{n,t}} (w_{n,t})^{\gamma_{n,t}^j (1-\alpha_{n,t})} (\xi_{n,t}^j)^{1-\gamma_{n,t}^j}$ & $\forall (n,j,t)$ \\
   (F3) & $P_{n,t}^j = \left[ \sum_i^N \left( \frac{\widetilde{c}_{i,t}^j b_{ni,t}^j}{A_{i,t}^j} \right)^{-\theta^j} \right]^{-\frac{1}{\theta^j}}$ & $\forall (n,j,t)$ \\
   (F4) & $P_{n,t}^K = \dfrac{1}{\kappa_{n,t}^K} \left[ \sum_j \kappa_{n,t}^{K,j} (P_{n,t}^j)^{1-\sigma^K} \right]^{\frac{1}{1-\sigma^K}}$ & $\forall (n,t)$ \\
   (F5) & $r_{n,t} = \dfrac{\alpha_{n,t}}{1-\alpha_{n,t}} \dfrac{w_{n,t} L_{n,t}}{K_{n,t}}$ & $\forall (n, t)$ \\
   (F6) & $\pi_{ni,t}^j = \left( \frac{\widetilde{c}_{i,t}^j b_{ni,t}^j}{A_{i,t}^j P_{n,t}^j} \right)^{-\theta^j}$ & $\forall (n,i,j,t)$ \\
   (F7) & $g_{n,t}^{j,j'} = \frac{\kappa_{n,t}^{j,j'}(P_{n,t}^{j'})^{1-\sigma^j} }{ \sum_{j''} \kappa_{n}^{j,j''}(P_{n,t}^{j''})^{1-\sigma^j} }$ & $\forall (n,j,j',t)$ \\
   (F8) & $g_{n,t}^{K,j} = \frac{\kappa_{n,t}^{K,j}(P_{n,t}^{K,j})^{1-\sigma^K} }{ \sum_{j'} \kappa_{n,t}^{K,j'}(P_{n,t}^{K,j'})^{1-\sigma^K} }$ & $\forall (n,j,t)$ \\
   (H1) & $NI_{n,t} = (1-\phi_{n,t})(w_{n,t} L_{n,t} + r_{n,t} K_{n,t} + \widetilde{T}_{n,t}) + L_{n,t}T_t^P$ & $\forall (n,t)$ \\
   (H2) & $E_{n,t} = (1-\rho_{n,t})NI_{n,t}$ & $\forall (n,t)$ \\
   (H3) & $E_{n,t} = L_{n,t}\left[\sum_{j}\Omega_{n,t}^{j}\left\{ \left(\frac{C_{n,t}}{L_{n,t}}\right)^{\epsilon^{j}}P_{n,t}^{j}\right\}^{1-\sigma} \right]^{\frac{1}{1-\sigma}}$ & $\forall (n,t)$ \\
   (H4) & $\omega_{n,t}^j = \frac{\Omega_{n,t}^{j}\left\{ \left(\frac{C_{n,t}}{L_{n,t}}\right)^{\epsilon^{j}}P_{n,t}^{j}\right\}^{1-\sigma}}{ \sum_{j'} \Omega_{n,t}^{j'}\left\{ \left(\frac{C_{n,t}}{L_{n,t}}\right)^{\epsilon^{j'}}P_{n,t}^{j'}\right\}^{1-\sigma} } \ \left( = \dfrac{P_{n,t}^j C_{n,t}^j}{E_{n,t}} \right)$ & $\forall (n,j,t)$ \\
   (H5) & $X_{n,t}^{j}=[\omega_{n,t}^j (1-\rho_{n,t}) + g_{n,t}^{K,j} \rho_{n,t}]NI_{n,t} + \sum_{j'}(1-\gamma_{n,t}^{j'})g_{n,t}^{j',j}\sum_i^N \pi_{in,t}^j X_{i,t}^j/\widetilde{\tau}_{in,t}^j$ & $\forall (n,j,t)$ \\
   (H6) & $Z_{n,t} = \left[ \beta \frac{\zeta_{n,t+1}}{\zeta_{n,t}} \frac{L_{n,t+1}}{L_{n,t}}  \frac{E_{n,t}}{E_{n,t+1}} \frac{\overline{\epsilon}_{n,t}}{\overline{\epsilon}_{n,t+1}} \frac{\left[ (1-\phi_{n,t+1})r_{n,t+1} - P_{n,t+1}^K \Phi_2(K_{n,t+2}, K_{n,t+1}) \right]}{P_{n,t}^K \Phi_1(K_{n,t+1}, K_{n,t})} \right]^{\frac{1}{\psi-1}}$ & \\    &\qquad \qquad \qquad \qquad $ - \frac{C_{n,t+1}}{C_{n,t}} \frac{L_{n,t}}{L_{n,t+1}}$ &$\forall (n,t)$ \\
   (H7) & $\Phi_1(K_{n,t+2}, K_{n,t+1}) = \frac{\delta_{n,t+1}^{1-\frac{1}{\lambda}}}{\lambda} \left( \frac{K_{n,t+2}}{K_{n,t+1}} - (1-\delta_{n,t+1})\right)^{\frac{1-\lambda}{\lambda}}$ & $\forall (n,t)$ \\
   (H8) & $\Phi_2(K_{n,t+2}, K_{n,t+1}) = \Phi_1(K_{n,t+2}, K_{n,t+1}) \left[ (\lambda-1)\frac{K_{n,t+2}}{K_{n,t+1}} - \lambda(1-\delta_{n,t+1})\right]$ & $\forall (n,t)$ \\
   (M1) & $\widetilde{T}_{n,t} = \sum_j \sum_i^N \tau_{ni,t}^j X_{n,t}^j \dfrac{\pi_{ni,t}^j}{\widetilde{\tau}_{ni,t}^j}$ & $\forall (n,t)$ \\
   (M2) & $T_t^{P}=\sum_{i}^{N}\phi_{i,t}(w_{i,t}L_{i,t}+r_{i,t}K_{i,t}+\widetilde{T}_{i,t})\big/ \sum_{i}^{N}L_{i,t}$ & $\forall (t)$ \\
   (M3) & $Y_{n,t}^j = \sum_{i}^N X_{i,t}^j\frac{\pi_{in,t}^j}{\widetilde{\tau}_{in,t}^j}$ & $\forall (n,j,t)$ \\
   (M4) & $w_{n,t} = (1-\alpha_{n,t})\sum_j \gamma_{n,t}^j Y_{n,t}^j/L_{n,t}$ & $\forall (n,j,t)$ \\ \hline \hline
\end{tabular} \\
{\footnotesize
\begin{flushleft}
    \textit{Note}: $b_{ni,t}^j = d_{ni,t}^j \widetilde{\tau}_{ni,t}^j$ and $\widetilde{\tau}_{ni,t}^j = 1+\tau_{ni,t}^j$. Roughly, (F*) is a condition for firms/production; (H*) for household; (M*) for market clearing. An alternative expression of (H5) is $X_{n,t}^{j}=\omega_{n,t}^j E_{n,t} + g_{n,t}^{K,j} P_{n,t}^K I_{n,t} +
        \sum_{j'}(1-\gamma_{n,t}^{j'})g_{n,t}^{j',j}\sum_i^N \pi_{in,t}^j X_{i,t}^j/\widetilde{\tau}_{in,t}^j$.
\end{flushleft} }
\end{table}

\section{Applying the Newton Method} \label{ap: newton}
In the system of steady state and equilibrium conditions, most economic variables are \textit{explicitly} expressed. Therefore they are directly computable given parameters and other economic variables. The only variable that is \textit{implicitly} expressed is aggregate consumption $C_{n,t}$ (in steady state, $C_{n}$). See equation (H3) in Table~\ref{tb: algorithm transition path} (in steady state, Table~\ref{tb: algorithm steady state}). In the following, we describe the computational method for transition paths, but a similar argument applies for steady states. Given parameters $\sigma$, $\epsilon^{j}$, $\Omega_{n,t}^{j}$, $L_{n,t}$, and economic variables $P_{n,t}^{j}$, we need to solve this equation for $C_{n,t}$. We apply the Newton--Raphson method to numerically solve this equation.

Observe that equation
\begin{equation}
    \left(\frac{E_{n,t}}{L_{n,t}} \right)^{1-\sigma}=\sum_{j=a,m,s}\Omega_{n,t}^{j}\left\{\left(\frac{C_{n,t}}{L_{n,t}} \right)^{\epsilon^{j}}P_{n,t}^{j} \right\}^{1-\sigma}.
\end{equation}
is equivalent to (H3).
Based on this equation, define real-valued function $\Delta$ by
\begin{equation}
    \Delta(x_{n,t})=\sum_{j=a,m,s}\Omega_{n,t}^{j}\left\{\left(\frac{x_{n,t}}{L_{n,t}} \right)^{\epsilon^{j}}P_{n,t}^{j} \right\}^{1-\sigma} - \left(\frac{E_{n,t}}{L_{n,t}} \right)^{1-\sigma}.
\end{equation}
$C_{n,t}$ is such that $\Delta(C_{n,t})=0$. The derivative of $\Delta$ is
\begin{equation}
    \Delta'(x_{n,t})=(1-\sigma)\sum_{j=a,m,s}\Omega_{n,t}^{j}(L_{n,t})^{-\epsilon^{j}(1-\sigma)}(P_{n}^{j})^{1-\sigma}\epsilon^{j}x_{n}^{\epsilon^{j}(1-\sigma)-1}.
\end{equation}
Using these expressions, we compute $C_{n,t}$ in the following iterative way.

Make an initial guess $x_{n,t}^{0}>0$. The superscript of $x$ keeps track of the number of updates in iteration. Then compute the updated value as
\begin{equation}
    x_{n,t}^{1}= x_{n,t}^{0} - \frac{\Delta(x_{n,t}^{0})}{\Delta'(x_{n,t}^{0})}.
\end{equation}
If $x_{n,t}^{1}$ is close enough to $x_{n,t}^{0}$, we got the solution. Otherwise, use $x_{n,t}^{1}$ as a new guess, and compute $x_{n,t}^{2}$ and compare these two. Repeat this process until $x_{n,t}^{k}$ and $x_{n,t}^{k+1}$ for some $k$.

\section{Baseline Parameter Values} \label{ap: baseline parameters}
\begin{figure}[H]    \centering
    \caption{Productivity evolution (1990=1)} \label{fig: prod}    \subfloat[\centering Canada]{{\includegraphics[width=0.34\textwidth]{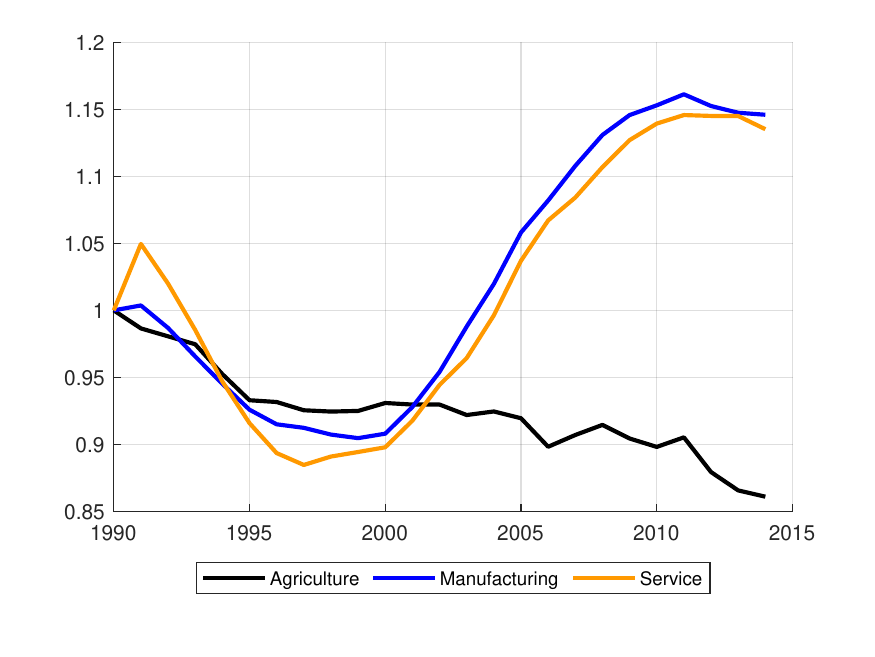} }}    \subfloat[\centering China]{{\includegraphics[width=0.34\textwidth]{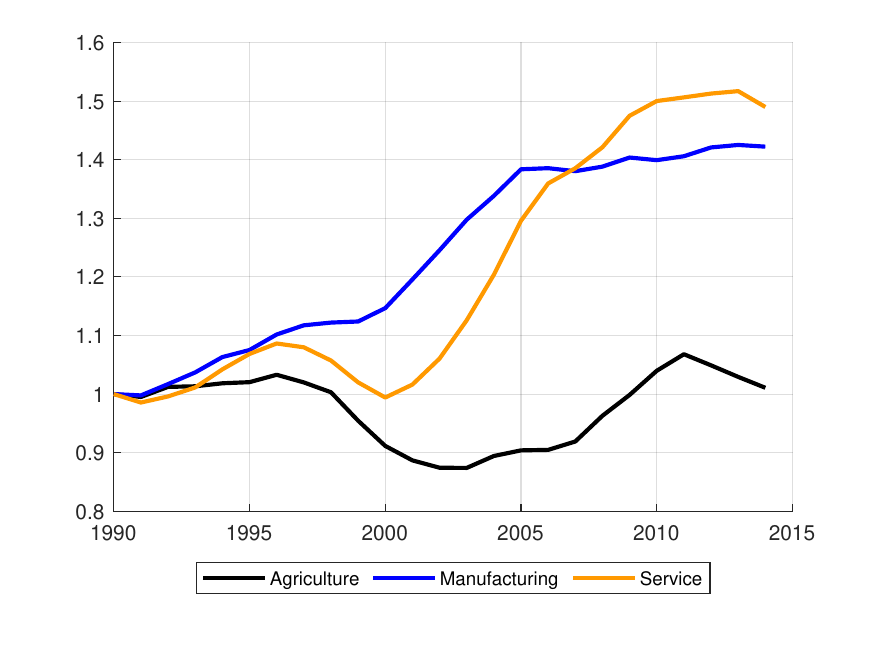} }}    \subfloat[\centering Germany]{{\includegraphics[width=0.34\textwidth]{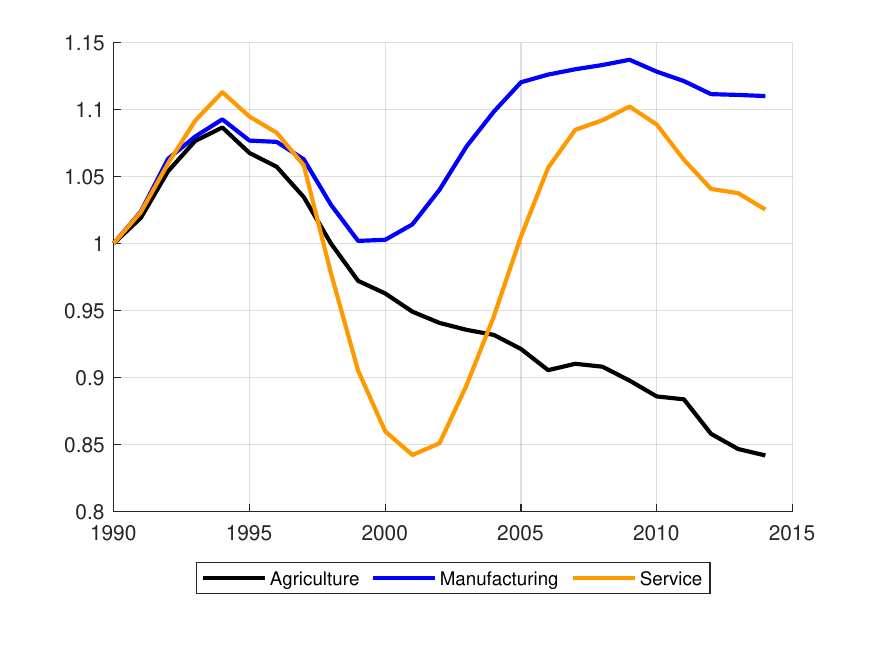} }}    \quad
    \subfloat[\centering Japan]{{\includegraphics[width=0.34\textwidth]{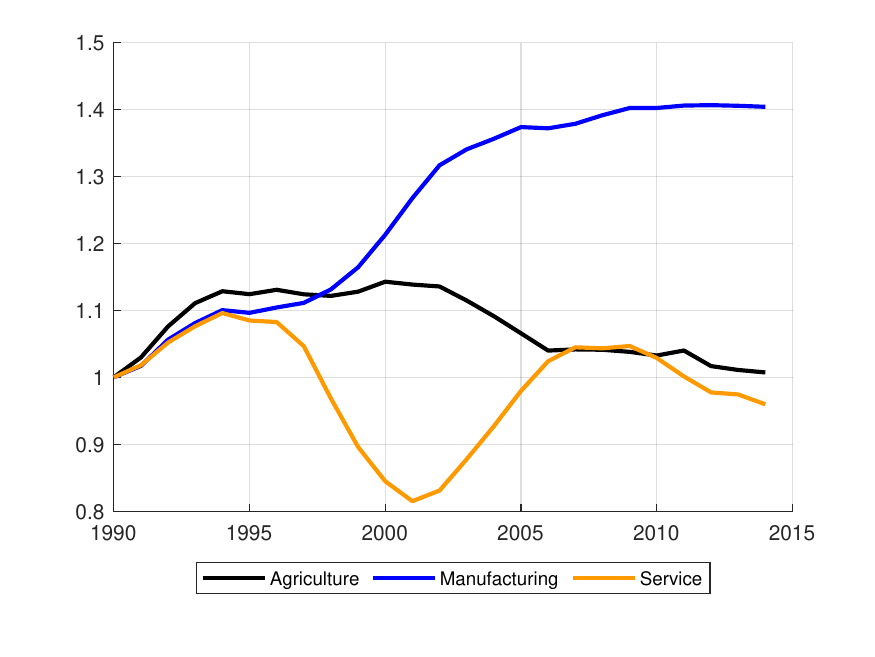} }}    \subfloat[\centering Mexico]{{\includegraphics[width=0.34\textwidth]{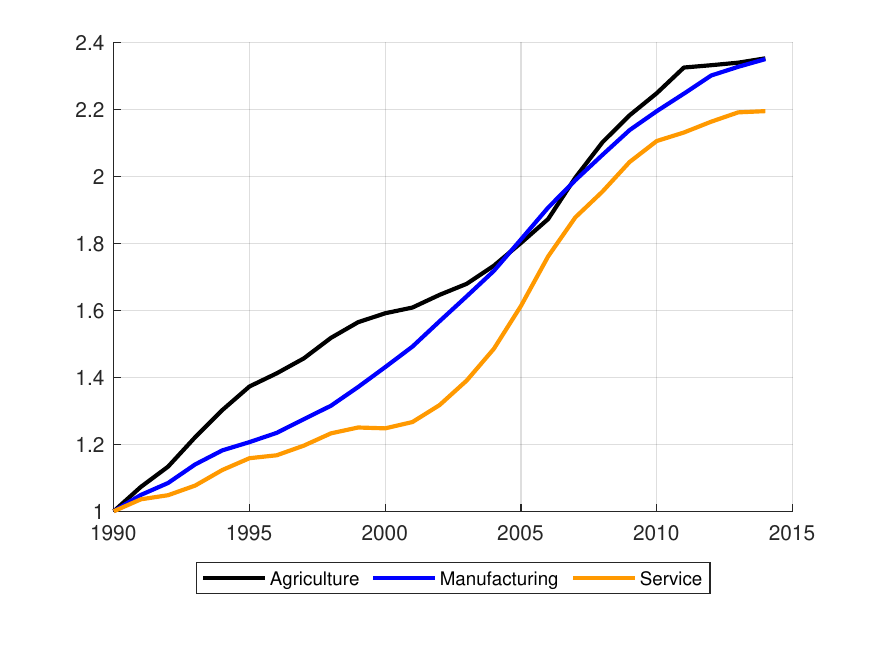} }}    \subfloat[\centering United States]{{\includegraphics[width=0.34\textwidth]{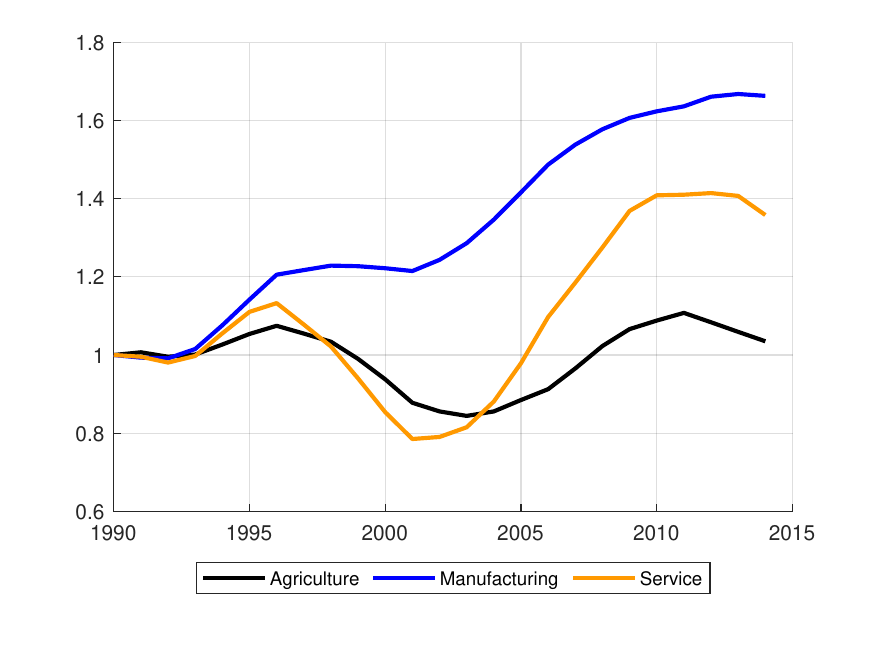} }}    {\footnotesize
    \begin{flushleft}
        \textit{Note:} This table shows the calibrated values of sectoral productivity, $A_{n,t}^j$. We take a five-year moving average.
    \end{flushleft}
    }
\end{figure}

We summarize the baseline fundamentals we calibrated in Section~\ref{ap: shock}. Figure~\ref{fig: prod} shows the evolution of sectoral productivity in six countries, Canada, China, Germany, Japan, Mexico, and the U.S. We normalize the productivity in 2000 to be one and take the moving average over five years to remove the noise. In every country other than Canada and Japan, the productivity of the manufacturing sector grows more than that of the service sector. In the U.S., the manufacturing productivity increased by a factor of 1.6 while the service sector productivity increased by a factor of 1.4. The productivity growth biased toward manufacturing implies that the expenditure share on manufacturing may drop due to the relative price effect, even if we do not take into account the impacts of international trade and non-homotheticity preferences-driven demand changes.
In Canada, productivity growth is similar across manufacturing and services. For Japan, the manufacturing productivity grew faster than the service sector until the mid-2000s, but the pattern reversed afterward.

\begin{figure}[H]    \centering
    \caption{Evolution of average tariff}\label{fig: tradecost_tau}    \subfloat[\centering Canada]{{\includegraphics[width=0.34\textwidth]{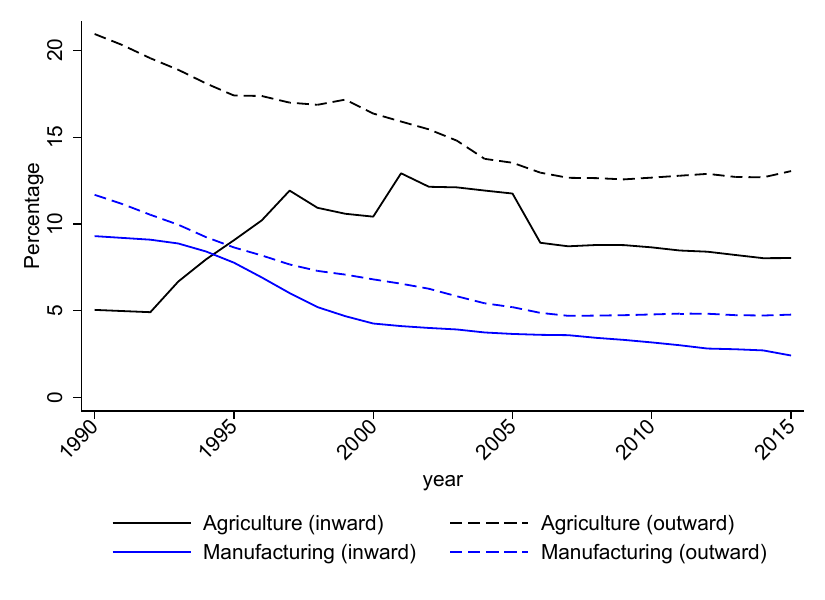} }}    \subfloat[\centering China]{{\includegraphics[width=0.34\textwidth]{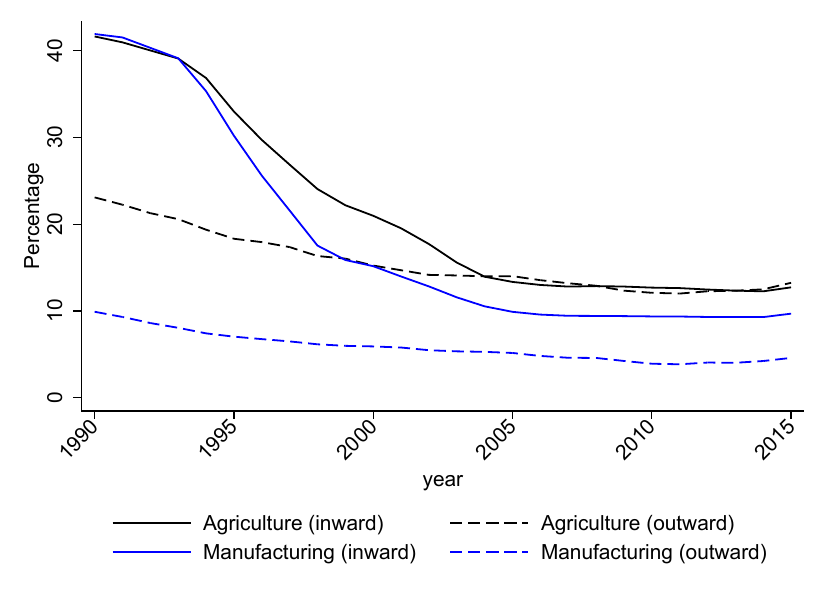} }}    \subfloat[\centering Germany]{{\includegraphics[width=0.34\textwidth]{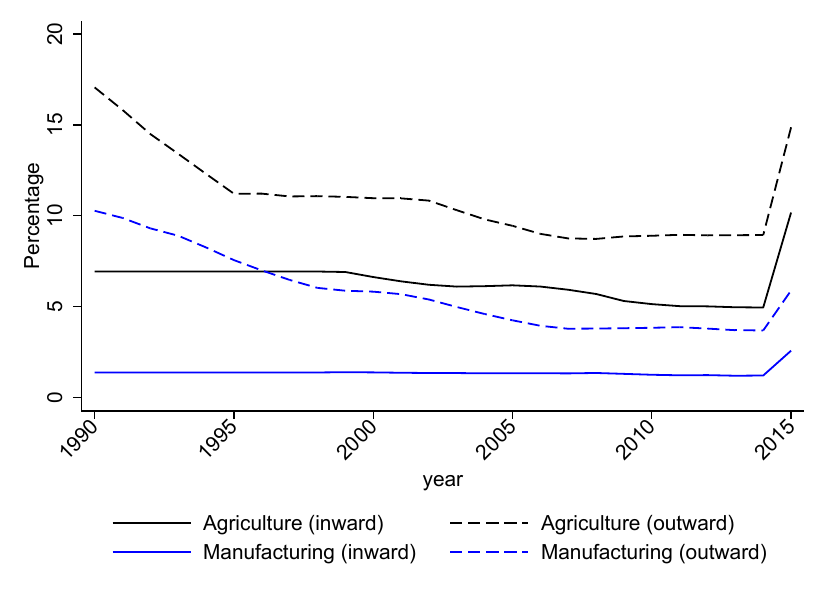} }}    \quad
    \subfloat[\centering Japan]{{\includegraphics[width=0.34\textwidth]{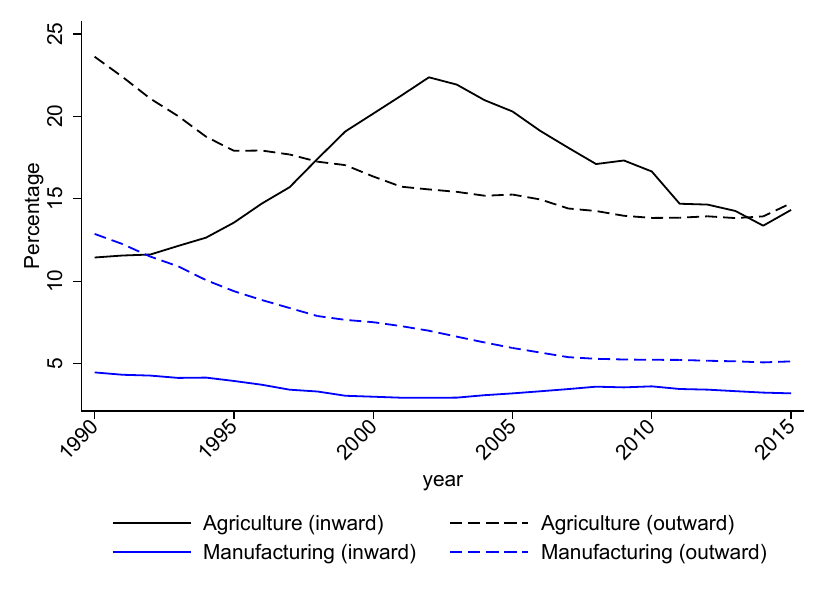} }}    \subfloat[\centering Mexico]{{\includegraphics[width=0.34\textwidth]{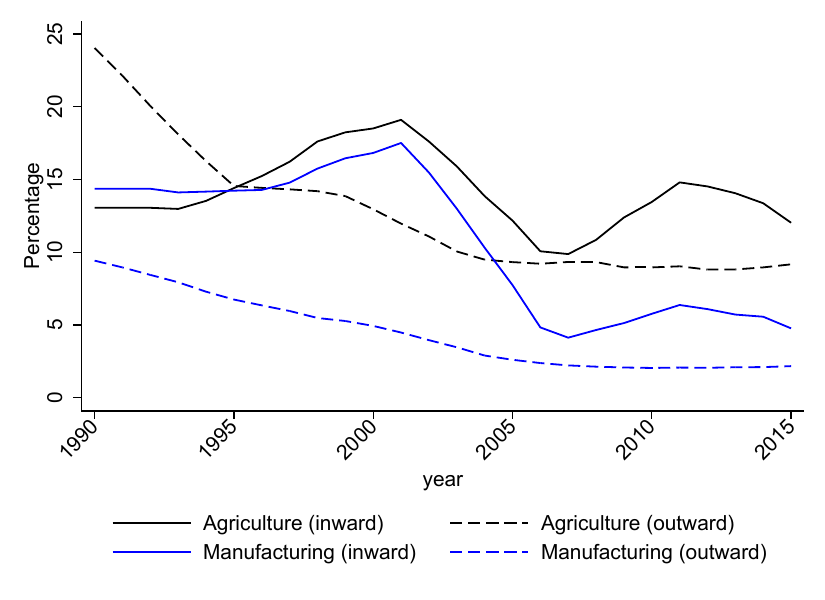} }}    \subfloat[\centering United States]{{\includegraphics[width=0.34\textwidth]{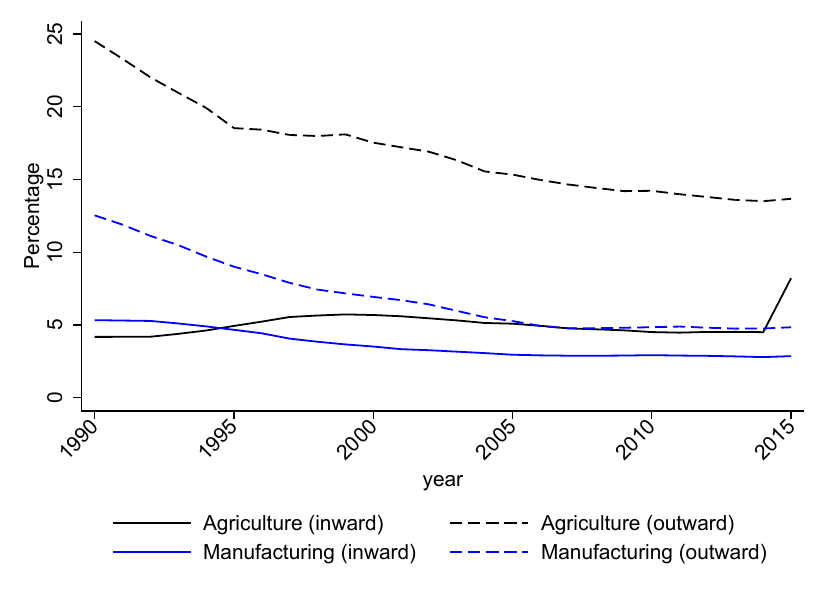} }}    {\footnotesize
    \begin{flushleft}
        \textit{Note:} This table shows the average sectoral tariffs a country imposes on the other countries (\verb|inward|, solid lines) and those imposed on the country by the others (\verb|outward|, dashed lines).
    \end{flushleft}
    }
\end{figure}

Figure~\ref{fig: tradecost_tau} shows the evolution of trade costs in the six countries. For each country and year, we compute the simple arithmetic average of the bilateral tariff rate (inward and outward) with all its trading partners. The source of the tariff data is the WITS, which provides the bilateral tariff after the late 1980s for the major countries. For the period prior to the year when the tariffs are reported for the first time, we apply the tariff rates in the first year. The figure confirms that the tariffs are continuously falling after the 1990s for the manufacturing sector in most of the countries. It is also worth noting that China's export tariffs drop more significantly in the late 1990s than in the 2000s when China joined the WTO.

\begin{figure}[H]    \centering
    \caption{Evolution of average non-tariff barrier}\label{fig: tradecost_d}    \subfloat[\centering Canada]{{\includegraphics[width=0.34\textwidth]{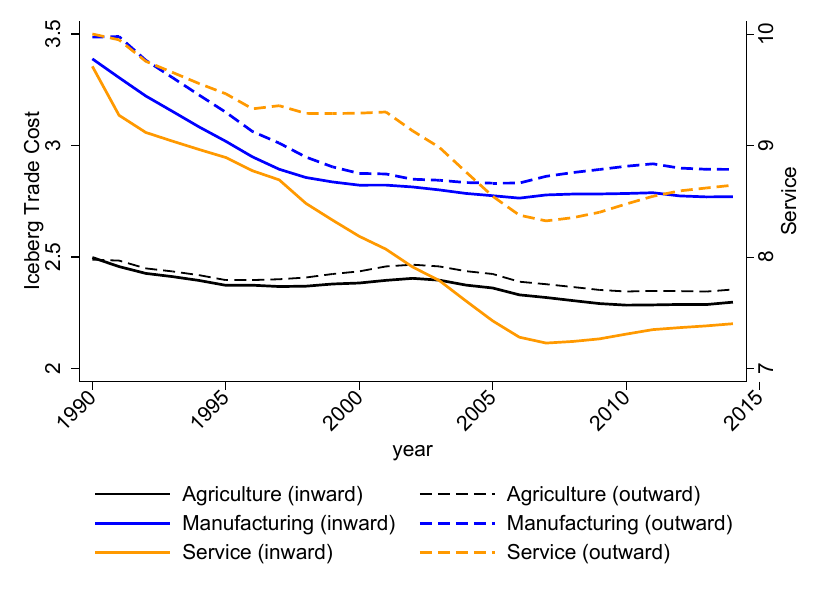} }}    \subfloat[\centering China]{{\includegraphics[width=0.34\textwidth]{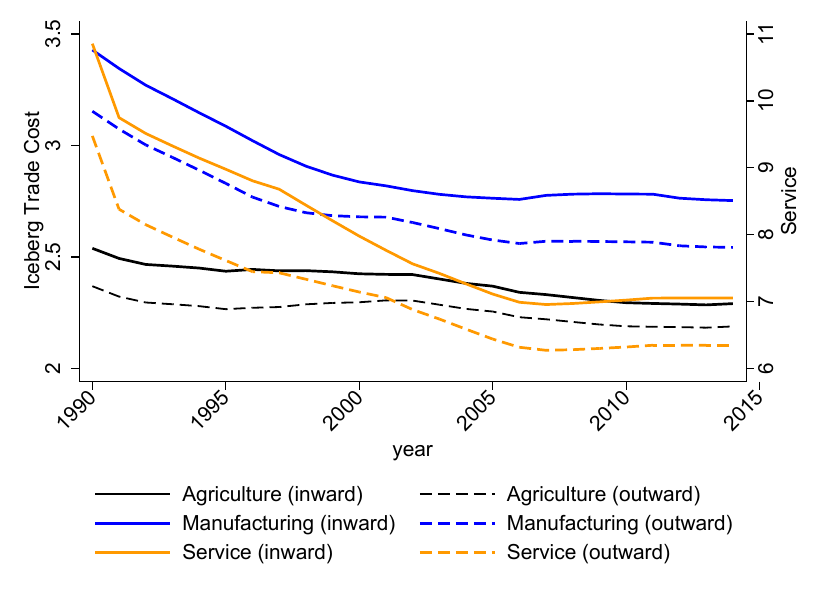} }}    \subfloat[\centering Germany]{{\includegraphics[width=0.34\textwidth]{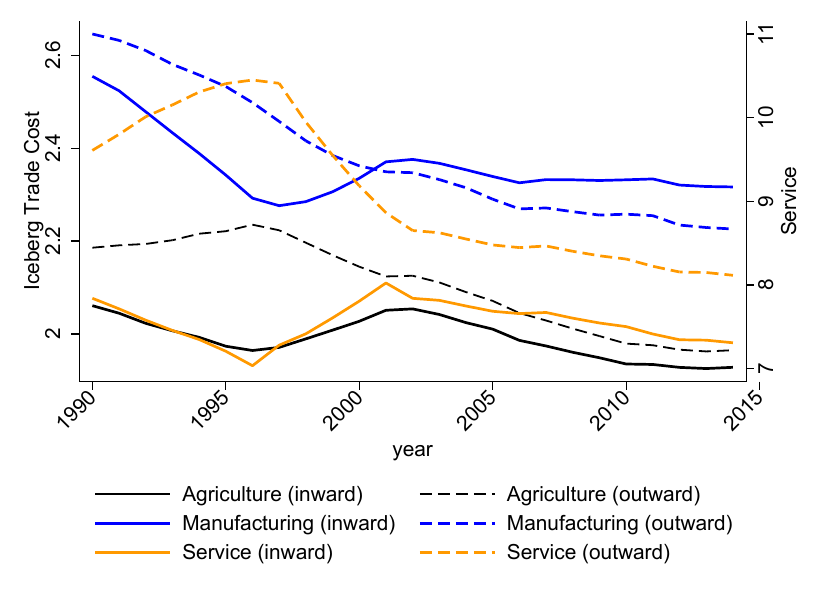} }}    \quad
    \subfloat[\centering Japan]{{\includegraphics[width=0.34\textwidth]{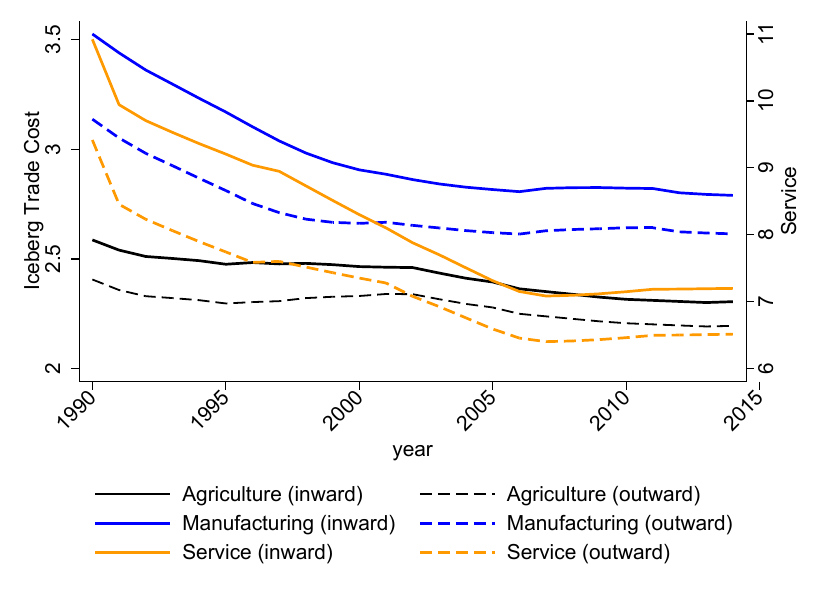} }}    \subfloat[\centering Mexico]{{\includegraphics[width=0.34\textwidth]{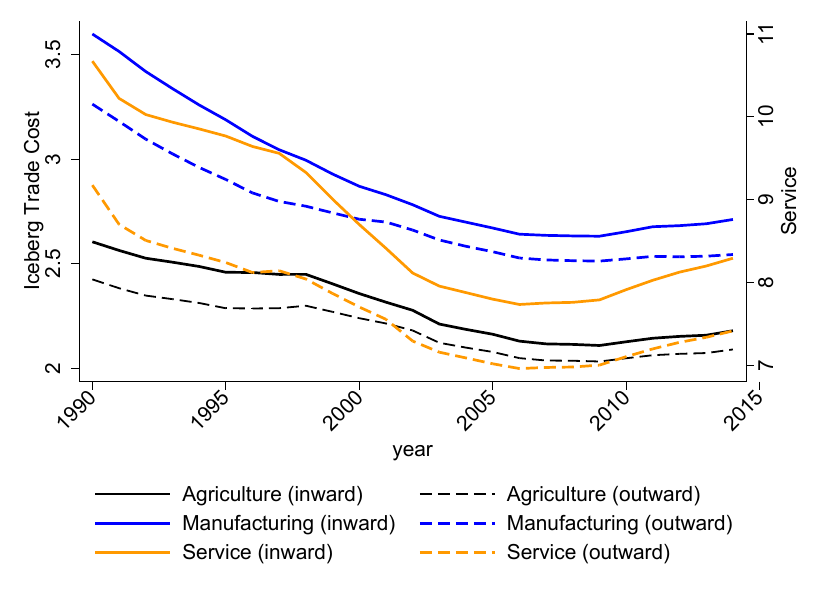} }}    \subfloat[\centering United States]{{\includegraphics[width=0.34\textwidth]{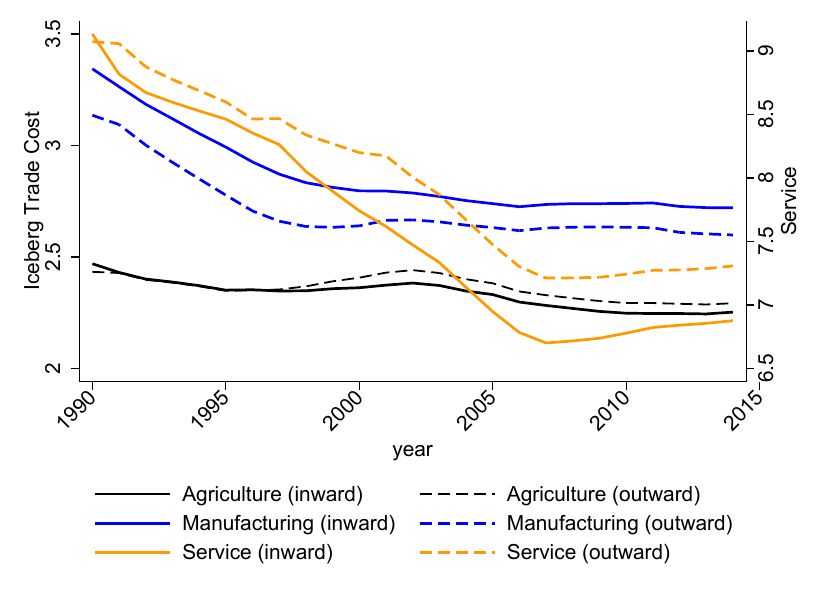} }}    {\footnotesize
    \begin{flushleft}
        \textit{Note:} This table shows the calibrated values of the average sectoral non-tariff barriers a country imposes on the other countries (\verb|inward|, solid lines) and those imposed on the country by the others (\verb|outward|, dashed lines).
    \end{flushleft}
    }
\end{figure}

Figure~\ref{fig: tradecost_d} exhibits the evolution of non-tariff barriers in the six countries, measured as the simple average of inward and outward iceberg trade costs. First, we see the fall in non-tariff barriers is more significant in magnitude compared to the tariff barriers. For instance, in the U.S., the non-tariff barriers in the U.S. dropped from 400 percent to 300 percent over the five decades while the tariff barriers dropped from five percent to three percent. Second, the non-tariff barriers for the service sector are much higher in level than the good sectors, but exhibits a significant drop over time. While the service trade is often overlooked in the quantitative trade analysis, the result suggests that the falling service trade cost is a crucial factor in understanding the sectoral reallocation in the global context.

\clearpage
\section{Model Fit and Baseline Results}\label{ap: fit}

\begin{figure}[H]\centering
    \caption{Model fit: Sectoral consumption expenditure share}\label{fig: bl_expshare}
    \subfloat[\centering Canada ]{{\includegraphics[width=0.34\textwidth]{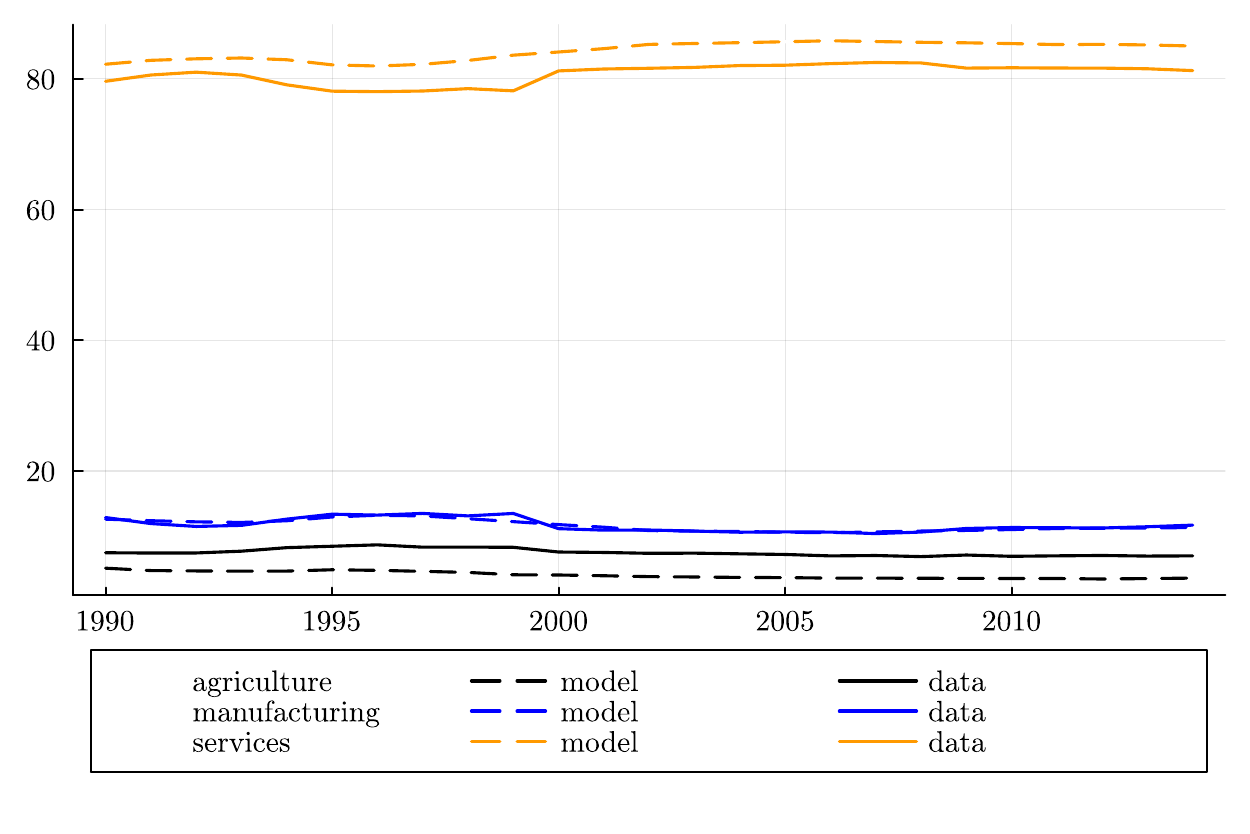} }}    \subfloat[\centering China ]{{\includegraphics[width=0.34\textwidth]{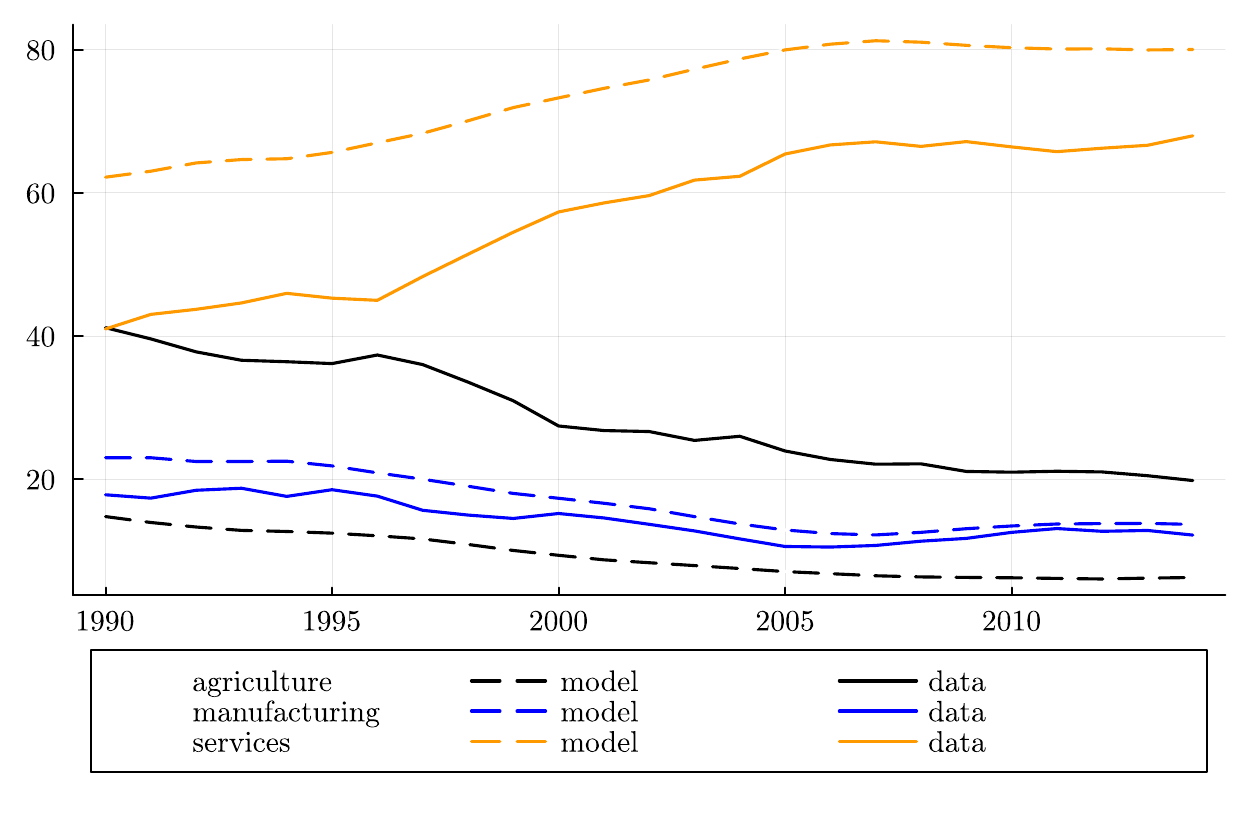} }}    \subfloat[\centering Germany ]{{\includegraphics[width=0.34\textwidth]{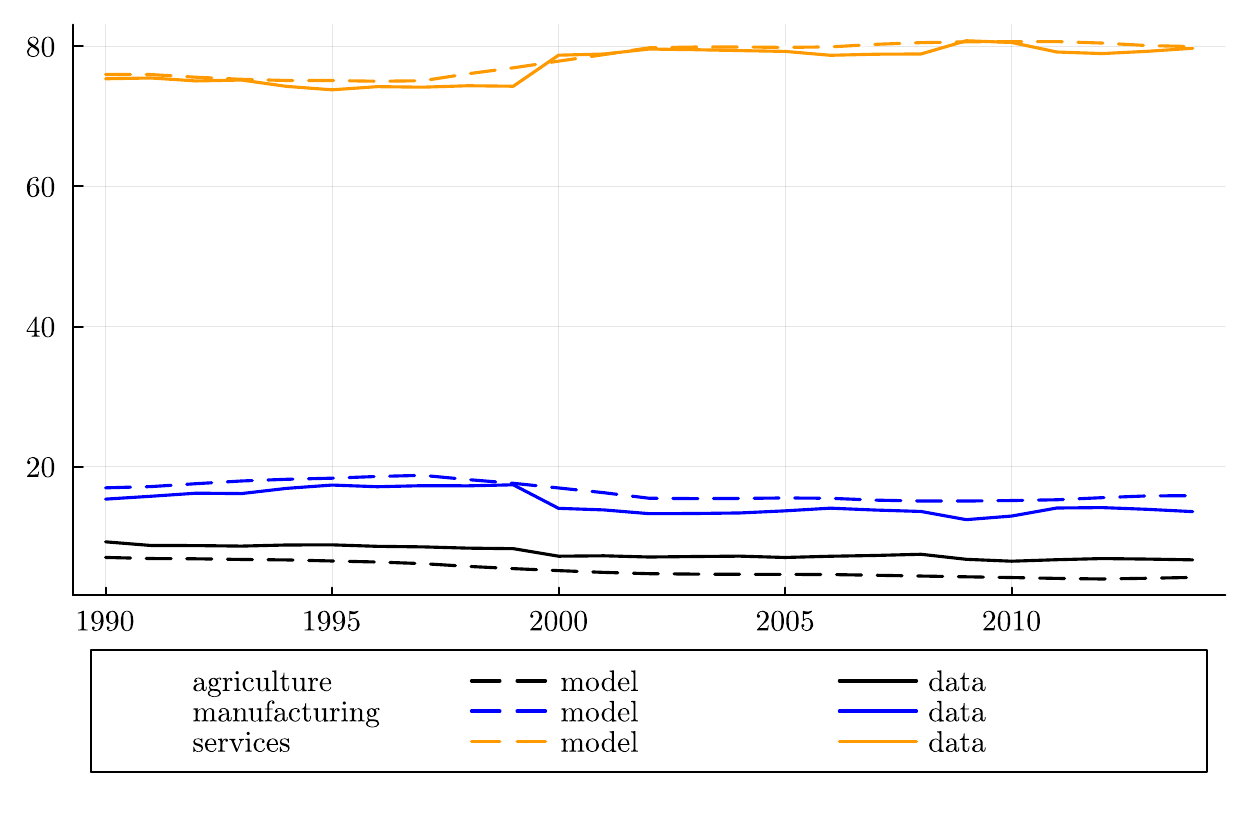} }}
    \quad
    \subfloat[\centering Japan ]{{\includegraphics[width=0.34\textwidth]{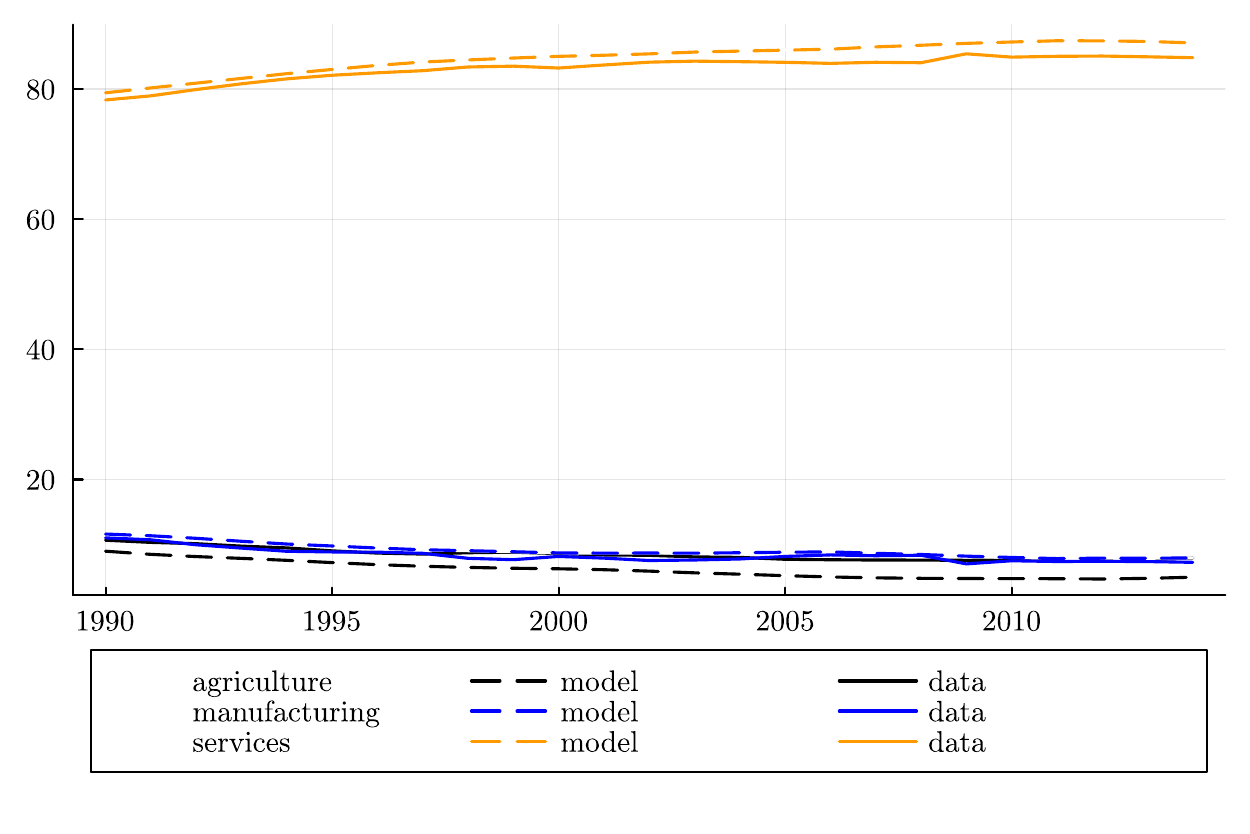} }}    \subfloat[\centering Mexico  ]{{\includegraphics[width=0.34\textwidth]{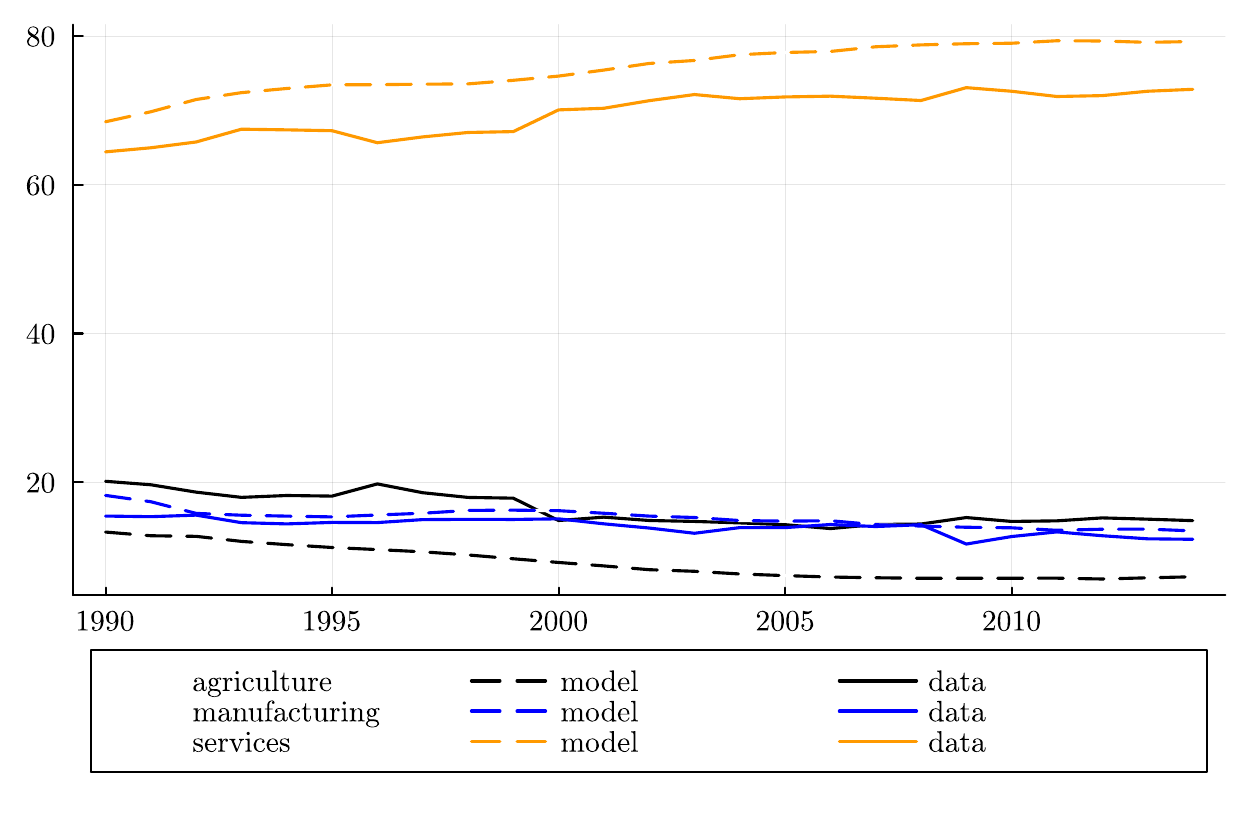} }}
    \subfloat[\centering U.S. ]{{\includegraphics[width=0.34\textwidth]{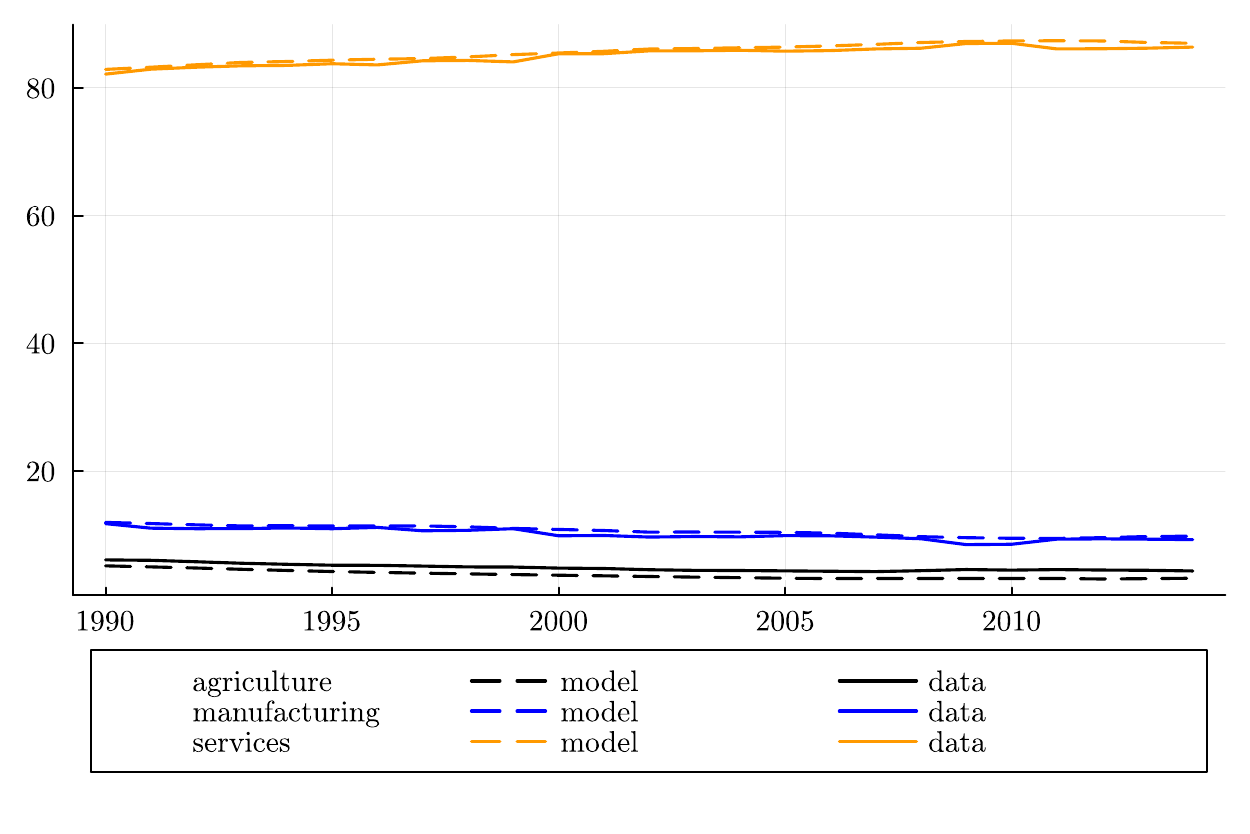} }}
    {\footnotesize
    \begin{flushleft}
    \textit{Note:} Each panel shows the sectoral consumption expenditure shares in the baseline equilibrium (dashed lines) and in the data (solid lines).
    \end{flushleft}
    }
\end{figure}

Figure~\ref{fig: bl_expshare} demonstrates the model fit of sectoral expenditure shares in final consumption, $\omega_{n,t}^j = P_{n,t}^j C_{n,t}^j/E_{n,t}$. The model-implied expenditure shares are shown in dashed lines, while the data counterparts are shown in solid lines, and the colors are the same as in Figure~\ref{fig: bl_vashare}. The model captures the shift of final expenditure from agriculture to manufacturing, and then to services over time. In all countries, the model overpredicts the service expenditure and underestimates the agricultural expenditure. As in the case of the value-added shares, the fit of the model is better for advanced economies than for emerging economies.

\begin{figure}[H]    \centering
    \caption{Model fit: Saving rate}\label{fig: bl_saving}    \subfloat[\centering Canada ]{{\includegraphics[width=0.34\textwidth]{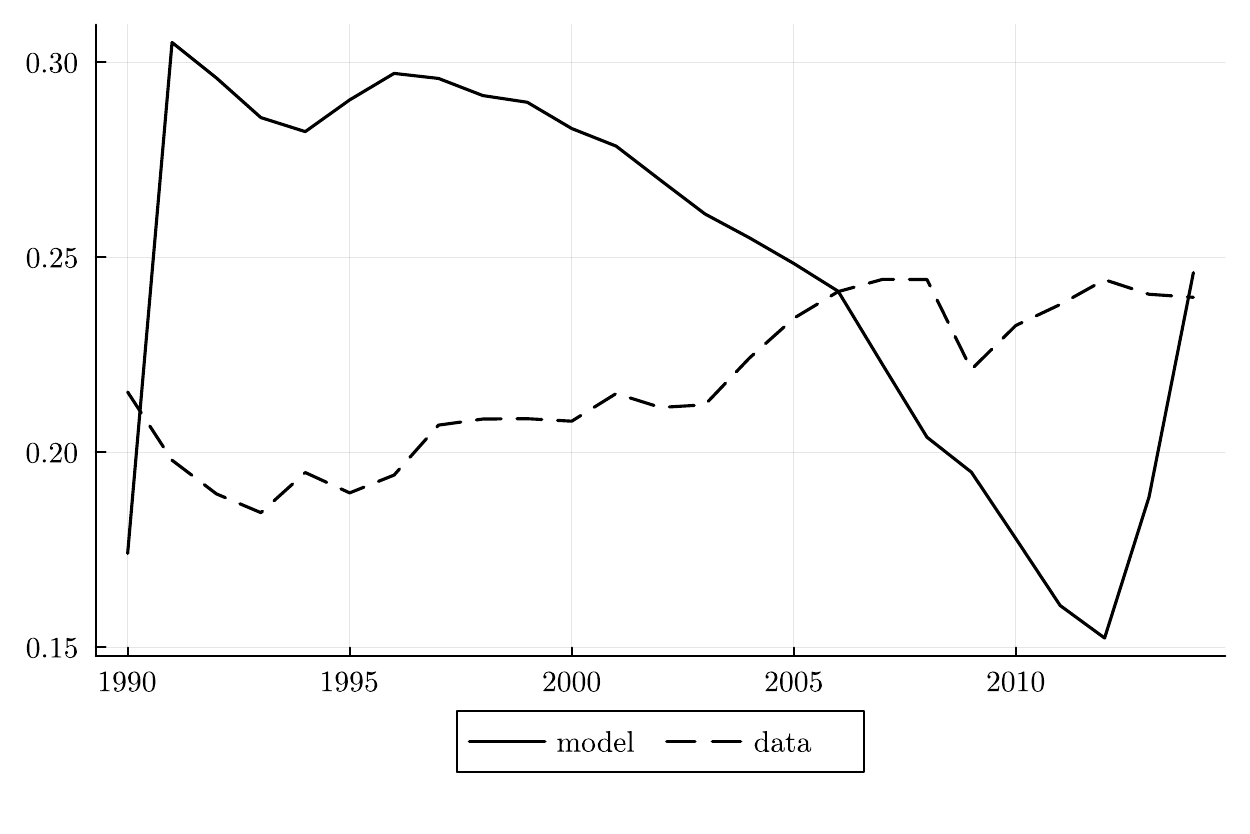} }}    \subfloat[\centering China ]{{\includegraphics[width=0.34\textwidth]{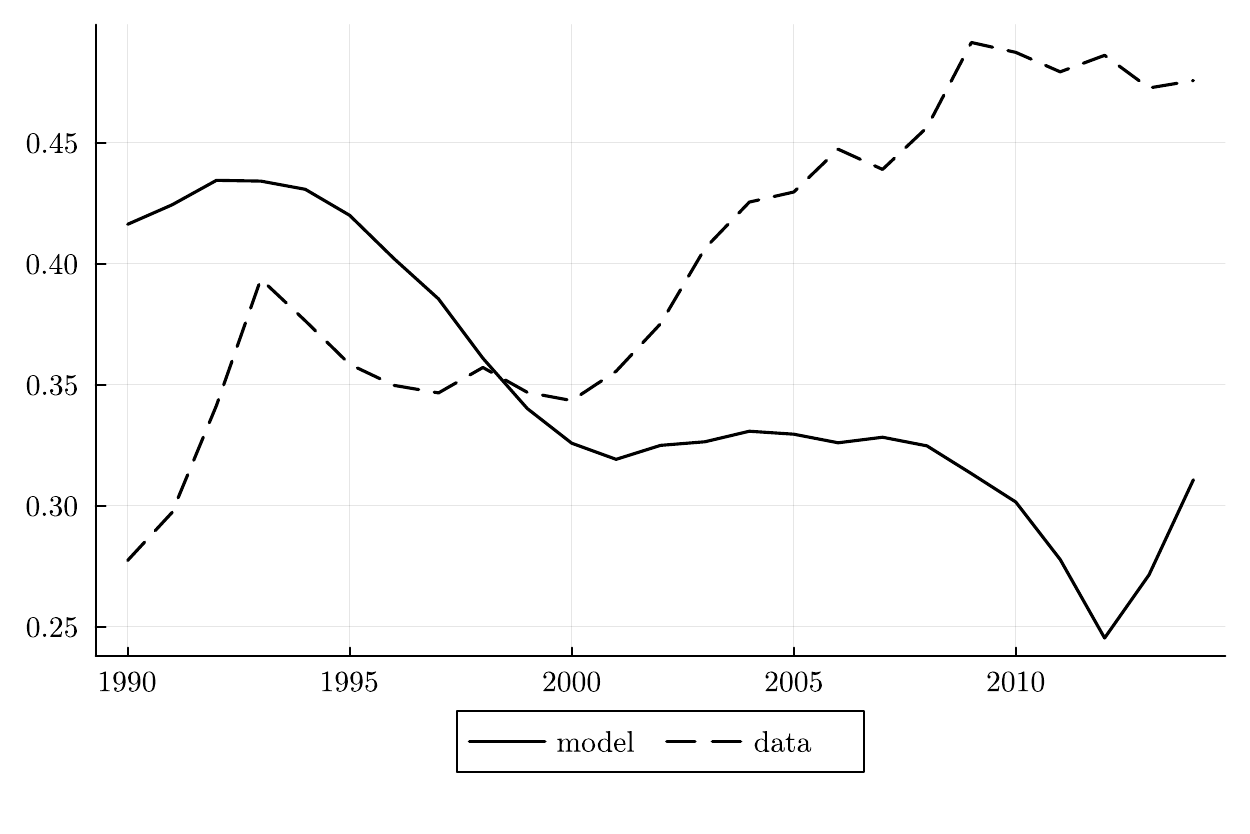} }}    \subfloat[\centering Germany ]{{\includegraphics[width=0.34\textwidth]{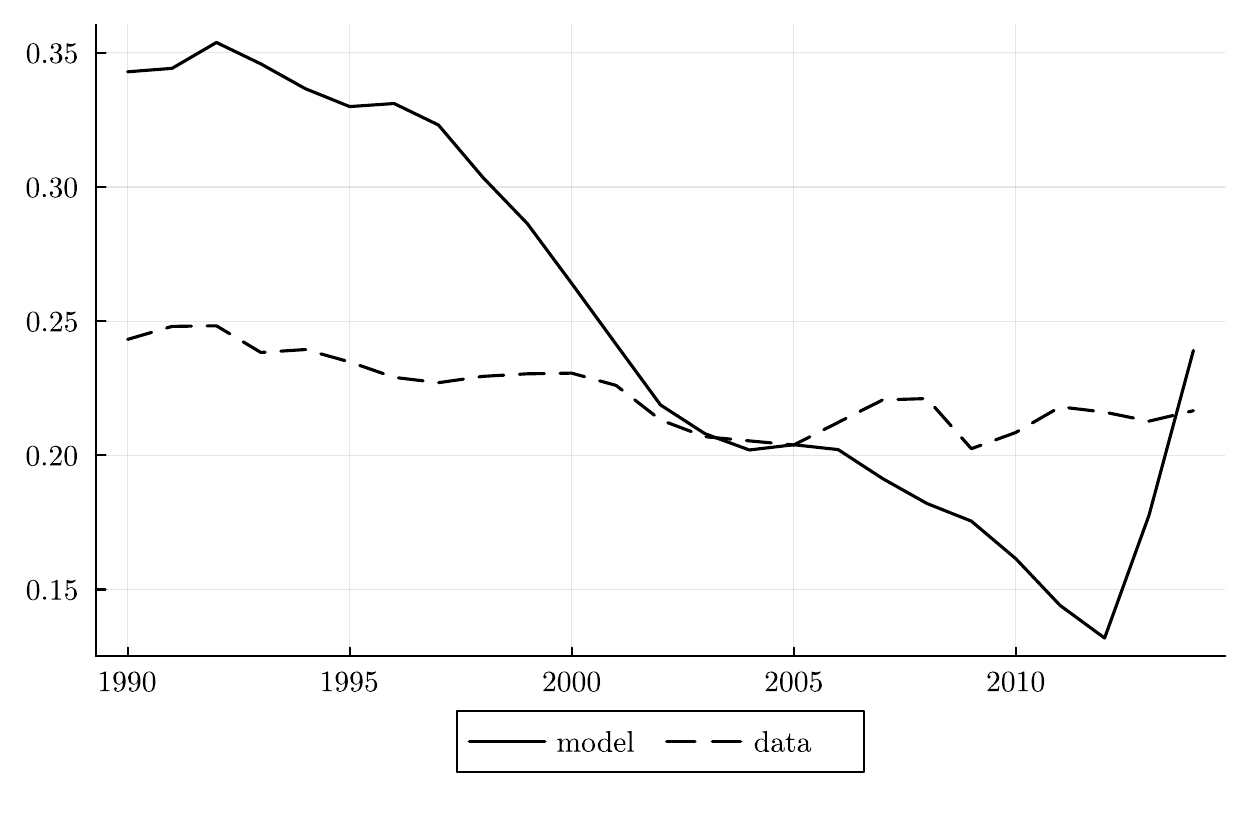} }}
    \quad
    \subfloat[\centering Japan ]{{\includegraphics[width=0.34\textwidth]{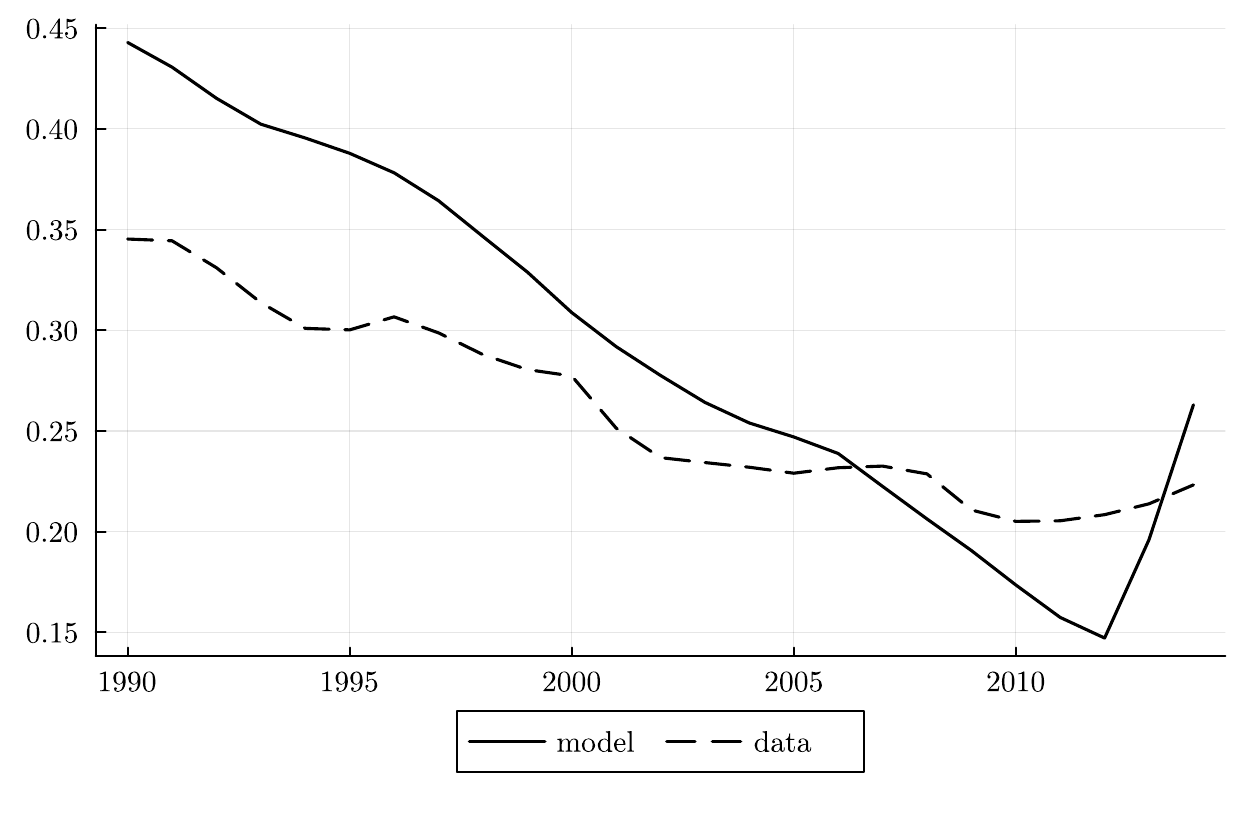} }}    \subfloat[\centering Mexico  ]{{\includegraphics[width=0.34\textwidth]{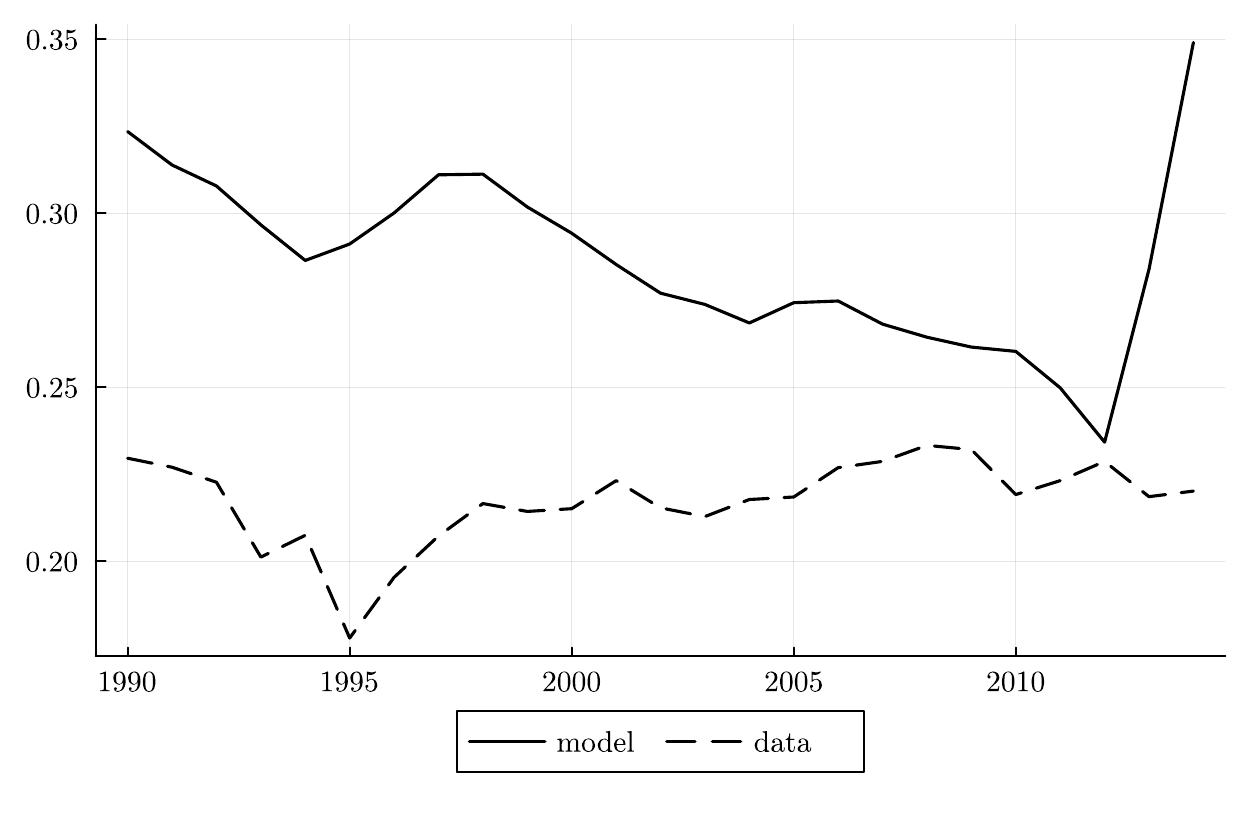} }}
    \subfloat[\centering U.S. ]{{\includegraphics[width=0.34\textwidth]{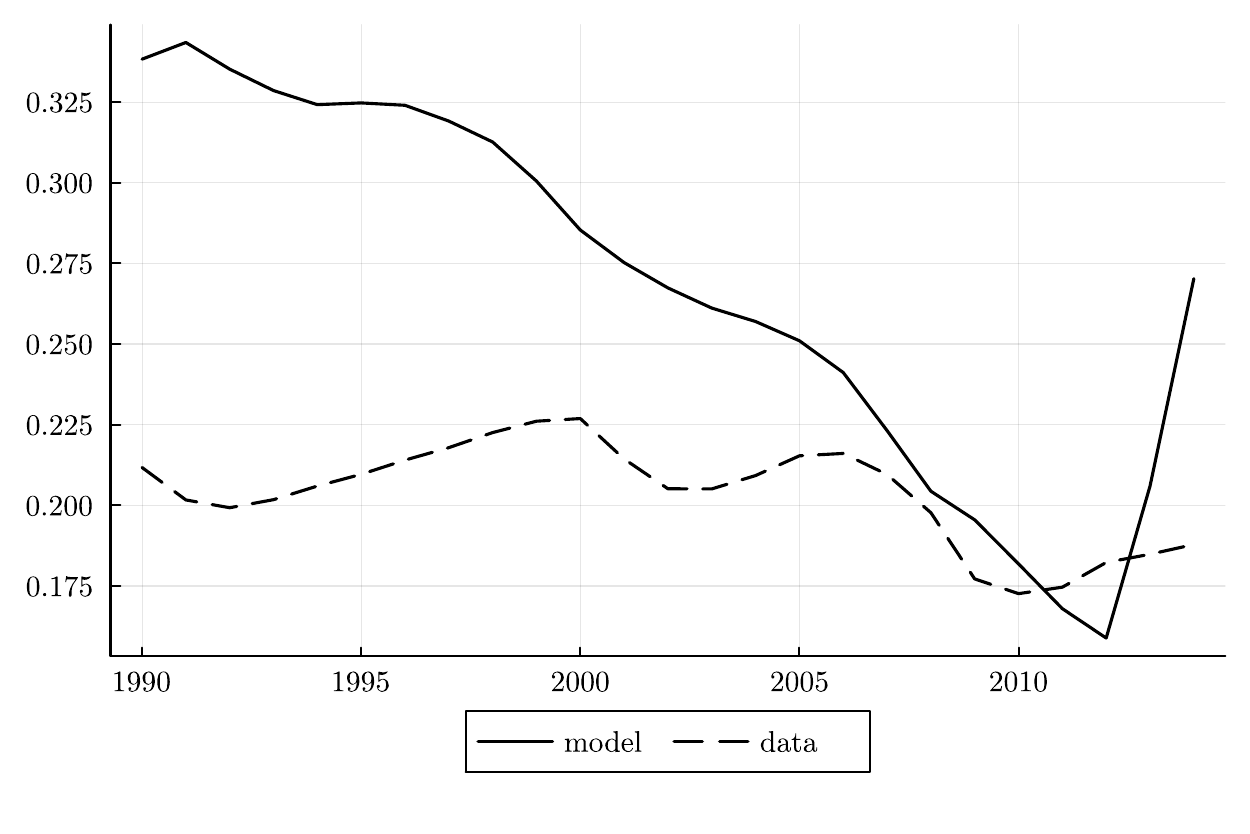} }}
    {\footnotesize
    \begin{flushleft}
        \textit{Note:} This table shows the saving rate, $\rho_{n,t}$ in \eqref{eq: rho}, in the baseline equilibrium (solid lines) and in the data (dashed lines).
    \end{flushleft}
    }
\end{figure}

Figure~\ref{fig: bl_saving} compares the saving rates in the baseline equilibrium to the data. In all six countries, the model predicts a higher saving rate than the data counterpart in earlier years. The model-implied saving rate gradually falls and converges to levels close to the data.

\clearpage

\section{More on Counterfactual Results} \label{ap: counterfactual additional}

\subsection{Trade War}\label{ap: trade war}

\begin{figure}[htb]    \centering
    \caption{Impacts of a trade war between the U.S. and the rest of the world}\label{fig: newtradepolicy_7_nh}    \subfloat[\centering Real per capita consumption]{{\includegraphics[width=0.45\textwidth]{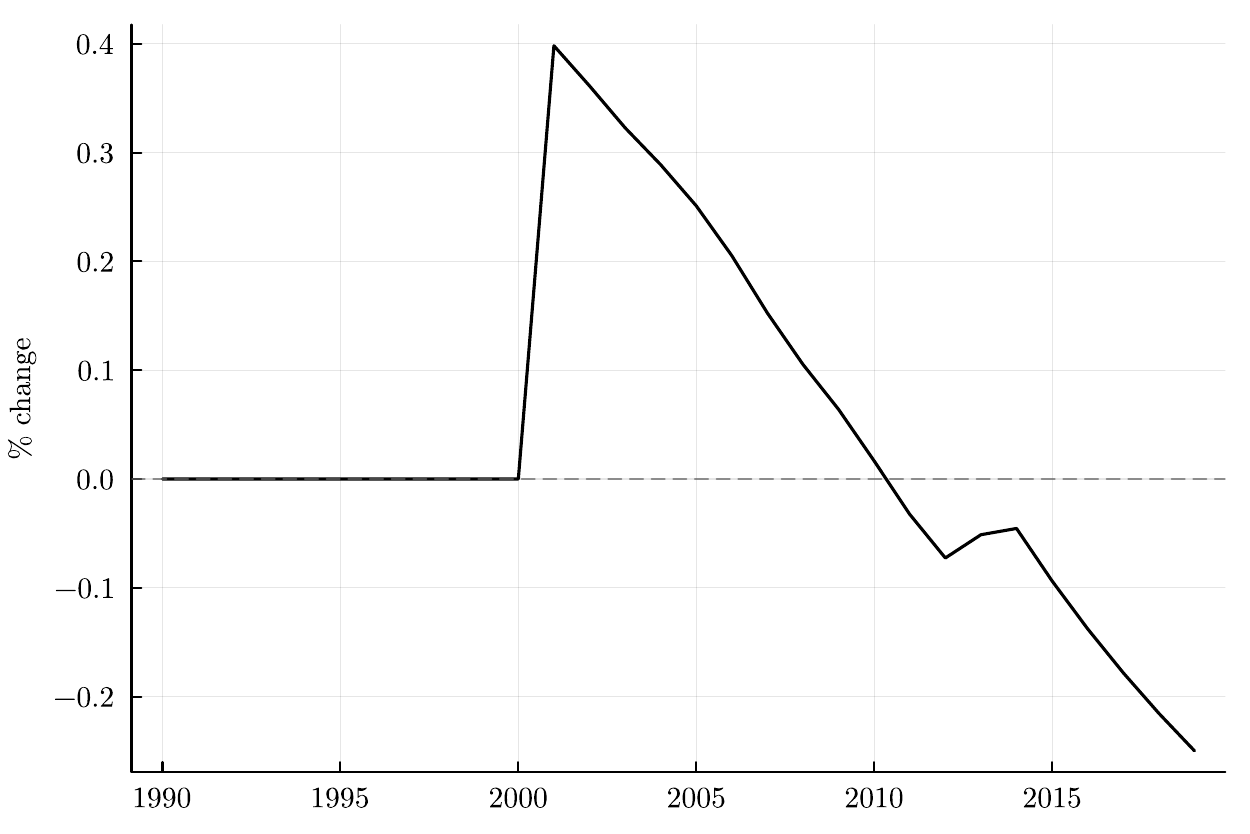} }}    \subfloat[\centering Saving rate]{{\includegraphics[width=0.45\textwidth]{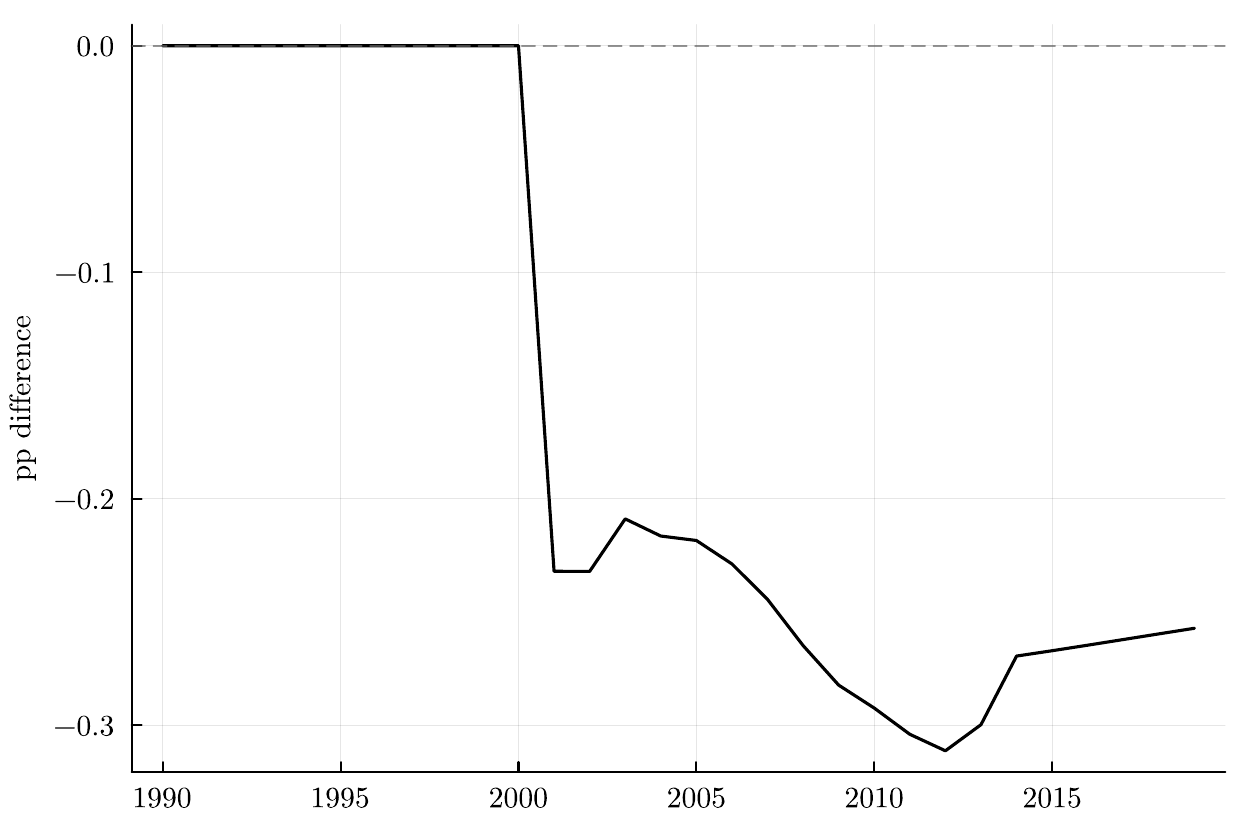} }}    \quad
    \subfloat[\centering Sectoral expenditure share]{{\includegraphics[width=0.45\textwidth]{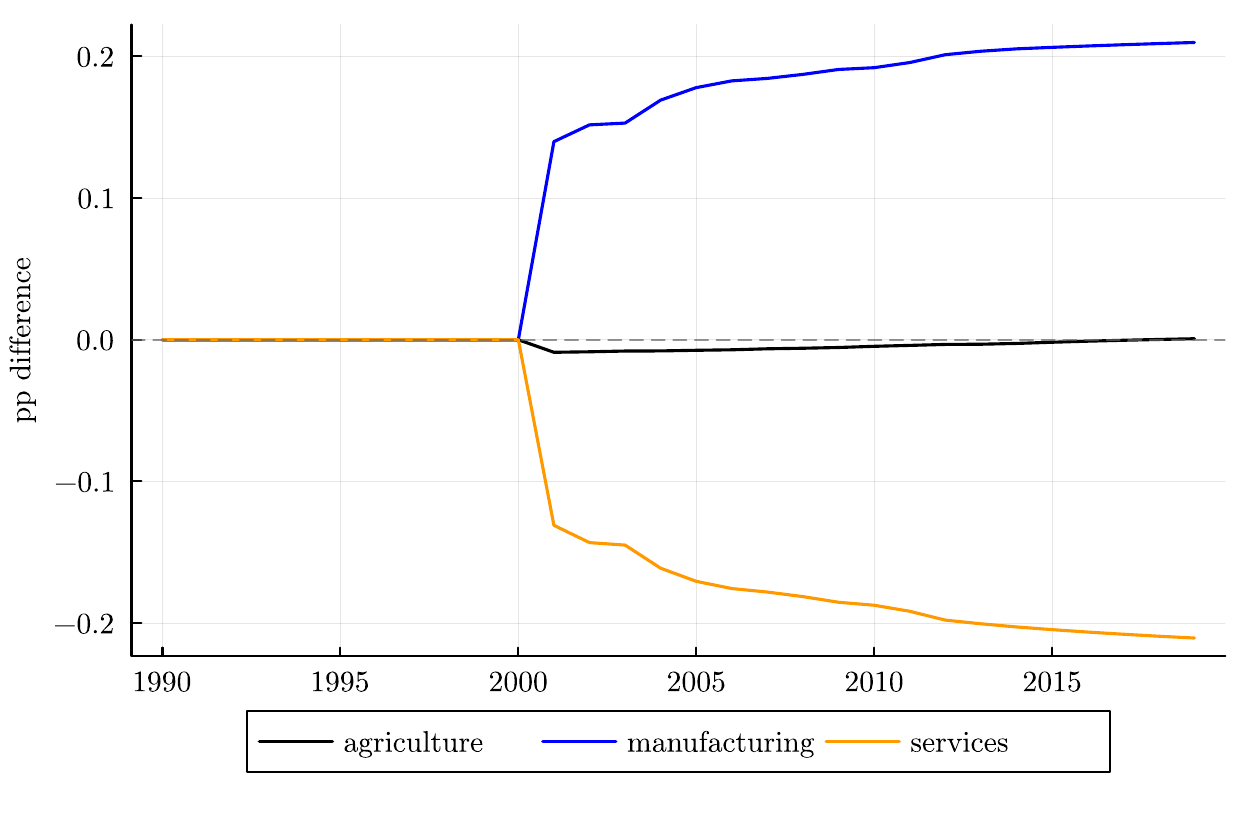} }}    \subfloat[\centering Sectoral value-added share]{{\includegraphics[width=0.45\textwidth]{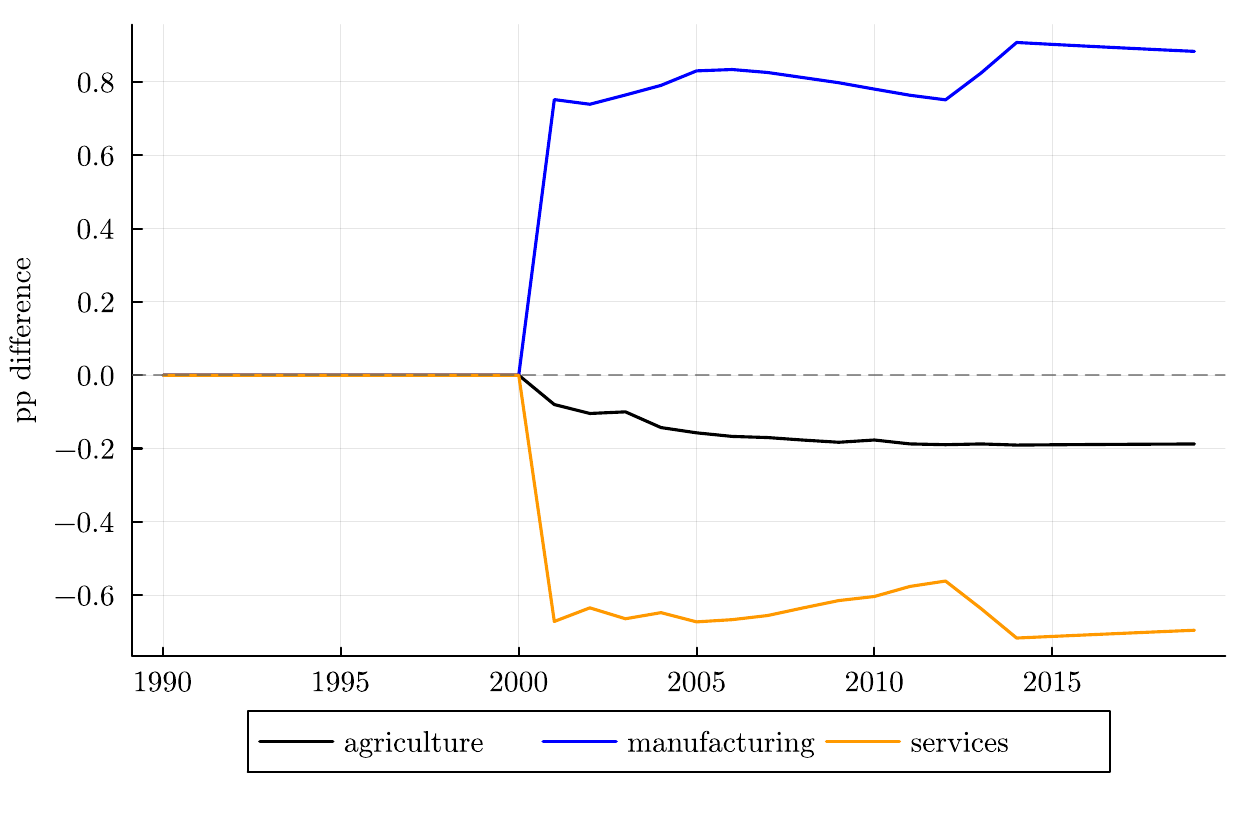} }}    {\footnotesize
\begin{flushleft}
    \textit{Note}: Each panel shows the impacts of a counterfactual trade war between the U.S. and the other countries since 2001 on the transition paths in the U.S. For real per capita consumption, the vertical axis represents the percent change from the baseline to the counterfactual equilibrium. For the other variables, the vertical axis measures the percentage point change from the baseline.
\end{flushleft} }
\end{figure}

Figure~\ref{fig: newtradepolicy_7_nh} shows how the U.S. economy evolves when the U.S. and all other countries impose manufacturing tariffs 20 percentage points above their observed levels from 2001 onward. Panel (A) shows the time path of real consumption per capita. In the first year of this counterfactual trade war, real consumption per capita is 0.40 percent higher than in the baseline. However, the gap between the counterfactual and baseline paths narrows rapidly. By 2010, counterfactual real consumption per capita is only 0.02 percent above the baseline level. From 2011 onward, it falls below the baseline, and the gap widens over time.

Panel (B) reports the path of the U.S. saving rate. The counterfactual saving rate is 0.21--0.32 percentage points lower than in the baseline. In general, the differences between the trade-war counterfactual and the baseline are smaller than those between the unilateral U.S. tariff counterfactual and the baseline.

Panel (C) shows the sectoral composition of consumption expenditure. The manufacturing expenditure share is approximately 0.2 percentage points higher than in the baseline. This increase is larger than that under the unilateral U.S. tariff counterfactual, where the corresponding difference is approximately 0.17 percentage points. This is because the U.S. is relatively poorer under the trade-war counterfactual than under the unilateral-tariff counterfactual. Because preferences are nonhomothetic, the decline in relative income leads U.S. households to allocate a larger share of their expenditure to manufactured goods.

Nevertheless, the manufacturing value-added share is only about 0.8 percentage points higher than in the baseline, compared with an increase of approximately one percentage point under the unilateral U.S. tariff counterfactual. Because other countries impose higher tariffs on U.S. manufactured goods, U.S. manufacturing production is weaker under the trade war than under the unilateral-tariff counterfactual.

\clearpage
\subsection{Optimal Tariff}

\begin{figure}[htb!]    \centering
    \caption{Optimal tariffs under different preferences: Fixed $\Omega$ and $\zeta$}
    \label{fig:optimal-tariffs-omega1-zeta1}        \includegraphics[width=0.49\textwidth]{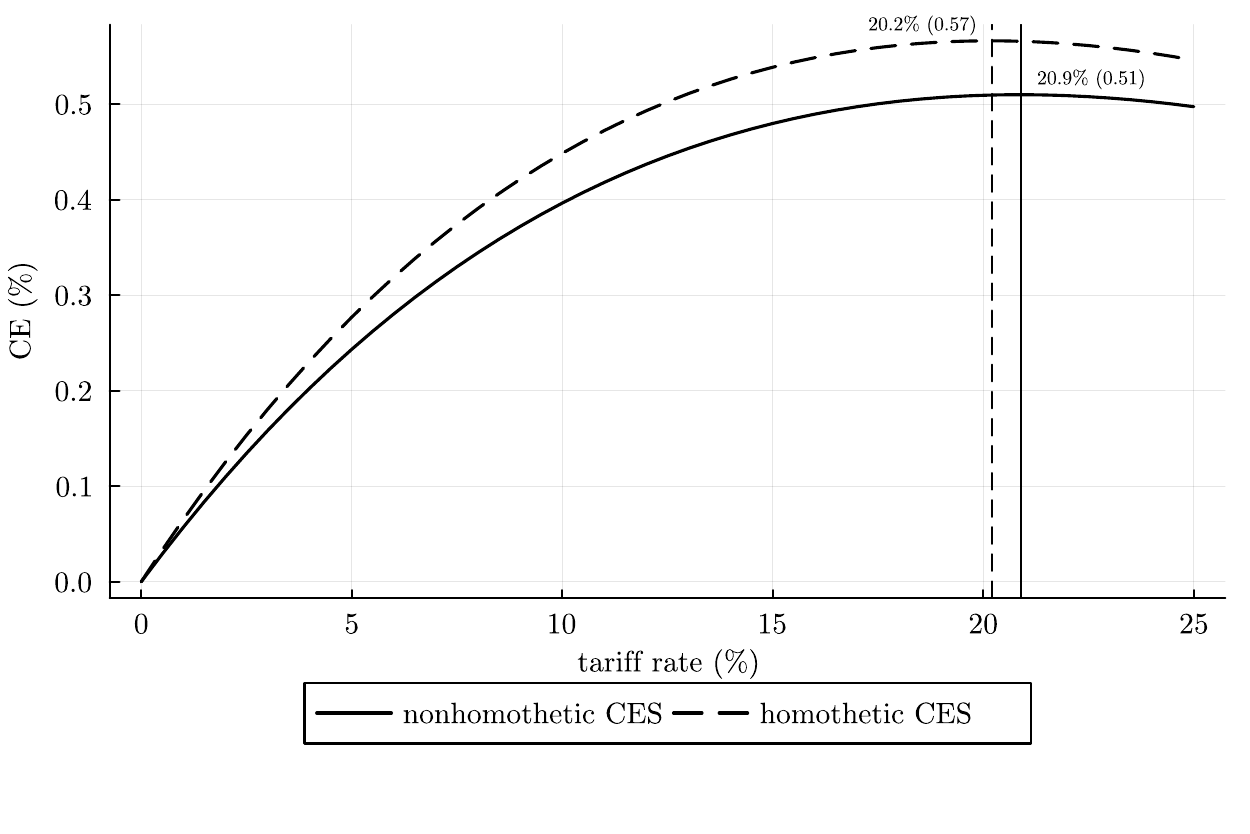}        {\footnotesize
\begin{flushleft}
    \textit{Note}: The figure plots welfare, measured as the consumption-equivalent change relative to a zero time-invariant manufacturing tariff, against alternative time-invariant manufacturing tariff rates for the U.S. $\Omega_{n,t}^j=1/3$ and $\zeta_{n,t}=1$ for all $n$, $t$, and $j$.
\end{flushleft} }
\end{figure}

\

\begin{figure}[htb]    \centering
    \caption{Expenditure share in the U.S. under the optimal tariff}\label{fig: expshare_optimum}    \subfloat[\centering $\Omega$ and $\zeta$ as in the calibrated values]{{\includegraphics[width=0.45\textwidth]{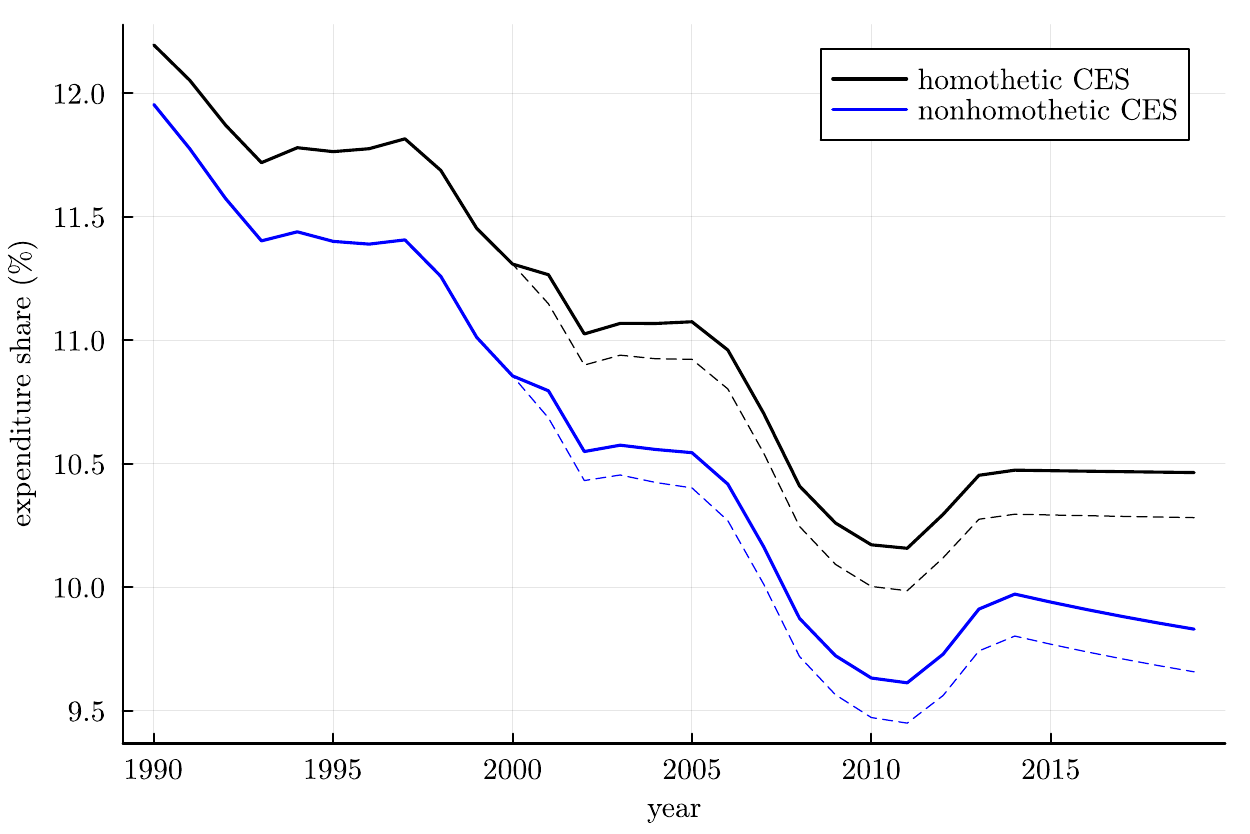} }}    \subfloat[\centering $\Omega$ and $\zeta$ fixed at 1]{{\includegraphics[width=0.45\textwidth]{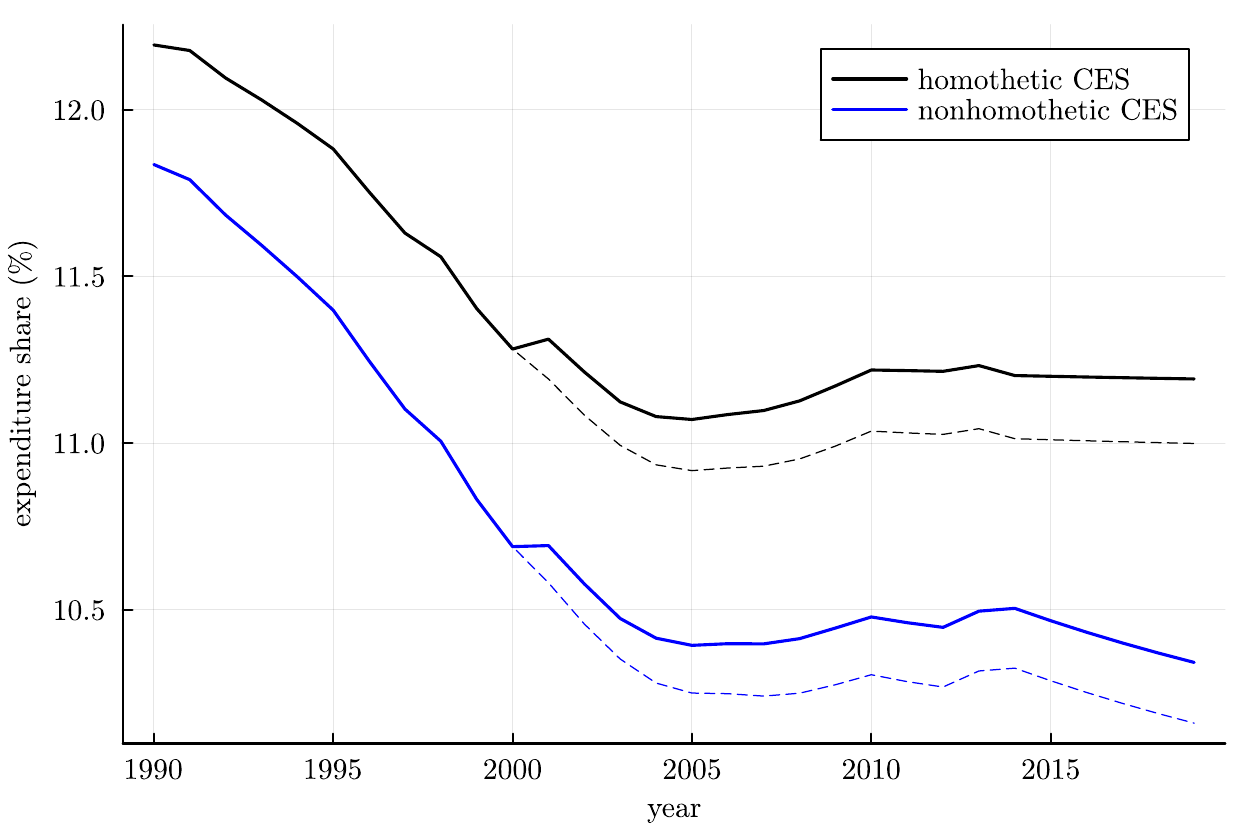} }}    {\footnotesize
\begin{flushleft}
    \textit{Note}: The figure plots the expenditure share $\omega_{n,t}^{j}$ for manufacturing in the U.S. when it imposes the optimal tariffs on manufacturing imports implied by Figures~\ref{fig: optimal tariffs} and \ref{fig:optimal-tariffs-omega1-zeta1}. Black lines are under homothetic CES, and blue lines are under nonhomothetic CES. Solid lines show the expenditure share with tariffs, and dashed lines show the expenditure share in the baseline equilibrium (tariffs as in the data).
\end{flushleft} }
\end{figure}

Figure~\ref{fig: optimal tariffs} in the main text shows the U.S. welfare against various manufacturing tariff rates. We relate this to the expenditure share on manufacturing. However, this is under the sectoral demand shifters $\{\Omega_{n,t}^{j}\}$ calibrated with the data. Moreover, such $\{\Omega_{n,t}^{j}\}$ directly affects sectoral expenditure shares $\{\omega_{n,t}^{j} \}$. Are the lower manufacturing expenditure share and the higher optimal tariff under nonhomothetic preferences than under homothetic preferences just caused by the particular $\{\Omega_{n,t}^{j}\}$ we have calibrated? To answer this question, in Figure~\ref{fig: expshare_optimum}, we redo a similar exercise to Figure~\ref{fig: optimal tariffs} under $\Omega_{n,t}^{j}=1/3$ and $\zeta_{n,t}=1$ for any sector $j$, country $n$, and year $t$. The results do not change substantially. Therefore, the order of the optimal tariffs across different preferences is not just determined by the parameter values of $\{\Omega_{n,t}^{j}\}$ we have assigned.

This is underpinned by Figure~\ref{fig: expshare_optimum}. No matter whether $\{\Omega_{n,t}^{j}\}$ are calibrated with the data or not, the expenditure share on manufacturing at the respective optimal tariffs is higher under homothetic CES preferences than under nonhomothetic CES preferences. Then, the expression in \eqref{ap_eq: optimal tariff} suggests that the optimal tariff under nonhomothetic CES preferences is higher than the one under homothetic CES preferences.

\end{document}